\documentclass[preprint]{elsarticle}
\usepackage{amsmath,amsfonts,amsthm,fullpage}
\usepackage{multirow}
\usepackage{hyperref}
\hypersetup{
    colorlinks=true,
    linkcolor=blue,
    citecolor=blue,
    urlcolor=orange
}
\newtheorem{definition}{Definition}

\title{A hierarchical splines-based $h$-adaptive isogeometric solver for all-electron Kohn--Sham equation\tnoteref{t1}}
\tnotetext[t1]{Submitted date: \today}
\author[JLU]{Tao Wang}
\ead{wangt21@mails.jlu.edu.cn}
\author[GDUT]{Yang Kuang \corref{cor1}}
\ead{ykuang@gdut.edu.cn}
\cortext[cor1]{Corresponding author. Email:{\tt ykuang@gdut.edu.cn}}
\author[JLU]{Ran Zhang}
\ead{zhangran@jlu.edu.cn}
\author[UM,UMZhuhai]{Guanghui Hu}
\ead{garyhu@um.edu.mo}
\affiliation[JLU]{organization={School of Mathematics},
                addressline={Jilin University},
                city={Changchun},
                country={China}}
\affiliation[GDUT]{organization={School of Mathematics and Statistics \& Center for Mathematics and Interdisciplinary Science (CMIS)},
                addressline={Guangdong University of Technology},
                country={China}}
\affiliation[UM]{organization={State Key Laboratory of Internet of Things for Smart City and Department of Mathematics},
                addressline={University of Macau},
                city={Macao SAR},
                country={China}}
\affiliation[UMZhuhai]{organization={Zhuhai UM Science and Technology Research Institute},
                city={Zhuhai},
                country={China}}
\date{\today}

\usepackage[ruled,linesnumbered]{algorithm2e}
\SetKwRepeat{Do}{do}{while}
\usepackage{array}

\usepackage{subcaption}
\usepackage{cleveref} 
\usepackage{tikz}
\usetikzlibrary{calc}
\usepackage{adjustbox}

\begin{document}

\begin{abstract}
In this paper, a novel $h$-adaptive isogeometric solver utilizing high-order hierarchical splines is proposed to solve the all-electron Kohn--Sham equation. In virtue of the smooth nature of Kohn--Sham wavefunctions across the domain, except at the nuclear positions, high-order globally regular basis functions such as B-splines are well suited for achieving high accuracy. To further handle the singularities in the external potential at the nuclear positions, an $h$-adaptive framework based on the hierarchical splines is presented with a specially designed residual-type error indicator, allowing for different resolutions on the domain. The generalized eigenvalue problem raising from the discretized Kohn--Sham equation is effectively solved by the locally optimal block preconditioned conjugate gradient (LOBPCG) method with an elliptic preconditioner, and it is found that the eigensolver's convergence is independent of the spline basis order. A series of numerical experiments confirm the effectiveness of the $h$-adaptive framework, with a notable experiment that the numerical accuracy $10^{-3} \mathrm{~Hartree/particle}$ in the all-electron simulation of a methane molecule is achieved using only $6355$ degrees of freedom, demonstrating the competitiveness of our solver for the all-electron Kohn--Sham equation.
\end{abstract}

\maketitle

\section{Introduction}
Kohn--Sham density functional theory (KSDFT) \cite{fiolhais2003primerDFT} serves as a fundamental tool in quantum physics and computational chemistry, providing an efficient framework for investigating the physical and chemical properties of many-body systems. The core concept of KSDFT is to represent the many-body problem in terms of the electron density instead of the many-body wavefunction, so as to resolve the curse of dimensionality. The Kohn--Sham (KS) equation provides a computable model for KSDFT, introduced by Kohn and Sham in 1965 \cite{kohn1965self}. In the KS equation, a system with non-interacting particles influenced by an effective potential that incorporates both external potentials and electron-electron interactions is established, such that its ground state electron density is equivalent to that of the real many-body problem. Compared to the pseudopotential methods, which approximate core electron contributions in the KS equation, all-electron calculations account for every electron in the system \cite{fiolhais2003primerDFT}, making them crucial for precisely describing the electronic structures, particularly in extreme conditions such as high pressure or temperature, where pseudopotential methods may fail \cite{oganov2003allpseudo, xiao2010firstpseudo, zhang2019equation}. As the analytical solution is typically unattainable, the numerical solution of the KS equation is essential for understanding the electronic structure of materials and serves as a foundation for further theoretical and experimental investigations \cite{lin2019numerical}. 

To numerically solve the all-electron KS equation, a variety of methods had been developed, including the augmented plane-wave expansion methods \cite{kresse1996DFTplanewave,mokrousov2006abplanewave,haule2024allplanewave}, linear combination of atomic orbitals \cite{andzelm1991dgaussLCAO,qu2022dftLCAO}, and real-space methods  \cite{fattebert1999DFTFDM,fiolhais2003primerDFT,pask2005DFTfiniteelememt}, such as the finite difference method, the finite element method, the finite volume method, the discontinuous Galerkin method, the spectral method, etc. Among these methods, finite element methods had been favored in recent decades due to their flexibility in handling non-periodic boundary conditions, their use of complete local basis functions, and their ability to allow spatially adaptive resolution. In the context of finite element methods, low-order and low-regularity basis functions, such as the standard linear Lagrange polynomial basis \cite{bao2012hadaptiveKS}, have been widely adopted for their ability to produce highly sparse linear systems that are computationally efficient to solve. Moreover, over the past decade, the significance of higher-order elements in achieving faster convergence rates—and thereby greater efficiency compared to lower-order elements—has become increasingly evident \cite{fang2012highorderKSkohn,motamarri2013higherorderKS,davydov2016adaptivehighorderKS,temizer2021radial,zhan2023spectralKS}.

Notably, based on the analytical wavefunction of the hydrogen atom and numerical observations for other atoms, the wavefunctions in the all-electron KS equation can be regarded as smooth except at the nuclear positions. Consequently, a high-order method with global regularity emerges as a promising approach for accurately capturing the behavior of wavefunctions in all-electron calculations. However, traditional finite element bases, including higher-order bases, have faced challenges in constructing basis spaces with global regularity. This issue persisted until the work by \cite{hu2023construction}, which made it possible to construct $C^p$ conforming finite element spaces of arbitrary order $p$. In contrast, such continuity properties are inherent to B-splines. Motivated by this, in this paper, we concentrate on constructing a high-order isogeometric method with a local basis set tailored for the Kohn--Sham equation. Among the range of basis sets available in the market, the B-spline basis is selected as the focal point of our research based on several key considerations. In addition to the previously mentioned global regularity of B-splines, studies have shown that they can offer higher accuracy per degree of freedom compared to classical high-order Lagrange elements, which exhibit discontinuous derivatives \cite{masud2018bsplinesbetter,temizer2020nurbsfordft}. This advantage is partially attributed to the unique $k$-refinement in the isogeometric analysis introduced by Huges \cite{hughes2005isogeometric}, which enables the construction of the spline space with fewer degrees of freedom and offers rigorous theoretical foundations.
There have been relevant studies on using B-splines and non-uniform rational B-splines (NURBS) in KSDFT, providing valuable references for our work.

A quality mesh is crucial for constructing an effective approximation space in Kohn--Sham density functional theory. In pseudopotential calculations, the smooth nature of the wavefunctions and potential allowed for the use of the uniform mesh, which was computationally feasible with the NURBS basis \cite{romanowski2011bsplineforallelectron, masud2012bsplinefordft,cimrman2018convergence,cimrman2018isogeometricfordft,masud2018bsplinesbetter}. However, in all-electron calculations, the presence of singularities made a uniform mesh unaffordable, as it would require a prohibitively large number of grid points to achieve high accuracy. To address this issue, a logarithmic mesh was employed in \cite{romanowski2007b} for one-dimensional atomic calculations and in \cite{kuang2024nurbsdft} for three-dimensional calculations. Additionally, \cite{temizer2020nurbsfordft} introduced a specialized mesh with a uniform inner region and a radial outer region. While these mesh generation strategies successfully created high-quality approximation spaces for the systems studied, achieving further accuracy generally required global refinement. Furthermore, their limited flexibility in controlling the mesh made them less suitable for dynamic simulations. For these reasons, adaptive mesh strategies are in demand in the solution of the Kohn--Sham equation using the NURBS basis.

There are three types of adaptive mesh strategies: the $h$-adaptive method, which allows local mesh refinement and coarsening; the $r$-adaptive method, which facilitates redistribution of the mesh grid; and the $p$-adaptive method, which adjusts the polynomial degree. The $r$-adaptive method using NURBS basis functions has been explored in \cite{wang2024MMIGM} for all-electron calculations, while studies employing $h$- or $p$-adaptive strategies with B-splines in KSDFT remain limited. The limitations of the $h$-adaptive strategy are primarily due to the tensor-product structure of B-splines in multidimensional space \cite{hughes2005isogeometric}, which restricts the local mesh refinement. For the $p$-adaptive strategy, the constraints arise from the construction of B-splines, which inherently produce the basis functions with the same order and lack the flexibility to enrich varying polynomial orders.

In this work, we focus on the $h$-adaptive method with B-splines for KSDFT. Indeed, there had been a series of suitable extensions for B-splines to facilitate the capability of local refinement. By allowing the hanging nodes, there were some developments based on T-splines such as the analysis-suitable T-splines (dual-compatible T-splines) \cite{da2012analysis,scott2012local}, and polynomial splines over T-meshes \cite{deng2008polynomial}. In \cite{buffa2016adaptive}, an adaptive isogeometric method (AIGM) based on hierarchical B-splines (HB-splines) was introduced with the admissibility of the hierarchical mesh configuration. The comprehensive analysis including the complexity and optimal convergence rates of AIGM was given in \cite{buffa2016adaptive,buffa2016complexity,buffa2017adaptiveall}. In engineering research, HB-splines and analysis-suitable T-splines, which possessed the capability of local refinement, were widely adopted in adaptive isogeometric methods due to their thorough theoretical analysis on optimal convergence of adaptive algorithms \cite{buffa2022mathematicalthb}. Furthermore, to the best of our knowledge, there was limited literature on using splines with local refinement capabilities to solve the eigenvalue problems, especially for the all-electron KS equation. Given the ability of HB-splines to preserve regularity and the comprehensive convergence analysis of HB-splines in the field of adaptive algorithms, we chose HB-splines as the basis functions in this work to explore their performance in all-electron calculations. Building on the isogeometric library GeoPDEs \cite{de2011geopdes,garau2018algorithmsGeoPDEs}, we aimed to extend the framework to enhance the solver for KSDFT, thereby improving the accuracy of numerical simulations in all-electron KS calculations.

In this paper, based on the hierarchical splines, a novel $h$-adaptive isogeometric solver is proposed for simulating the all-electron KS equation towards high computational accuracy. The solver comprises four modules. In the first \textit{Solve} module, the self-consistent field (SCF) iteration is employed to numerically resolve the nonlinearity of the Kohn--Sham equation. Within this iteration, the Kohn--Sham equation is restricted in a closed computational domain and discretized using the hierarchical splines function space on the corresponding hierarchical mesh. The homogeneous Dirichlet boundary condition is applied to the Kohn--Sham equation. Simultaneously, a Poisson equation is solved to obtain the Hartree potential with an inhomogeneous boundary condition generated by the multipole expansion method. The wavefunctions are obtained by solving the generalized eigenvalue problem by the eigensolver LOBPCG with an elliptic preconditioner, where the convergence is found to be independent of the basis order. The electron density is updated via a simple linear mixing scheme. In the second \textit{Estimate} module, the residual-type error indicator is specially designed to assess the residual of the Kohn--Sham equation on the hierarchical mesh. We highlight that the jump term is ignored due to the global regularity of hierarchical splines. For the final \textit{Mark} and \textit{Refine} modules, the maximum strategy is adopted for marking the cells guaranteed by the indicator in the \textit{Estimate} module. The marked cells are subsequently refined to enhance mesh quality. We note that the initial input for the wavefunctions on the refined hierarchical mesh is obtained by projecting the wavefunctions from the previous mesh in the \textit{Refine} module. As a result, an acceleration to the convergence of the SCF iteration in the \textit{Solve} module is achieved. The capability to achieve high accuracy is validated through a range of numerical experiments, spanning from one-dimensional atomic calculations to three-dimensional molecular simulations.

This paper is organized as follows. In Section \ref{sec:KSequation_and_discretization}, the Kohn--Sham equation is demonstrated, followed by the hierarchical splines and its isogeometric discretization using the hierarchical spline function space. An $h$-adaptive isogeometric solver, along with its four modules and the implementation details, is mainly demonstrated in Section \ref{sec:hadaptive_alg}. Then in Section \ref{sec:numerical_experiments}, various numerical experiments are examined for the effectiveness of the $h$-adaptive isogeometric solver. Finally, the paper ends up in Section \ref{sec:concluding_remarks}.

\section{The Kohn--Sham density functional theory and its isogeometric discretization}\label{sec:KSequation_and_discretization}
\subsection{The Kohn--Sham equation}
The Kohn--Sham equation for a $N_{\mathrm{ele}}$-electron system is summarized as the following eigenvalue problem
\begin{equation}\label{KSeq_infty}
\begin{cases}{H} \psi_i(\mathbf{r})=\varepsilon_i \psi_i(\mathbf{r}), & i=1,2, \ldots, N_{\mathrm{occ}}, \\ \int_{\mathbb{R}^3} \psi_i \psi_{i^{\prime}} \mathrm{d} \mathbf{r}=\delta_{i i^{\prime}}, & i, i^{\prime}=1,2, \ldots, N_{\mathrm{occ}},\end{cases}
\end{equation}
where $N_{\mathrm{occ}}=\operatorname{ceil}(N_\mathrm{ele}/2)$ describes the number of occupied orbitals, $\psi_i(\mathbf{r})$ is the $i$-th wavefunction, and $\varepsilon_i$ is the corresponding eigenvalue, $\delta_{i i^\prime}$ is the Kronecker delta, and ${H}$ denotes the Hamiltonian operator, consisting of the following four terms
\begin{equation}
    \label{Hamiltonian_strong_operator}
    {H}([\rho] ; \mathbf{r})=-\frac{1}{2} \nabla_{\mathbf{r}}^2+V_{\mathrm{ext}}(\mathbf{r})+V_{\mathrm{Har}}([\rho] ; \mathbf{r})+V_{\mathrm{xc}}([\rho] ; \mathbf{r}),
\end{equation}
where $\rho(\mathbf{r})$ stands for the electron density defined by
\begin{equation}\label{rhodensity}
\rho(\mathbf{r}) = \sum_{i=1}^{N_{\mathrm{occ}}} f_i\left|\psi_i(\mathbf{r})\right|^2,
\end{equation}
in which $f_i$ is the occupation number. Note that the atomic units are adopted hereafter. The symbol $V([\rho];\mathbf{r})$ implies that $V$ is a functional of the electron density $\rho(\mathbf{r})$. Specifically, in \eqref{Hamiltonian_strong_operator}, the first term $-\frac{1}{2}\nabla_\mathbf{r}^2$ is the kinetic operator. The second term $V_{\mathrm{ext}}(\mathbf{r})$ is the Coulomb external potential, describing the attraction between the nuclei and the electron. In the all-electron calculation, $V_{\mathrm{ext}}$ takes the following form
\begin{equation}
    \label{externalpotential}
    V_{\mathrm{ext}}(\mathbf{r}) = -\sum_{j=1}^M \frac{Z_j}{\left|\mathbf{r}-\mathbf{R}_j\right|},
\end{equation}
where $M$ is the number of nuclei in the system, $Z_j$ is the charge of the $j$-th nuclei, and $\mathbf{R}_j$ stands for the position of the $j$-th nuclei. The third term $V_{\mathrm{Har}}(\mathbf{r})$ is the Hartree potential which illustrates the Coulomb repulsion among the electrons in the system. The specific expression is denoted as
\begin{equation}
    \label{Hartreepotentialintegral}
    V_{\mathrm{Har}}(\mathbf{r}) = \int_{\mathbb{R}^3} \frac{\rho(\mathbf{r})}{\left|\mathbf{r}-\mathbf{r}^\prime\right|} \mathrm{~d}\mathbf{r}^\prime.
\end{equation}
Suppose $N$ is the number of degrees of freedom associated with the scale of the numerical method, directly solving the above integral would result in the inefficient $\mathcal{O}(N^2)$ computational complexity. An alternative way for obtaining the Hartree potential $V_{\mathrm{Har}}(\mathbf{r})$ is to solve the following Poisson equation
\begin{equation}\label{Hartree_Poisson_intfy}
\begin{cases} &-\nabla^2 V_{\mathrm{Har}}(\mathbf{r}) = 4\pi\rho(\mathbf{r}), \\ 
& V_{\mathrm{Har}}(\mathbf{r})  = 0, \quad |\mathbf{r}|\rightarrow \infty.
\end{cases}
\end{equation}
Compared with the $\mathcal{O}(N^2)$ complexity, we can reduce the complexity into $\mathcal{O}(N)$ by solving \eqref{Hartree_Poisson_intfy} with a robust and efficient solver such as the algebraic multigrid solver.

Finally, the last term $V_{\mathrm{xc}}$ describes the exchange-correlation potential, caused by the Pauli exclusion principle and other non-classical Coulomb interactions. It is worth noting that there is no analytical expression for the exchange-correlation potential. Consequently, an approximation is needed, such as the local density approximation (LDA) and the generalized gradient approximation (GGA). In this paper, we adopt LDA which can be obtained from the library \texttt{Libxc} \cite{marques2012libxc} in practical simulations. 

\subsection{Hierarchical splines}
\label{sec:HBspace}
In this section, we will introduce the definition of hierarchical splines. Following \cite{garau2018algorithmsGeoPDEs}, the parametric domain for hierarchical splines will be chosen as $\widehat{\Omega} = [0,1]^d\subset \mathbb{R}^d$, where $d$ is the dimension of the physical domain.
\subsubsection{Nested sequence of tensor-product spline spaces}
Based on the construction of B-spline space, we consider a nested sequence $\{\widehat{\mathcal{S}}_\ell\}_{\ell=0}^{N-1}$ of $N$ tensor-product $d$-variate spline spaces such that
\begin{equation}
    \label{nestedBspline}
    \widehat{\mathcal{S}}_0\subset \widehat{\mathcal{S}}_1\subset \cdots \subset \widehat{\mathcal{S}}_{N-1},
\end{equation}
which are defined by their corresponding knot vectors $\Xi_\ell$ in $\widehat{\Omega}$. The knot vector $\Xi_0$ is assumed to be open such that the B-spline functions are standard with the attractive properties of local support, non-negativity, linearly independent, partition of unity and global-regularity, see e.g. \cite{piegl1996nurbs,hughes2005isogeometric}. Furthermore, for each level $\ell=0,1,\cdots,N-1$, we denote the B-spline basis as $\widehat{\mathcal{B}}_\ell:=\{\widehat{\beta}_{i,\ell}~|~ i=1,2,\cdots,n_\ell\}$, where $n_\ell$ is the dimension of the space $\widehat{\mathcal{S}}_\ell$. Based on the nested sequence \eqref{nestedBspline}, for the two adjacent levels, there exists a \textbf{two-scale} relation such that any basis $\widehat{\beta}_{i,\ell}$ could be represented as the linear combinations of B-splines basis of level $\ell+1$, which means that
\begin{equation}
    \label{twoscalerelation}
    \widehat{\beta}_{i,\ell} = \sum_{k=1}^{N_{\ell+1}} c_{k,\ell+1}(\widehat{\beta}_{i,\ell}) \widehat{\beta}_{k,\ell+1},
\end{equation}
where all of the coefficients are non-negative. 

On the level $\ell$, the Cartesian mesh is denoted by $\widehat{\mathcal{G}}_\ell$ and any $\widehat{Q}\in \widehat{\mathcal{G}}_\ell$ is called as a cell at level $\ell$. Based on the assumption in \cite{garau2018algorithmsGeoPDEs}, all the cells are closed sets in $\widehat{\Omega}$.

\subsubsection{Hierarchical B-splines}
We follow \cite{kunoth2018foundationsTHB, garau2018algorithmsGeoPDEs, bracco2019adaptive} to illustrate the definition of hierarchical B-splines.
\begin{definition}[Hierarchy of subdomains]
    \label{hierarchysubdomain}
    The set $\widehat{\mathbf{\Omega}}_N:=\{\widehat{\Omega}_0,\widehat{\Omega}_1,\cdots,\widehat{\Omega}_N\}$ is said to be a hierarchy of subdomains of depth $N$ if
\begin{equation}
    \label{nestedsubdomain}
    \widehat{\Omega} = \widehat{\Omega}_0 \supset \widehat{\Omega}_1 \supset \cdots \supset \widehat{\Omega}_{N-1}\supset \widehat{\Omega}_N = \emptyset,
\end{equation}
and each subdomain $\widehat{\Omega}_\ell$ is the union of cells of level $\ell-1$.
\end{definition}
Then, an underlying hierarchical mesh $\widehat{\mathcal{Q}}$ associated with the hierarchy of subdomains $\mathbf{\widehat{\Omega}}_N$ is denoted by
$$
\widehat{\mathcal{Q}}:=\bigcup_{\ell=0}^{N-1}\widehat{\mathcal{Q}}_\ell,\quad 
\widehat{\mathcal{Q}}_\ell = \left\{\widehat{Q}\in \widehat{\mathcal{G}}_\ell:\widehat{Q}\subset \widehat{\Omega}_\ell\wedge \widehat{Q}\not\subset \widehat{\Omega}_{\ell+1}\right\}.
$$

In general, we say $\widehat{Q}$ is an active cell if $\widehat{Q}\in \widehat{\mathcal{Q}}$, and an active cell of level $\ell$ if $\widehat{Q}\in \widehat{\mathcal{Q}}_\ell$.
\begin{definition}[Hierarchical B-splines]
    \label{def:hierarchicalBspline}
    Suppose that $\{\widehat{\mathcal{S}}_\ell\}_{\ell = 0}^N$ is the nested sequence of space \eqref{nestedBspline} with B-spline basis $\{\widehat{\mathcal{B}}_\ell\}_{\ell=1}^N$, and $\mathbf{\widehat{\Omega}}_N=\{\widehat{\Omega}_0,\widehat{\Omega}_1,\cdots,\widehat{\Omega}_N\}$ is a hierarchy of subdomains with depth $N$. The Hierarchical B-spline (HB-spline) basis $\widehat{\mathcal{H}} :=  \widehat{\mathcal{H}}_{N-1}$ is defined by the following recursive algorithm:
    \begin{equation*}
\left\{\begin{aligned}
\widehat{\mathcal{H}}_0:= & \{\widehat{\beta}\in\widehat{\mathcal{B}}_0: \operatorname{supp}\widehat{\beta}\not\subset \widehat{\Omega}_0\}, \\
\widehat{\mathcal{H}}_{\ell+1}:= & \left\{\widehat{\beta} \in \widehat{\mathcal{H}}_{\ell}: \operatorname{supp} \widehat{\beta} \not \subset \widehat{\Omega}_{\ell+1}\right\} \cup \\
& \left\{\widehat{\beta} \in \widehat{\mathcal{B}}_{\ell+1} : \operatorname{supp} \widehat{\beta} \subset \widehat{\Omega}_{\ell+1}\right\}, \quad \ell=0, \ldots, N-2 .
\end{aligned}\right.
\end{equation*}
\end{definition}

\begin{figure}[h]
    \centering
    \begin{tabular}{cc}
         \begin{minipage}{6cm}
            \centering
            \includegraphics[width=1\linewidth]{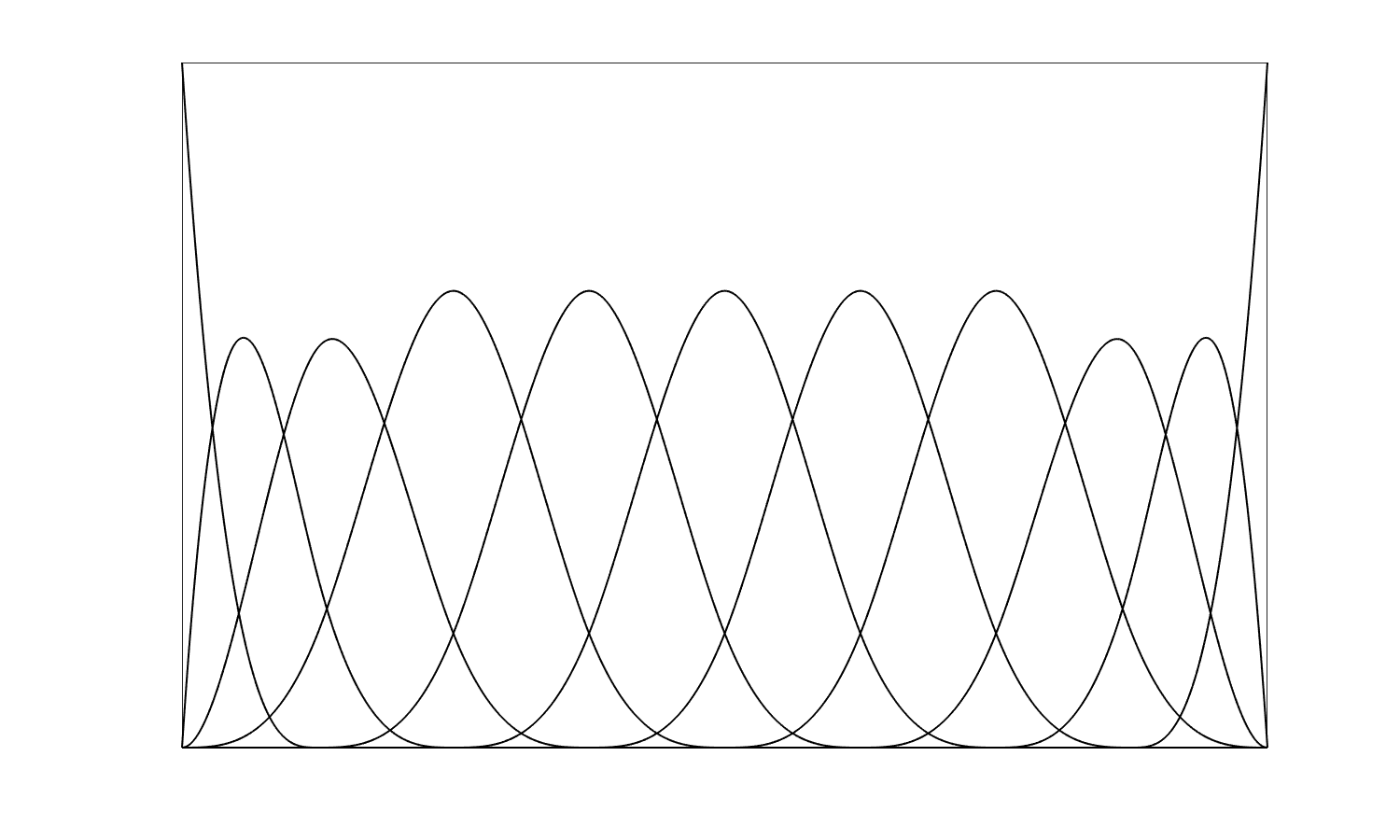}
            \centerline{(a1)~ $\widehat{\mathcal{B}}_0$}
        \end{minipage} & 
        \begin{minipage}{6cm}
            \centering
            \includegraphics[width=1\linewidth]{figure/THB/HB1.png}
            \centerline{(b1)~ $\widehat{\mathcal{H}}_0$}
        \end{minipage} \\
         \begin{minipage}{6cm}
            \centering
            \includegraphics[width=1\linewidth]{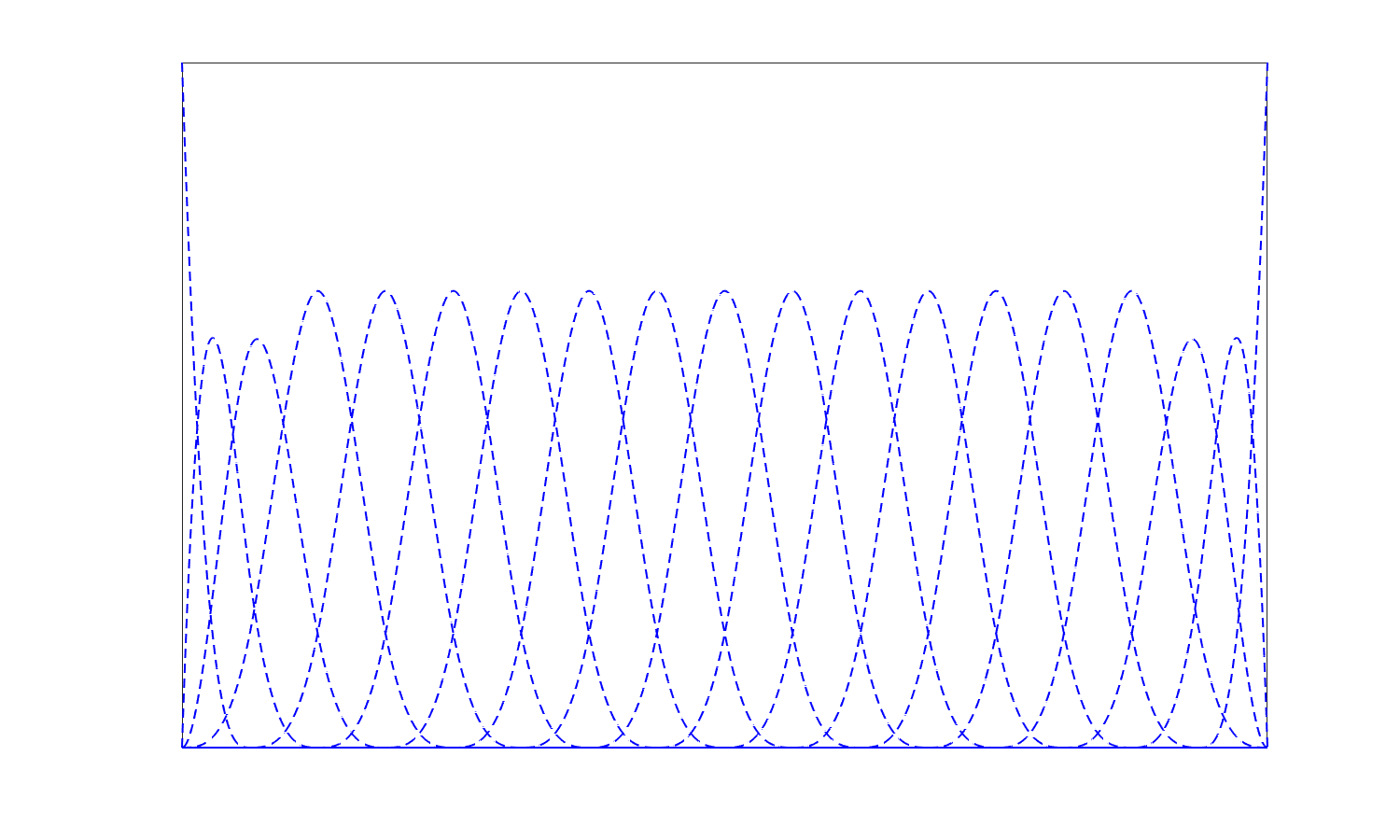}
            \centerline{(a2)~ $\widehat{\mathcal{B}}_1$}
        \end{minipage} & 
        \begin{minipage}{6cm}
            \centering
            \includegraphics[width=1\linewidth]{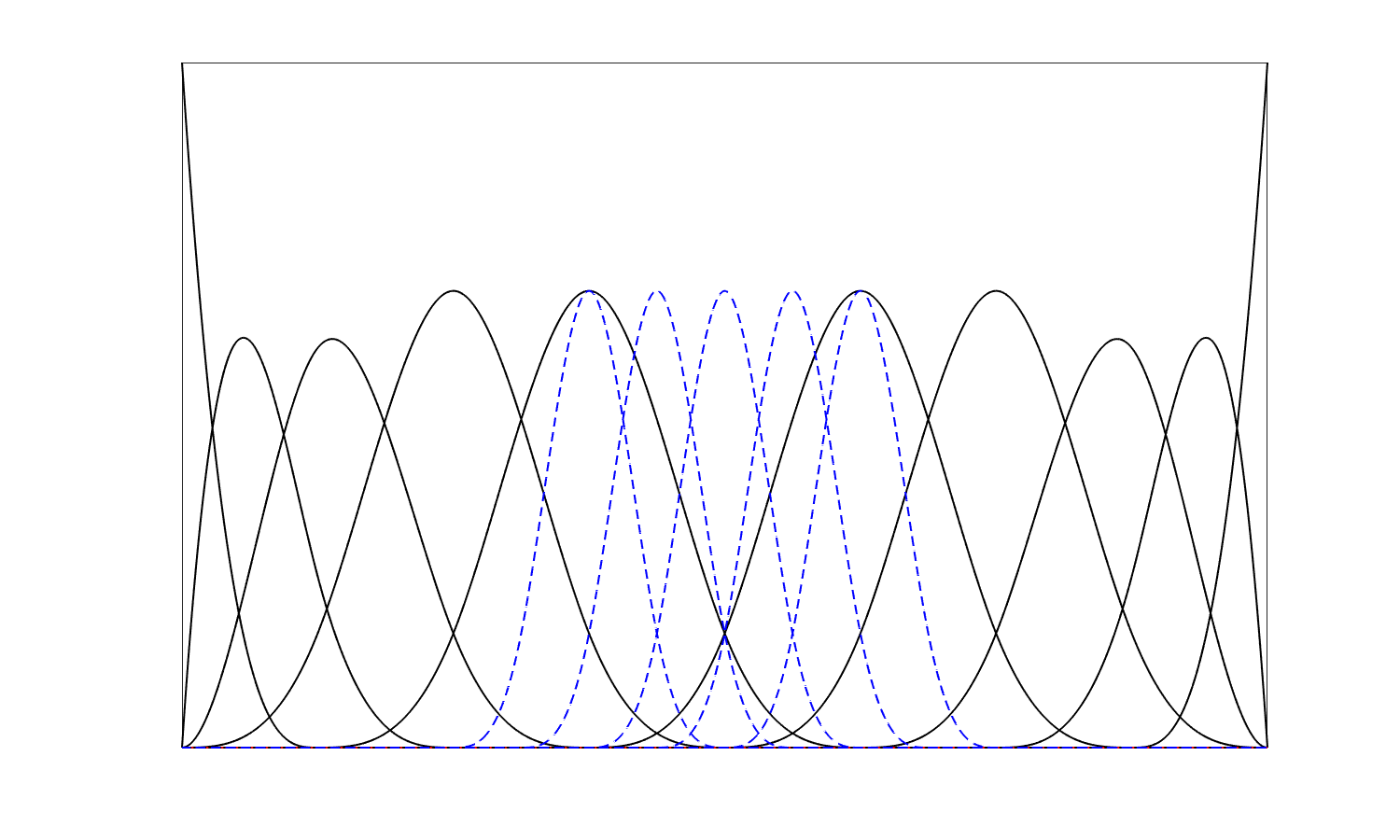}
            \centerline{(b2)~ $\widehat{\mathcal{H}}_1$}
        \end{minipage} \\
         \begin{minipage}{6cm}
            \centering
            \includegraphics[width=1\linewidth]{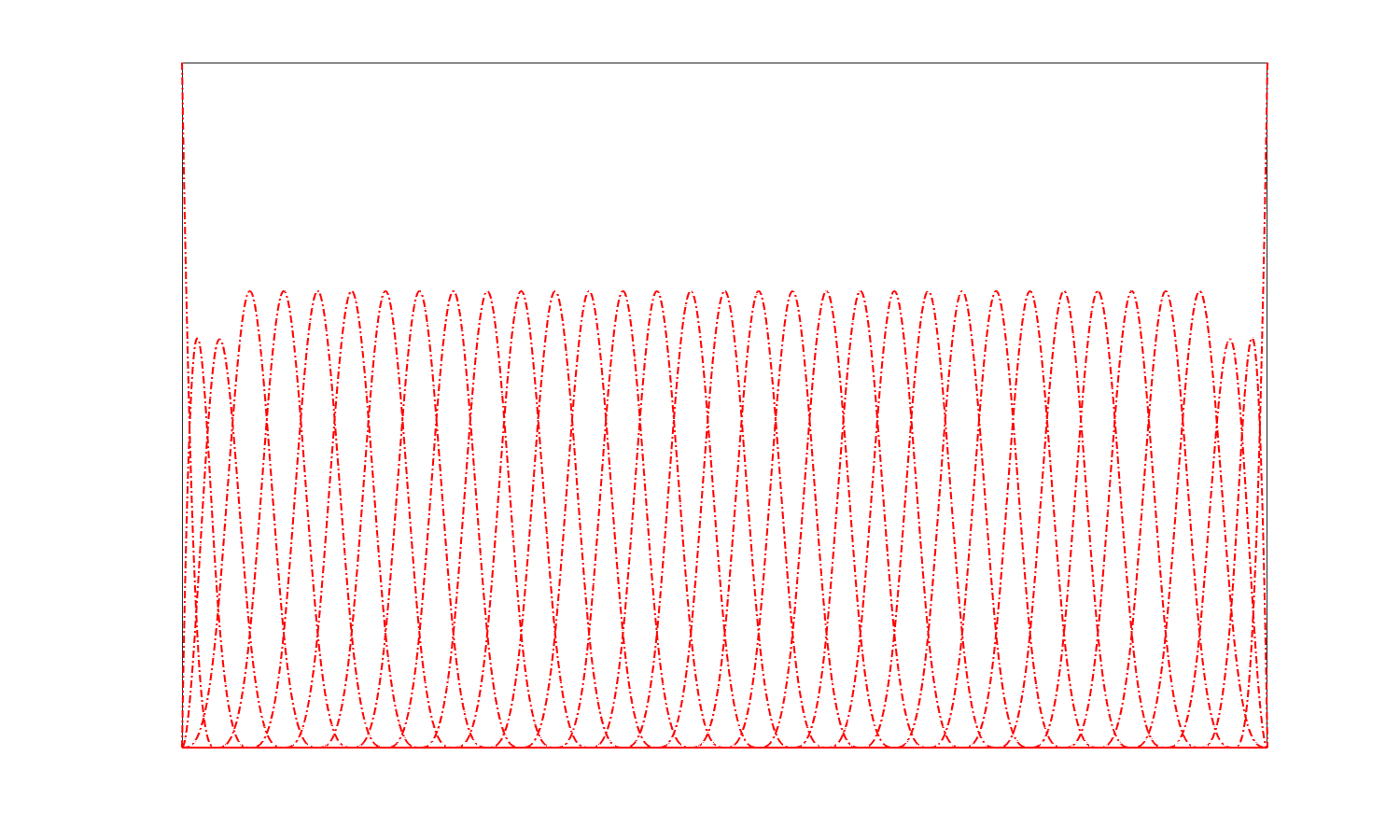}
            \centerline{(a3)~ $\widehat{\mathcal{B}}_2$}
        \end{minipage} & 
        \begin{minipage}{6cm}
            \centering
            \includegraphics[width=1\linewidth]{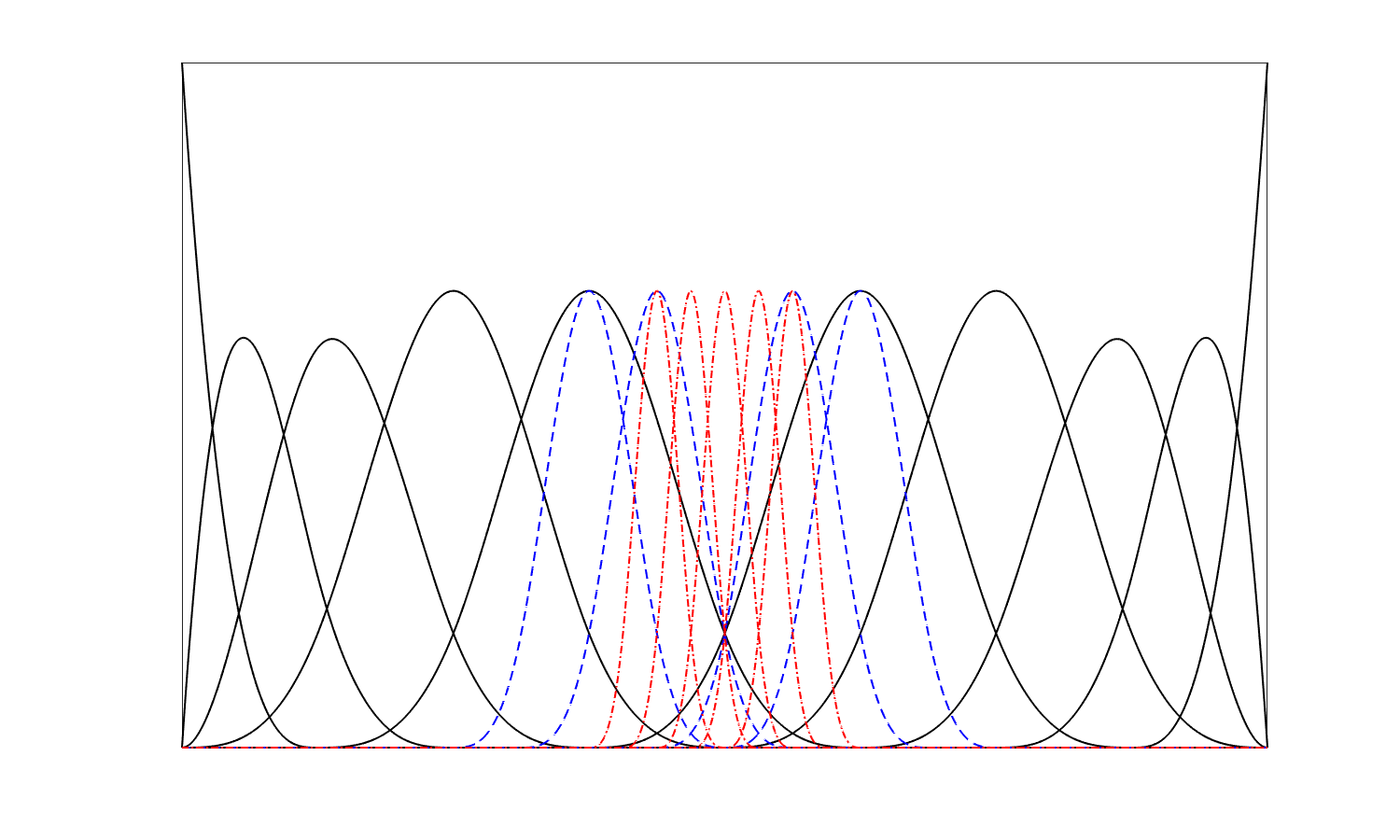}
            \centerline{(b3)~ $\widehat{\mathcal{H}}_2$}
        \end{minipage}
    \end{tabular}
    \caption{An example of cubic HB-spline with depth $3$. Refined B-spline basis (a1-a3) and Refined HB-spline basis (b1-b3) are marked by the black solid, blue dashed line, and red dashed-dotted line from level $0$ to level $2$, respectively.}
    \label{fig:Bspline_and_HBspline}
\end{figure}

We present an example of cubic HB-spline for the recursive algorithm in Definition \ref{def:hierarchicalBspline} in Figure \ref{fig:Bspline_and_HBspline} with a hierarchy of subdomains $\widehat{\mathbf{\Omega}}_3  = \{[0,1],[1/4,3/4],[3/8,5/8],\emptyset\}$.
Three nested sequence of B-spline spaces $\widehat{\mathcal{B}}_0,\widehat{\mathcal{B}}_1$ and $\widehat{\mathcal{B}}_2$ are considered in Figure \ref{fig:Bspline_and_HBspline} (a1-a3). Based on $\widehat{\mathbf{\Omega}}_3$, the corresponding HB-spline basis is illustrated in Figure \ref{fig:Bspline_and_HBspline} (b1-b3). Meanwhile, the newly added basis is remarked with the blue dashed line in level $2$ and the red dashed-dotted line in level $3$.

In \cite{kunoth2018foundationsTHB}, it is noted that the hierarchical B-spline basis inherits most of the favorable properties of the B-spline basis, such as local support, non-negativity, linear independence, and global regularity. However, a main drawback of the HB-spline is its failure to maintain the partition of unity, which distinguishes it from the standard B-spline. Fortunately,
as discussed in \cite{vuong2011hierarchicalrefinement}, this feature can be restored by proper scaling of the basis functions with non-negative weights $\omega({\widehat{\beta}})$. Consequently, the normalized hierarchical B-spline basis $\widehat{\mathcal{H}}_\omega$, which forms a partition of unity, is defined as follows:
$$
\widehat{\mathcal{H}}_\omega = \left\{\widehat{\beta} = \omega({\widehat{\beta}})\widehat{\beta}:~\widehat{\beta}\in \widehat{\mathcal{H}}\wedge 1 = \sum_{\widehat{\beta}\in \widehat{\mathcal{H}}}\omega({\widehat{\beta}}) \widehat{\beta}\right\}.
$$

Furthermore, in the hierarchical configuration, some of the coefficients associated with the basis at a high level tend to be zero if there are no additional restrictions for the hierarchy of subdomains \cite{kiss2015theoryTHB}. Therefore, the contributions of basis at a high level have disappeared, which is not expected for the practical simulations. Fortunately, the difficulties can be handled by using the truncated mechanism. Thus we can construct the truncated hierarchical spline basis which reduces the overlap of the supports for the hierarchical basis between the two adjacent levels by the following truncation operator.

\begin{definition}[truncation]\label{def:truncation}
    For any basis $\widehat{\beta_\ell}\in \widehat{\mathcal{B}}_\ell$, the truncation operator $\operatorname{trunc}_{\ell+1}$ is
    $$
    \operatorname{trunc}_{\ell+1}(\widehat{\beta}_\ell):=\sum_{k=1}^{N_{\ell+1}}c_{k,\ell+1}^{\tau}(\widehat{\beta}_\ell)\widehat{\beta}_{k,\ell+1},
    $$
    where
    $$
    c_{k, \ell+1}^\tau(\widehat{\beta}_{\ell})= \begin{cases}
    c_{k, \ell+1}(\widehat{\beta}_{\ell}), & \text { if }\operatorname{supp}\widehat{\beta}_\ell\not\subset\widehat{\Omega}_{\ell+1},\\
    0, & \text { otherwise, }\end{cases}
    $$
    and $c_{k,\ell}(\widehat{\beta}_\ell)$ is defined in the two-scale relation \eqref{twoscalerelation}.
\end{definition}

\begin{figure}[h]
    \centering
    \begin{tabular}{cc}
         \begin{minipage}{6cm}
            \centering
            \includegraphics[width=1\linewidth]{figure/THB/HB1.png}
            \centerline{(b1)~ $\widehat{\mathcal{H}}_0$}
        \end{minipage} & 
        \begin{minipage}{6cm}
            \centering
            \includegraphics[width=1\linewidth]{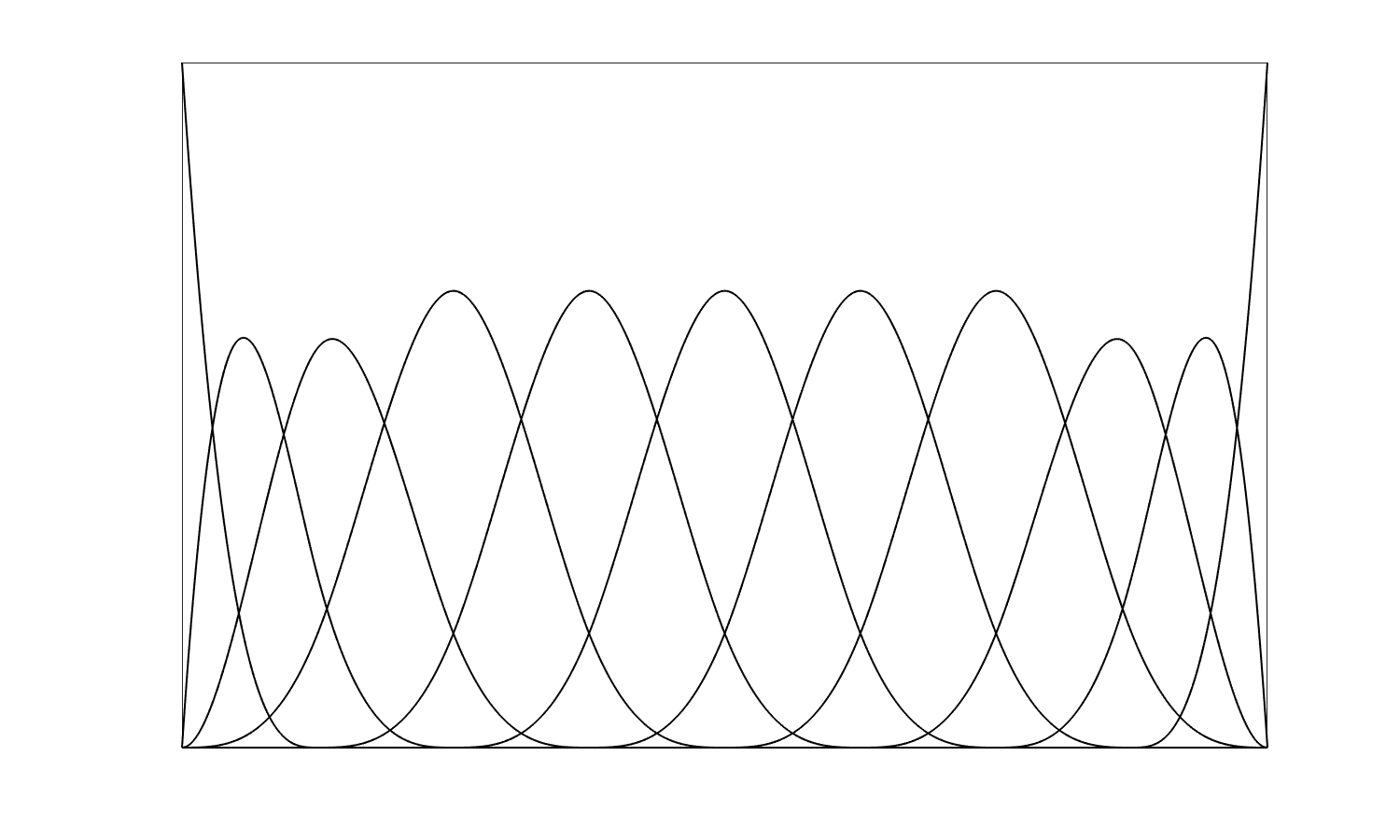}
            \centerline{(c1)~ $\widehat{\mathcal{T}}_0$}
        \end{minipage} \\
         \begin{minipage}{6cm}
            \centering
            \includegraphics[width=1\linewidth]{figure/THB/HB2.png}
            \centerline{(b2)~ $\widehat{\mathcal{H}}_1$}
        \end{minipage} & 
        \begin{minipage}{6cm}
            \centering
            \includegraphics[width=1\linewidth]{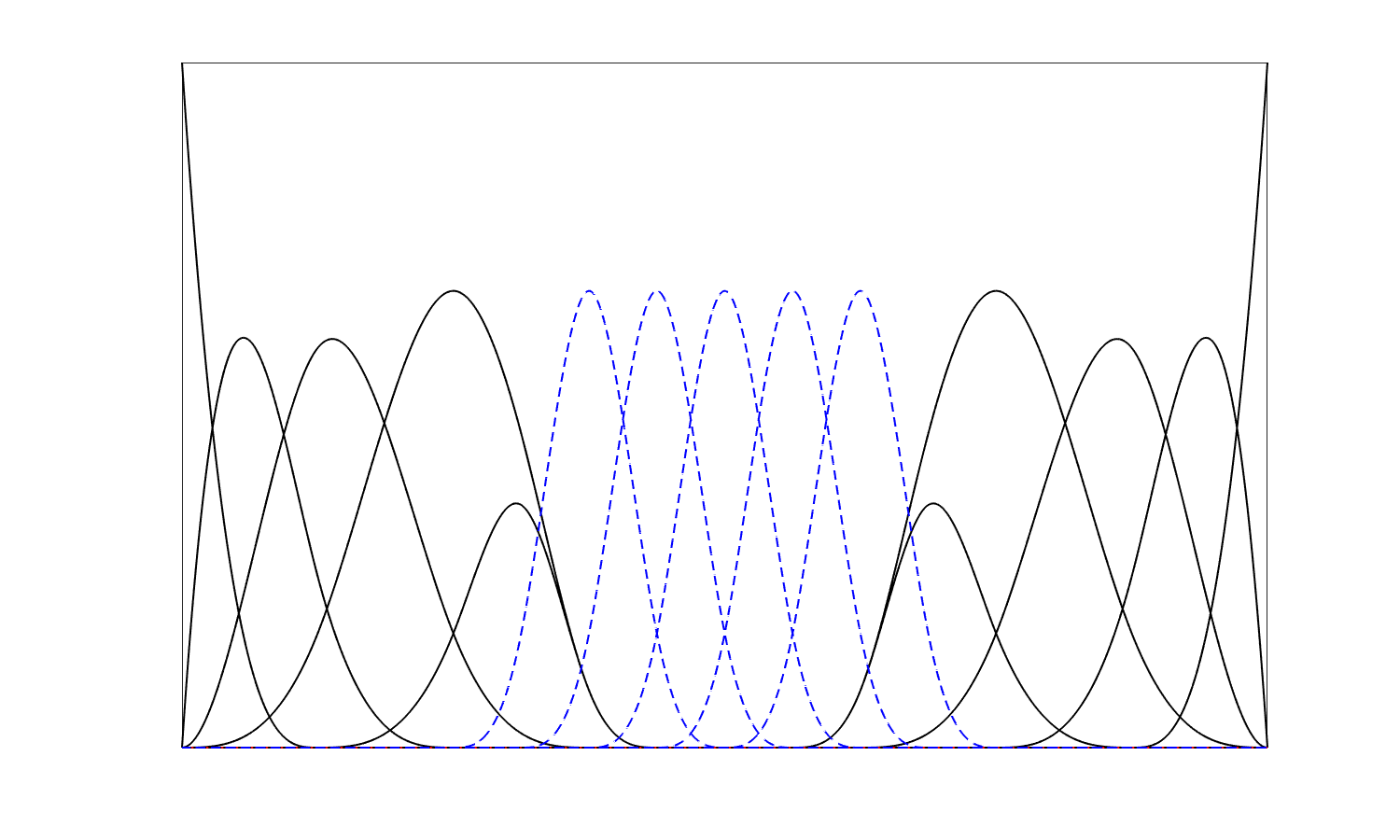}
            \centerline{(c2)~ $\widehat{\mathcal{T}}_1$}
        \end{minipage} \\
         \begin{minipage}{6cm}
            \centering
            \includegraphics[width=1\linewidth]{figure/THB/HB3.png}
            \centerline{(b3)~ $\widehat{\mathcal{H}}_2$}
        \end{minipage} & 
        \begin{minipage}{6cm}
            \centering
            \includegraphics[width=1\linewidth]{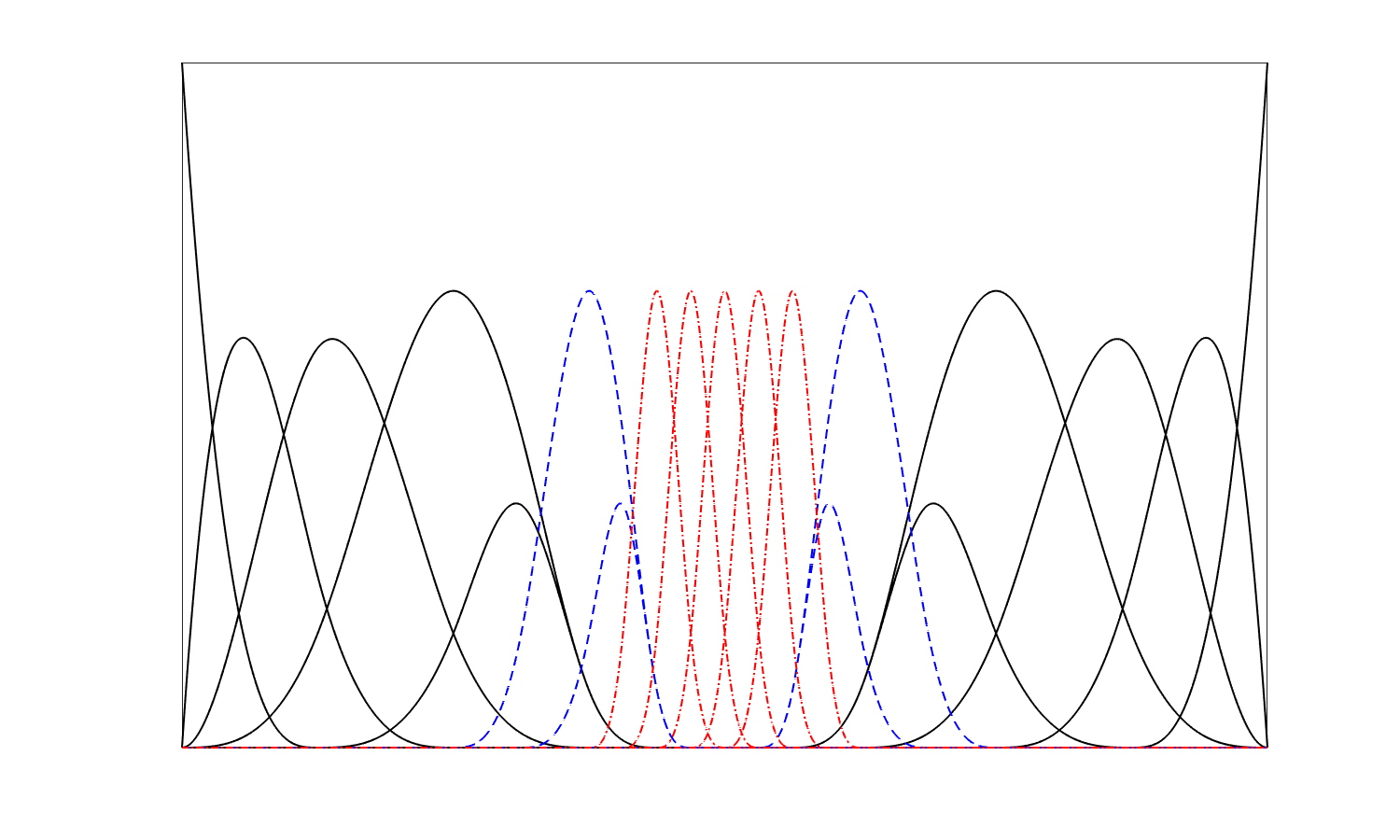}
            \centerline{(c3)~ $\widehat{\mathcal{T}}_2$}
        \end{minipage}
    \end{tabular}
    \caption{An example of cubic THB-spline with depth $3$. Refined HB-spline basis (b1-b3) and Refined THB-spline basis (c1-c3) are marked by the black solid, blue dashed line, and red dashed-dotted line from level $0$ to level $2$, respectively.}
    \label{fig:HBspline_and_THBspline}
\end{figure}

By recursively using the truncation operator, following \cite{kunoth2018foundationsTHB}, we introduce the definition of truncated hierarchical basis:
\begin{definition}[Truncated Hierarchical B-splines]
    With the same assumption in Definition \ref{def:hierarchicalBspline}, the Truncated Hierarchical B-spline(THB-spline) basis $\widehat{\mathcal{T}}:=\widehat{\mathcal{T}}_{N-1}$ is
    \begin{equation*}
\left\{\begin{aligned}
\widehat{\mathcal{T}}_0:= & \{\widehat{\beta}\in\widehat{\mathcal{B}}_0: \operatorname{supp}\widehat{\beta}\not\subset \widehat{\Omega}_0\}, \\
\widehat{\mathcal{T}}_{\ell+1}:= & \left\{\operatorname{trunc}(\widehat{\beta}) : \operatorname{supp} \widehat{\beta} \not \subset \widehat{\Omega}_{\ell+1}\wedge \widehat{\beta}\in\widehat{\mathcal{T}}_\ell\right\} \cup \\
& \left\{\widehat{\beta} \in \widehat{\mathcal{B}}_{\ell+1} : \operatorname{supp} \widehat{\beta} \subset \widehat{\Omega}_{\ell+1}\right\}, \quad \ell=0, \ldots, N-2 .
\end{aligned}\right.
\end{equation*}
\end{definition}
In \cite{kunoth2018foundationsTHB}, it states that $\operatorname{span}\widehat{\mathcal{T}}=\operatorname{\widehat{\mathcal{H}}}$. The same configuration is set for the example of THB-spline in Figure \ref{fig:HBspline_and_THBspline}. Some low-level basis crossing the intersection of subdomains have become short since the contributions of the basis in high level have been removed with the truncation operator in Definition \ref{def:truncation}.

THB-splines are non-negative, linearly independent, globally regular, and locally supported. Additionally, THB-splines contribute basis functions at higher levels while reducing overlap between adjacent basis functions, improving the sparsity of the stiffness matrix in practical simulations. It is important to note that a basis function in level $\ell$ ceases to belong to the basis set $\widehat{\mathcal{B}}_\ell$ after truncation and it is expressed as the linear combinations of the basis in $\widehat{\mathcal{B}}_{\ell+1}$. Given the advantageous properties of the THB-spline basis, we adopt it to construct the isogeometric function space for the Kohn--Sham equation. In conjunction with this, the (strictly) admissible meshes are required to leverage the reduced support of THB-splines relative to the standard HB-splines for the adaptive analysis. For more implemented details, please refer to  \cite{garau2018algorithmsGeoPDEs,bracco2019adaptive}.

\subsection{The isogeometric discretization}
A bounded physical domain $\Omega\subset\mathbb{R}^3$ is chosen as the computation domain for all-electron calculation in numerical simulations. The weak form of Kohn--Sham equation \eqref{KSeq_infty} with homogeneous Dirichlet boundary conditions is: Find $\left\{(\varepsilon_i,\psi_i)\right\}_{i=1,2,\cdots,N_{\mathrm{occ}}}\in \mathbb{R}\times \mathbb{V}_0$ such that
\begin{equation}
    \label{KohnSham_weak_form}
    a(\psi_i,\phi)=\int_\Omega H([\rho];\mathbf{r})\psi_i \phi~\mathrm{d}\mathbf{r} = \varepsilon_i\int_{\Omega} \psi_i\phi~\mathrm{d}\mathbf{r},\quad \forall \phi\in \mathbb{V}_0,
\end{equation}
where function space $\mathbb{V}_0 = H_0^1(\Omega) = \{v\in H^1(\Omega):v\mid_{\partial\Omega = 0}\}$.

To give the discretization of the above weak form, the isogeometric function space in the physical domain $\Omega$ should be constructed. Suppose that $\widehat{\mathcal{T}}$ is the THB-spline function space defined on the hierarchical mesh $\widehat{\mathcal{Q}}$, we assume that the standard isogeometric geometry mapping $\mathbf{F}: \widehat{\Omega}\rightarrow \Omega$ is given by
\begin{equation}
    \label{Geometry_function}
    \mathbf{F}(\widehat{\mathbf{x}}) = \sum_{\widehat{\beta}\in\widehat{\mathcal{T}}} \mathbf{C}_{\widehat{\beta}} \widehat{\beta}(\widehat{\mathbf{x}}),\quad \forall \widehat{\mathbf{x}}\in \widehat{\Omega} = [0,1]^3,
\end{equation}
where $\mathbf{C}_{\widehat{\beta}}$ are the Control Points in $\mathbb{R}^3$. Furthermore, we assume that the geometry mapping $\mathbf{F}$ is a bi-Lipschitz homeomorphism. The discrete function space is given by
\begin{equation*}
    \label{Isogemetric_discrete_space}
    \mathbb{V}_h = \operatorname{span}\mathcal{T} = \operatorname{span} \left\{\widehat{\beta}\circ \mathbf{F}^{-1}:\widehat{\beta}\in \widehat{\mathcal{T}}\right\}.
\end{equation*}

We remark that most of the notations in the parametric domain $\widehat{\Omega}$ can be directly converted into the notations in the physical domain $\Omega$ with the geometry mapping $\mathbf{F}$, such as the truncated basis $\beta$, the hierarchical mesh $\mathcal{Q}$, the subdomain $\Omega_\ell$, the B-spline space $\mathcal{B}_\ell$, and the THB-spline space $\mathcal{T}$ in the physical domain.

The discretization of the weak form \eqref{KohnSham_weak_form} is: Find $\left\{(\varepsilon_{i,h},\psi_{i,h})\right\}_{i=1,2,\cdots,N_{\mathrm{occ}}}\in \mathbb{R}\times \mathbb{V}_{0,h}$ such that
\begin{equation}
    \label{KohnSham_discrete_weak_form}
    a_h(\psi_{i,h},\phi_h)= (\varepsilon_{i,h}\psi_{i,h},\phi_h),\quad \forall \phi_h\in \mathbb{V}_{0,h},
\end{equation}
where $\mathbb{V}_{0,h} = \{v_h\in \mathbb{V}_h:v_h\mid_{\partial\Omega = 0}\}$ and the discrete bilinear operator is written as
\begin{equation*}
    a_h(\psi_h,\phi_h) = \int_\Omega \left(\frac{1}{2}\nabla\psi_h\cdot\nabla\phi_h+\left(V_{\mathrm{ext}}(\mathbf{r})+V_{\mathrm{Har}}(\mathbf{r})+V_{\mathrm{xc}}(\mathbf{r})\right)\psi_h\phi_h\right)~\mathrm{d}\mathbf{r}.
\end{equation*}
Consequently, the discrete systems \eqref{KohnSham_discrete_weak_form} forms the following generalized eigenvalue problem
\begin{equation}
    \label{Generalized_eigenvalue_problem}
    AX=MX\Lambda,
\end{equation}
where $A_{ij}=a_h(\psi_{i,h},\phi_{j,h}),~M_{i,j} = (\psi_{i,h},\phi_{j,h}),~\Lambda = \mathrm{diag}(\varepsilon_{1,h},\varepsilon_{2,h},\cdots,\varepsilon
_{N_{\mathrm{occ}},h})$. In this paper, following \cite{bracco2019adaptive}, we assume that $\mathbb{V}_{0,h}\subset C^1(\Omega)$ for simplicity, which distinguishes it from the $C^0$ finite element basis.

Note that the Hartree potential $V_{\mathrm{Har}}(\mathbf{r})$ is evaluated by the Poisson equation \eqref{Hartree_Poisson_intfy}. Similarly, the computation domain is restricted to the same bounded physical domain $\Omega$ for the Kohn--Sham equation. However, the boundary condition cannot be simply chosen as the homogeneous Dirichlet boundary condition since the Hartree potential decays linearly but the wavefunction in density functional theory exponentially decreases associated with the nuclear position. Therefore, in the physical domain $\Omega$, the Poisson equation with non-homogeneous Dirichlet boundary condition is solved
\begin{equation*}
    \label{Hartree_Poisson_Omega}
\left\{\begin{aligned}
-\nabla^2 V_{\mathrm{Har}}(\mathbf{r}) & =4 \pi \rho(\mathbf{r}),~~\quad \text { in } \Omega, \\
V_{\mathrm{Har}}(\mathbf{r}) & = V_{\mathrm{Har},\partial \Omega}(\mathbf{r}), \text { on } \partial \Omega,
\end{aligned}\right.
\end{equation*}
where the boundary value of $V_{\mathrm{Har}}$ is calculated by the multipole expansion method in \cite{bao2012hadaptiveKS,kuang2024towards} due to the high efficiency. The monopole moment, dipole moment, and quadrupole moment are used in the approximation of $V_{\mathrm{Har,\partial\Omega}}$
\begin{equation*}
    \label{Multipole_expansion}
    \begin{aligned}
    V_{\mathrm{Har,\partial\Omega}}(\mathbf{r}) = \frac{q_0}{\left|\mathbf{r}-\mathbf{r}^{\prime\prime}\right|}+\sum_{i=1}^3 p_i\cdot\frac{r_i-r_i^{\prime\prime}}{\left|\mathbf{r}-\mathbf{r}^{\prime\prime}\right|^3}+\sum_{i,j=1}^3 q_{ij}\cdot\frac{3(r_i-r_i^{\prime\prime})(r_j-r_{j}^{\prime\prime})-\delta_{ij}\left|\mathbf{r}-\mathbf{r}^{\prime\prime}\right|^2}{\left|\mathbf{r}-\mathbf{r}^{\prime\prime}\right|^5},
    \end{aligned}
\end{equation*}
where
\begin{equation*}
    \begin{aligned}
    & q_0 = \int_{\Omega} \rho(\mathbf{r}^\prime)~\mathrm{d}\mathbf{r}^\prime, \quad
    p_i = \int_\Omega \rho(\mathbf{r}^\prime)(r_i^\prime - r_i^{\prime\prime})~\mathrm{d}\mathbf{r}^\prime,\\
    & q_{ij} = \int_{\Omega} \frac{1}{2}\rho(\mathbf{r}^\prime)(r_i^\prime-r_{i}^{\prime\prime})(r_j^\prime-r_{j}^{\prime\prime})~\mathrm{d}\mathbf{r}^\prime, \quad
    \mathbf{r}^{\prime\prime}=\frac{\int \mathbf{r}\rho(\mathbf{r})~\mathrm{d}\mathbf{r}}{\int \rho(\mathbf{r})~\mathrm{d}\mathbf{r}}.
    \end{aligned}
\end{equation*}
In the formula of multipole expansion, $\mathbf{r}^{\prime\prime}$ describes the barycenter of electron density. We remark that the dipole moment $p_i = 0$ in the practical simulations due to the symmetry of the systems.

Analogously, the above Poisson equation for Hartree potential $V_{\mathrm{Har}}(\mathbf{r})$ can be discretized into the following linear system
\begin{equation*}
    \label{Hartree_linear_system}
    SY = b,
\end{equation*}
where
\begin{equation*}
    S_{ij} = \int_\Omega \nabla\psi_{i,h}\cdot \nabla\phi_{j,h}~\mathrm{d}\mathbf{r},\quad
    b_j = \int_\Omega 4\pi\rho(\mathbf{r})\phi_{j,h}~\mathrm{d}\mathbf{r},\quad \forall \psi_{i,h},\phi_{j,h}\in \mathbb{V}_{0,h}.
\end{equation*}

Observe that the Kohn--Sham equation in \eqref{KSeq_infty} is highly nonlinear since the Hamiltonian operator $H([\rho];\mathbf{r})$ depends on the electron density $\rho(\mathbf{r})$. Meanwhile, the electron density $\rho(\mathbf{r})$ is associated with the wavefunctions $\psi_i(\mathbf{r})$. In the practical computations, we assume that the electron density in $\eqref{Hamiltonian_strong_operator}$ is known and the SCF iteration is used for the numerical simulations of the Kohn--Sham equation. Specifically, in Algorithm \ref{alg:SCFiteration}, $(\varepsilon_{i}^0,\psi_{i}^0),~i=1,2,\cdots,N_{\mathrm{occ}}$ is the initial input with random initialization. The $MaxIter$ and $TOL$ denote the maximum number of SCF iterations and the tolerance for the criterion associated with the updated electron density, respectively. Furthermore, the linear mixing scheme for updating the electron density $\rho(\mathbf{r})$ is taken for the convergence of SCF iteration.

\begin{algorithm}[H]
	\SetAlgoLined
	\KwIn{$(\varepsilon_i^0,\psi_i^0)$, \textit{MaxIter}, \textit{TOL}.}
	\KwOut{$(\varepsilon_i,\psi_i),~i=1,2,\cdots,N_{\mathrm{occ}}.$}
	Evaluate $\rho^0(\mathbf{r})$ with $\{\psi_i^0\}$ by using \eqref{rhodensity}\;
	\While{ $\left\| \rho^{k+1}(\mathbf{r}) - \rho^{k}(\mathbf{r}) \right\|_2 >$ \text{TOL} or $k <$ \textit{MaxIter}}{
    Evaluate $V_{\mathrm{Har}}(\mathbf{r})$ and $V_{\mathrm{xc}}(\mathbf{r})$ by using $\rho^k(\mathbf{r})$ in \eqref{rhodensity}\;
		Form the generalized eigenvalue system \eqref{Generalized_eigenvalue_problem}\;
		Update $(\varepsilon_i^{k+1},\psi_i^{k+1})$ by LOBPCG\;
		Evaluate $\rho^{k+1}(\mathbf{r})$ by using \eqref{rhodensity}\;
		$\rho^{k+1}(\mathbf{r}) := \alpha\rho^{k+1}(\mathbf{r}) + (1-\alpha)\rho^{k}(\mathbf{r})$\;
        $k = k+1$\;
	}
	\caption{self-consistent field iteration}
	\label{alg:SCFiteration}
\end{algorithm}

\section{An $h$-adaptive isogeometric solver} \label{sec:hadaptive_alg}
In this section, we will introduce the ingredients of the $h$-adaptive isogeometric solver for improving the computational efficiency in all-electron calculations. In general, there are four necessary modules named \textit{Solve}, \textit{Estimate}, \textit{Mark}, and \textit{Refine} for each iterative loop. Following \cite{garau2018algorithmsGeoPDEs}, we will focus on the \textit{Solve} and \textit{Estimate} modules and give a brief introduction to the \textit{Mark} and \textit{Refine} modules.

\subsection{Solve and Estimate}
In \textit{Solve} module, the eigenpairs $\{(\varepsilon_i,\psi_i)\},~i=1,2,\cdots,N_{\mathrm{occ}}$ are achieved by solving \eqref{Generalized_eigenvalue_problem} in \Cref{alg:SCFiteration}, which provides the prerequisite to estimate the residual while solving the generalized eigenvalue problem \eqref{Generalized_eigenvalue_problem} on the current hierarchical mesh $\mathcal{Q}$. The core of the adaptive mesh solver is to capture the cells with dramatic errors so that it can obtain high numerical accuracy with less computational effort.

\subsubsection{The residual-type indicator}
To design an effective error indicator for the Kohn--Sham equation, we first review the typical finite element indicator with linear polynomial basis for the Poisson equation in \cite{verfurth2013posterioriindicator}. Suppose $u_{\mathrm{p},h}$ is the numerical solution of the Poisson equation with homogeneous boundary conditions, then the residual with integration by parts element-wise is defined by
\begin{equation}\label{eq:indicator_Poisson}
\begin{aligned}
&\int_\Omega \left( f v  - \nabla u_h\cdot \nabla v \right)\mathrm{~d}\boldsymbol{x} \\
 = & \int_\Omega fv \mathrm{~d}\boldsymbol{x} - \sum_{K\in\mathcal{T}}\int_{K} \nabla u_h \cdot \nabla v \mathrm{~d}\boldsymbol{x} \\
 = & \int_\Omega fv \mathrm{~d}\boldsymbol{x} + \sum_{K\in\mathcal{T}}\int_K \Delta u_h  v \mathrm{~d}\boldsymbol{x} - \sum_{K\in \mathcal{T}}\sum_{e\in \partial K} \int_e \left( \nabla u_h \big |_{K_e} - \nabla u_h \big |_{K_l} \right)\cdot \mathbf{n}_e v\mathrm{~d}\mathbf{s} \\
 =:& \int_\Omega fv \mathrm{~d}\boldsymbol{x} + \sum_{K\in\mathcal{T}}\int_K \Delta u_h  v \mathrm{~d}\boldsymbol{x} - \sum_{K\in \mathcal{T}}\sum_{e\in \partial K} \int_e \mathbb{J}_e(u_h)\cdot \mathbf{n}_e v\mathrm{~d}\mathbf{s}
\end{aligned}
\end{equation}
where $f$ is the right-hand-side function of the Poisson equation, $v$ is the trial function, $K$, $e$ denote the elements and the edges of elements,  $K_l$ represents the common element with $K_e$, $\mathbb{J}_e(u_h)$ is the jump term of the element $K$, and $\mathbf{n}_e$ describes the out normal vector on the common face $e$. Furthermore, with \eqref{eq:indicator_Poisson}, the residual-type indicator $\mathcal{R}_{\mathrm{P},K}^{\mathrm{poly}}$ for the Poisson equation in the element $K$ can be denoted as
\begin{equation}
    \label{eq:indicator_Poisson_element}
    \mathcal{R}_{\mathrm{P},K}^{\mathrm{poly}} = \left(h_K^2\int_K\left|f + \Delta u_h\right|^2\mathrm{~d}\boldsymbol{x} + \sum_{e\in\partial K}\frac{1}{2}h_e\int_e \left|\mathbb{J}_e(u_h)\right|^2 \mathrm{~d}\mathbf{s}\right)^{1/2},
\end{equation}
where $h_K$ is the length of the element $K$ and $h_e$ is the length of the common face $e$.

We note that the Laplacian term $\Delta u_h$ in \eqref{eq:indicator_Poisson_element} vanishes to zero using the linear polynomial basis, which is a main drawback for this indicator. To overcome this disadvantage, high-order basis functions such as the high-order polynomial function (global $C^0$) and the high-order splines (global $C^k$) have been widely utilized in the construction of the residual-type error indicator for the Poisson equation. The high-order version of residual-type error indicator in \cite{verfurth2013posterioriindicator} is also constructed by \eqref{eq:indicator_Poisson_element} with high-order polynomial functions, restoring the contributions of the Laplacian term $\Delta u_h$ to the estimation at each element. However, the jump term is still a non-negligible quantity due to the $C^0$ property of the basis function at the common face of the elements. Based on the global $C^{p-1}$ continuity property of splines, a specialized residual-type error indicator \eqref{eq:indicator_Poisson_element_splines} for the Poisson equation is defined in \cite{buffa2016adaptive,garau2018algorithmsGeoPDEs}, which not only enables precise computation of the Laplacian term but also ensures that the jump term vanishes to zero, significantly reducing the computational cost in numerical simulations.
\begin{equation}
    \label{eq:indicator_Poisson_element_splines}
    \mathcal{R}_{\mathrm{p},K}^{\mathrm{sp}} = \left(h_K^2\int_K\left|f + \Delta u_h\right|^2\mathrm{~d}\boldsymbol{x}\right)^{1/2}.
\end{equation}

To develop an appropriate indicator in \textit{Estimate} module for the Kohn--Sham equation, a residual-type indicator depending on a posterior error estimate is designed. Recalling the weak form \eqref{KohnSham_weak_form}, the corresponding residual $r_i$ for $i$-th eigenpair $(\varepsilon_{i,h},\psi_{i,h})$ is given by
\begin{equation}
    \label{residual_weak}
    (r_i,\phi) = (\varepsilon_{i,h}\psi_{i,h},\phi) - a(\psi_{i,h},\phi),\quad \forall \phi\in\mathbb{V}.
\end{equation}
By taking $\phi = \phi_{h}\in \mathbb{V}_{0,h}$,
\begin{equation}
    \label{residual_discrete_weak}
    \begin{aligned}
      (r_i,\phi_{h}) & = (\varepsilon_{i,h}\psi_{i,h},\phi_{h}) - a_h(\psi_{i,h},\phi_{h}) \\
      & = (\varepsilon_{i,h}\psi_{i,h} + \frac{1}{2}\nabla^2\psi_{i,h} - \left(V_{\mathrm{ext}}+V_{\mathrm{Har}}+V_{\mathrm{xc}}\right)\psi_{i,h},\phi_{h}).
    \end{aligned}
\end{equation}

The element-based indicator and the function-based indicator could be defined with \eqref{residual_discrete_weak}, respectively. The element-based indicator on the hierarchical mesh $\mathcal{Q}$ in the \textit{Estimate} module is given by
\begin{equation}
    \label{indicator_element_based}
    \mathcal{R}_Q = \left(\sum_{i=1}^{N_{\mathrm{occ}}} h_Q^2 \|r_i\|_{L^2(Q)}^2\right)^{\frac12},
\end{equation}
where $h_Q$ denotes the diameter of an active cell $Q\in\mathcal{Q}$ and
\begin{equation*}
    \|r_i\|_{L^2(Q)}^2 = \int_Q \big| \varepsilon_{i,h}\psi_{i,h} + \frac{1}{2}\nabla^2\psi_{i,h} - \left(V_{\mathrm{ext}}+V_{\mathrm{Har}}+V_{\mathrm{xc}}\right)\psi_{i,h}\big|^2 \mathrm{~d}\mathbf{r}.
\end{equation*}

Analogously, the function-based indicator on the THB-spline function space $\mathcal{T}$ in the \textit{Estimate} module is written as
\begin{equation}
    \label{indicator_function_based}
    \mathcal{R}_\beta = \left(\sum_{i=1}^{N_{\mathrm{occ}}} h_\beta^2 \alpha_\beta \|r_i\|_{L^2(\operatorname{supp}\beta)}^2\right)^{\frac12},
\end{equation}
where $h_\beta$ is the diameter of the support of basis $\beta\in\mathcal{T}$, $\alpha_\beta$ is the non-negative coefficient for the partition of unity in THB-splines, and
\begin{equation*}
    \|r_i\|_{L^2(\operatorname{supp}\beta)}^2 = \int_{\operatorname{supp}(\beta)}\big| \varepsilon_{i,h}\psi_{i,h} + \frac{1}{2}\nabla^2\psi_{i,h} - \left(V_{\mathrm{ext}}+V_{\mathrm{Har}}+V_{\mathrm{xc}}\right)\psi_{i,h} \big|^2\beta\mathrm{~d}\mathbf{r}.
\end{equation*}

Compared with the residual-type error indicator under traditional finite element framework (e.g. \cite{kuang2024towards}), the jump term in the indicator \eqref{indicator_element_based} and \eqref{indicator_function_based} vanishes to zero and the Laplacian term can be strictly calculated due to the assumption $\mathbb{V}_{0,h}\subset C^1(\Omega)$. Therefore, a considerable portion of computational resources will be saved in the whole numerical simulations. Furthermore, in the work, the function-based indicator \eqref{indicator_function_based} is applied in the subsequent numerical experiments.

\subsection{Mark and Refine}
In the \textit{Mark} module, with the \textit{maximum strategy}, a set of functions $\mathcal{M}\subset\mathcal{Q}$ is marked by the indicator \eqref{indicator_function_based}. To be specific, a set of functions is marked by
\begin{equation*}
    \mathcal{R}_\beta \geq \theta \max_{\beta^\prime\in \mathcal{T}} \mathcal{R}_{\beta^\prime}, 
\end{equation*}
where $\theta\in [0,1)$ is the marking parameter. In practical simulations, the parameter $\theta$ is always chosen as $0.5$ for the adaptive mesh refinement.

\begin{figure}[h!]
    \centering
    \begin{minipage}[h]{0.24\textwidth}
        \includegraphics[width=1\linewidth]{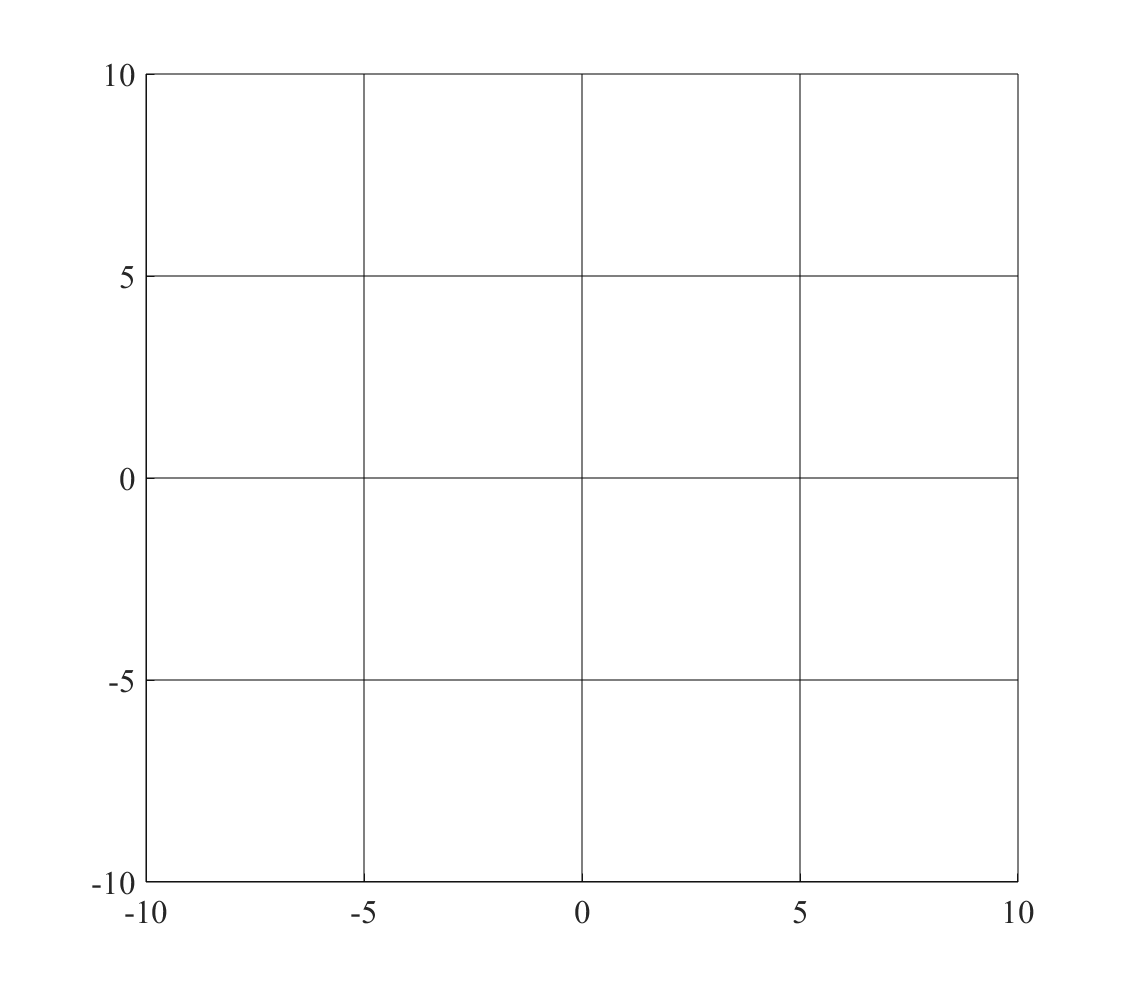}
    \end{minipage}
    \begin{minipage}[h]{0.24\textwidth}
        \includegraphics[width=1\linewidth]{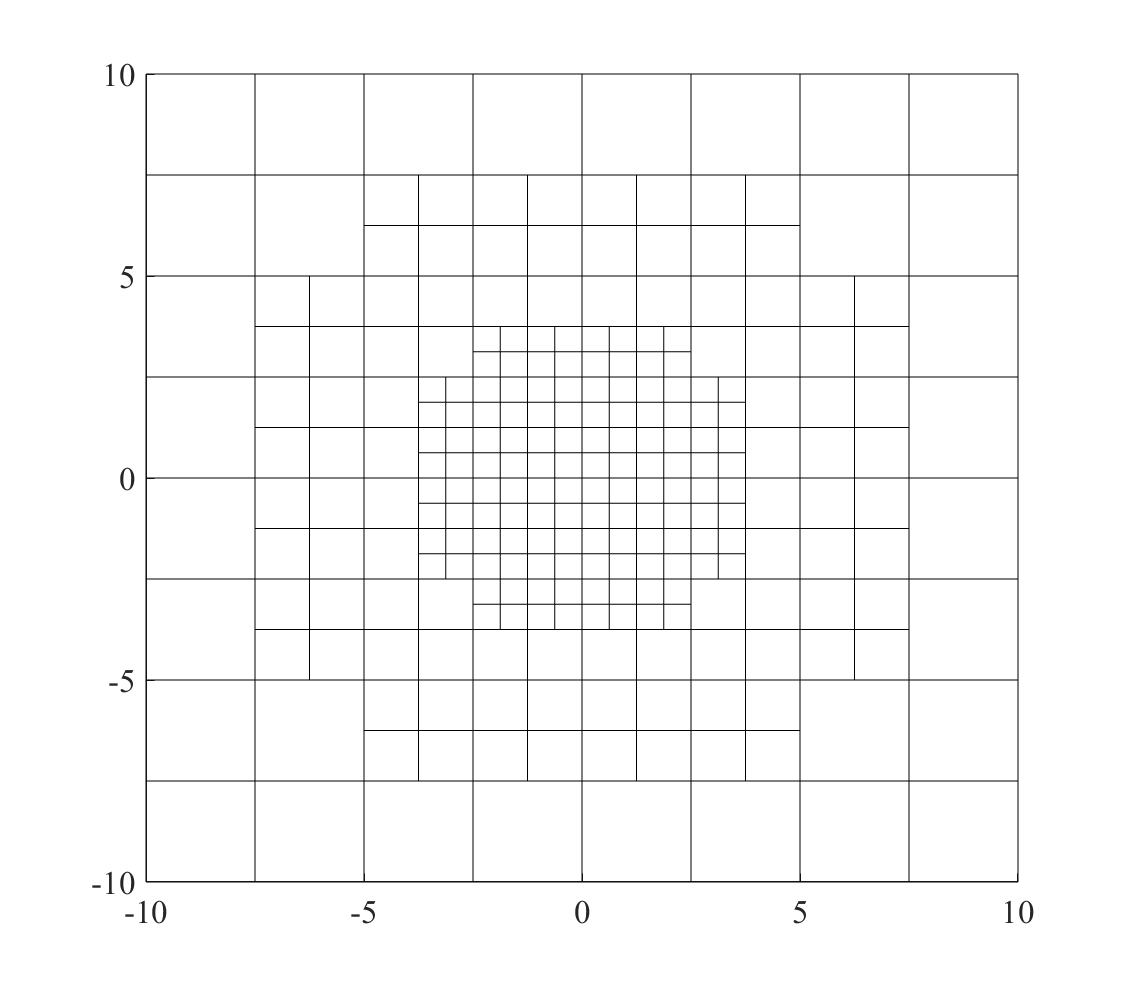}
    \end{minipage}
    \begin{minipage}[h]{0.24\textwidth}
        \includegraphics[width=1\linewidth]{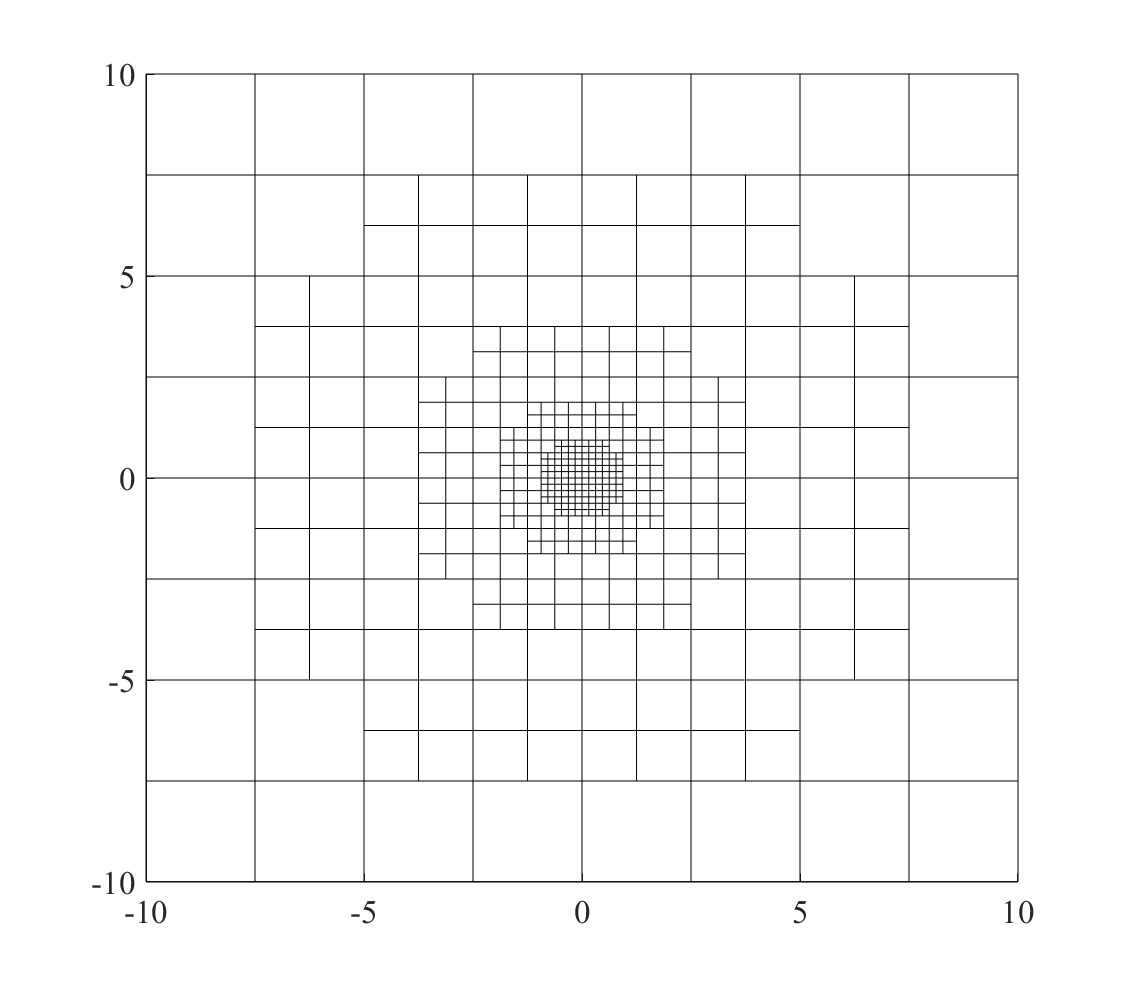}
    \end{minipage}
    \begin{minipage}[h]{0.24\textwidth}
        \includegraphics[width=1\linewidth]{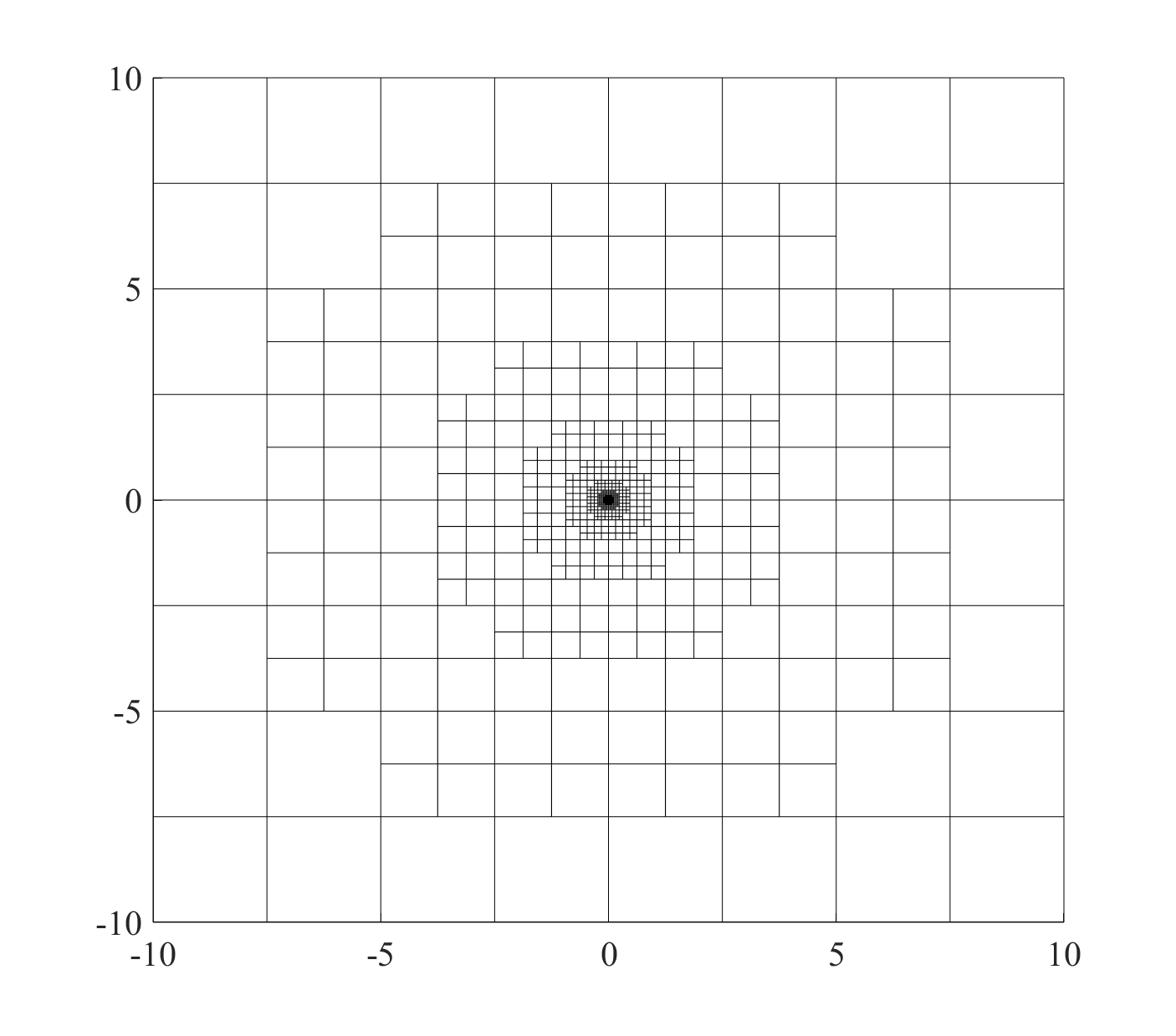}
    \end{minipage}
    \caption{Refined sliced mesh in the $X$-$Y$ plane for a helium atom ($p=3$).}
    \label{fig:Helium_2D_mesh_refine}
\end{figure}

Finally, in the \textit{Refine} module, the marked set $\mathcal{M}$ is used to update the hierarchy of subdomains, then generate the refined hierarchical mesh and its corresponding truncated hierarchical splines basis. For more details, please refer to \cite{garau2018algorithmsGeoPDEs}. To verify the effectiveness of the \textit{Refine} module, we present the sliced mesh in the $X$-$Y$ plane of the helium atom experiment in \Cref{fig:Helium_2D_mesh_refine}. The left subfigure shows the initial sliced mesh, the middle two subfigures present the sliced mesh at the intermediate iterations of the $h$-adaptive \Cref{alg:adaptivemesh}, and the right subfigure shows the final refined sliced mesh. It is obvious that the mesh has been successfully refined around the origin (i.e., the position of the helium nuclei), demonstrating the effectiveness of our algorithm.

In addition, the relation between the previous THB-splines function space and the refined THB-splines function space can be constructed based on the two-scale relation \eqref{twoscalerelation}. By projecting the wavefunctions from the previous THB-splines function space into the refined THB-splines space, we can obtain a good initial guess for SCF iteration in the \textit{Solve} module, accelerating the convergence in \Cref{alg:SCFiteration}.

As discussed above, we have presented all the modules in the $h$-adaptive isogeometric solver for solving the Kohn--Sham density functional theory. We end this section with \Cref{alg:adaptivemesh}, showing the outline of the whole algorithm.

\begin{algorithm}[H]
  \SetAlgoLined
  \KwIn{Initial mesh $\mathcal{Q}^{0}$, initial energy $E^0=0$, \textit{MaxIter}, \textit{TOL}.}
  \KwOut{Ground state energy $E$}
  \While{ $\left| E^{k+1}-E^k \right| >$ \text{TOL} or $k <$ \textit{MaxIter}}{
    \textit{Solve}: Update the eigenpair $(\varepsilon_i,\psi_i)$ by \Cref{alg:SCFiteration}\;
    \textit{Estimate}: Evaluate the error indicator $\mathcal{R}$ in \eqref{indicator_function_based}\;
    \textit{Mark}: Mark the elements by $\mathcal{R}$ with a marking parameter $\theta$\;
    \textit{Refine}: Refine the marked element $\mathcal{M}$ and get the adaptive mesh $\mathcal{Q}^{k+1}$\;
    Calculate the ground state energy $E^{k+1}$\;
    Update the wavefunctions from the previous mesh $\mathcal{Q}^{k}$ to the mesh $\mathcal{Q}^{k+1}$\;
    $k = k+1$;
    }
  \caption{$h$-adaptive isogeometric solver}
  \label{alg:adaptivemesh}
\end{algorithm}

\subsection{Implementation details}
In this subsection, several implementation details of \Cref{alg:adaptivemesh} are introduced in solving the all-electron KS equation. We first introduce the GeoPDEs \cite{garau2018algorithmsGeoPDEs} package on which our implementation is based. Then the eigensolver is presented and discussed in detail. 

\subsubsection{GeoPDEs}
We implemented our algorithm using GeoPDEs 3.4.2 \cite{de2011geopdes, vazquez2016new}, an open-source and free software package designed for research and teaching in Isogeometric Analysis. Written in Octave and fully compatible with MATLAB, GeoPDEs offers a versatile and standardized framework for developing and testing innovative isogeometric methods for solving partial differential equations. In particular, it includes implementations of adaptive methods with hierarchical B-splines, encompassing both refinement and coarsening capabilities.

Building on GeoPDEs 3.4.2, we have specifically implemented the $h$-adaptive solver for the all-electron Kohn--Sham equation in the \textit{Solve} module. The standard THB spline basis functions are employed. In the \textit{Estimate} module, we have designed a residual-type \textit{a posterior} error indicator for the all-electron Kohn--Sham equation. The remaining modules, such as \textit{Mark} and \textit{Refine}, are fully implemented based on the adaptive capabilities provided by the GeoPDEs library. 
\subsubsection{Eigensolver}
As we discussed earlier, an effective solver for the eigenvalue problem \eqref{Generalized_eigenvalue_problem} is required to be constructed associated with the all-electron Kohn--Sham equation. In this study, we utilize the locally optimal block preconditioned conjugate gradient (LOBPCG) method \cite{knyazev2007blockLOBPCG}, which has proven to be effective in electron structure calculations \cite{castro2006octopus, gonze2009abinit}. The LOBPCG method offers several advantages over traditional Krylov subspace methods, including reduced memory requirements and enhanced accuracy. It works by minimizing or maximizing the trace of the projected Hamiltonian in a larger subspace, allowing for more precise eigenvalue computations. A critical consideration for the LOBPCG method is the selection of the preconditioner, as it plays a significant role in accelerating the convergence, particularly in cases where the matrices possess high condition numbers. In this work, we follow our previous study \cite{bao2012hadaptiveKS} to design the elliptic preconditioner for the generalized eigenvalue problem.

The elliptic preconditioner takes the form $T = S/2 -\varepsilon M$, where $S$ is the discretized kinetic operator in \eqref{Hamiltonian_strong_operator}, and $\varepsilon$ is the approximated eigenvalue. For each eigenpair we will construct a particular preconditioner to accelerate the calculation, as a result, the preconditioner for the $i$-th eigenpair at the $j$-th iteration is designed as 
\begin{equation} \label{eq:precon}
T_{i}^{(j)}=\left\{\begin{aligned}
\frac{1}{2} S-\varepsilon_i^{(j)} M & \text {, if } \varepsilon_{i}^{(j)} <0, \\
\mathrm{Diag}(H)~ & \text {, otherwise }
\end{aligned} \quad \text { for } i=1, \ldots, N_{\mathrm{occ}}.\right.
\end{equation}
When $\varepsilon_i^{j}>0$, we use a diagonal preconditioner to ensure its positive definiteness.
In the practical computations, the preconditioning process involves the solution of linear systems $WT^{(j)}=(T_i^{(j)})^{-1}W^{(j)}$, which serves as the preconditioner of LOBPCG solver for the eigenvalue system \eqref{Generalized_eigenvalue_problem}.

To validate the effectiveness of the preconditioner, we compare the number of iterations required by the LOBPCG method to solve the generalized eigenvalue problems arising from the helium atom and methane molecule, both with and without the elliptic preconditioner. To be specific, we assess performance on two types of mesh during the SCF iteration, i.e, the uniform and the adaptive mesh. On the uniform mesh, we test different splines order from $2$ to $5$ for the two cases. As to the  adaptive type of mesh,  we test the helium example with $5093$ Dofs, and  the methane with $6355$ Dofs, where the spline order is set as $p=3$ and the computational domain is $[-10, 10]^3$. The adaptive mesh is generated by \Cref{alg:adaptivemesh}. The iteration for LOBPCG stops when the residual falls below the accuracy $10^{-12}$ or the iteration number achieves the maximum number $1000$.

\begin{figure}[!h]
    \centering
    \begin{minipage}[h]{0.49\textwidth}
        \includegraphics[width=1\linewidth]{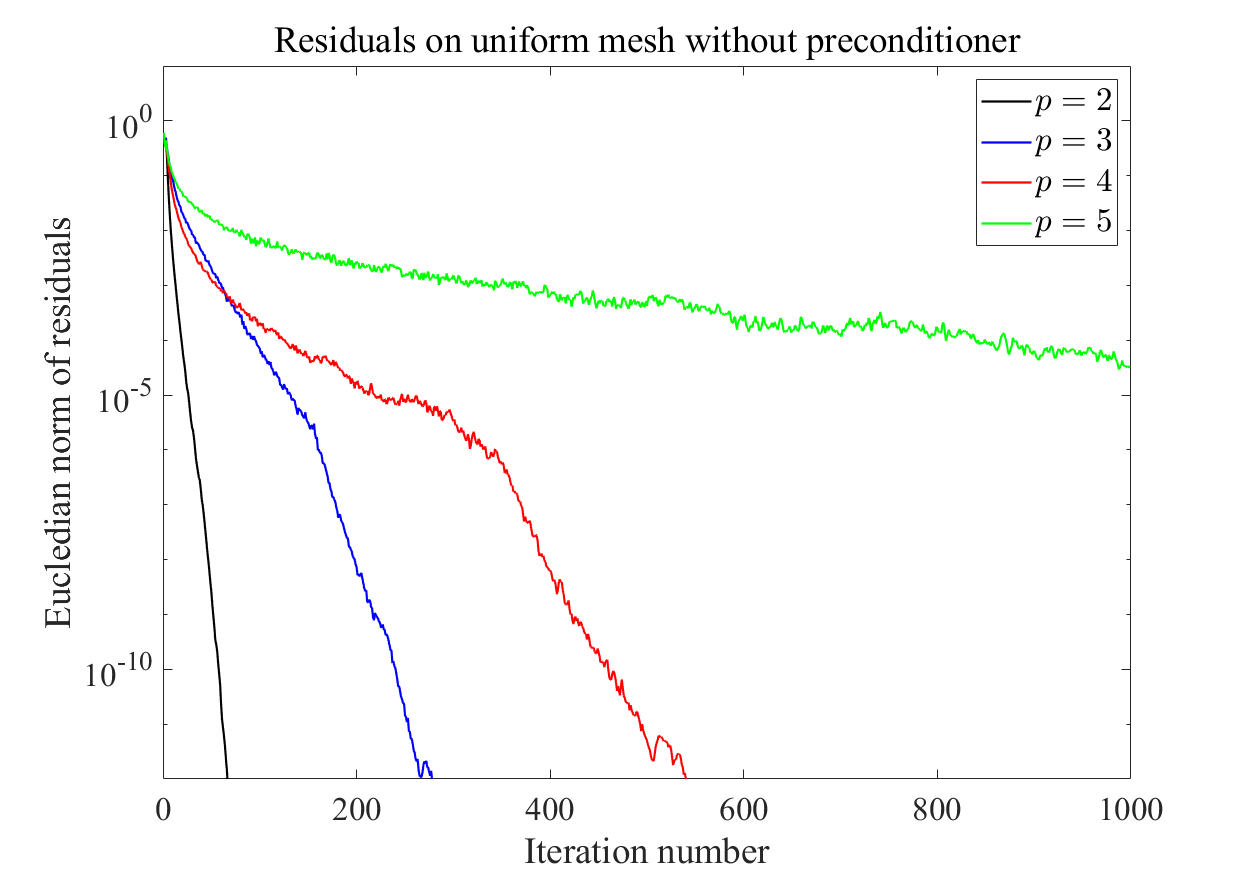}
    \end{minipage}
    \begin{minipage}[h]{0.49\textwidth}
        \includegraphics[width=1\linewidth]{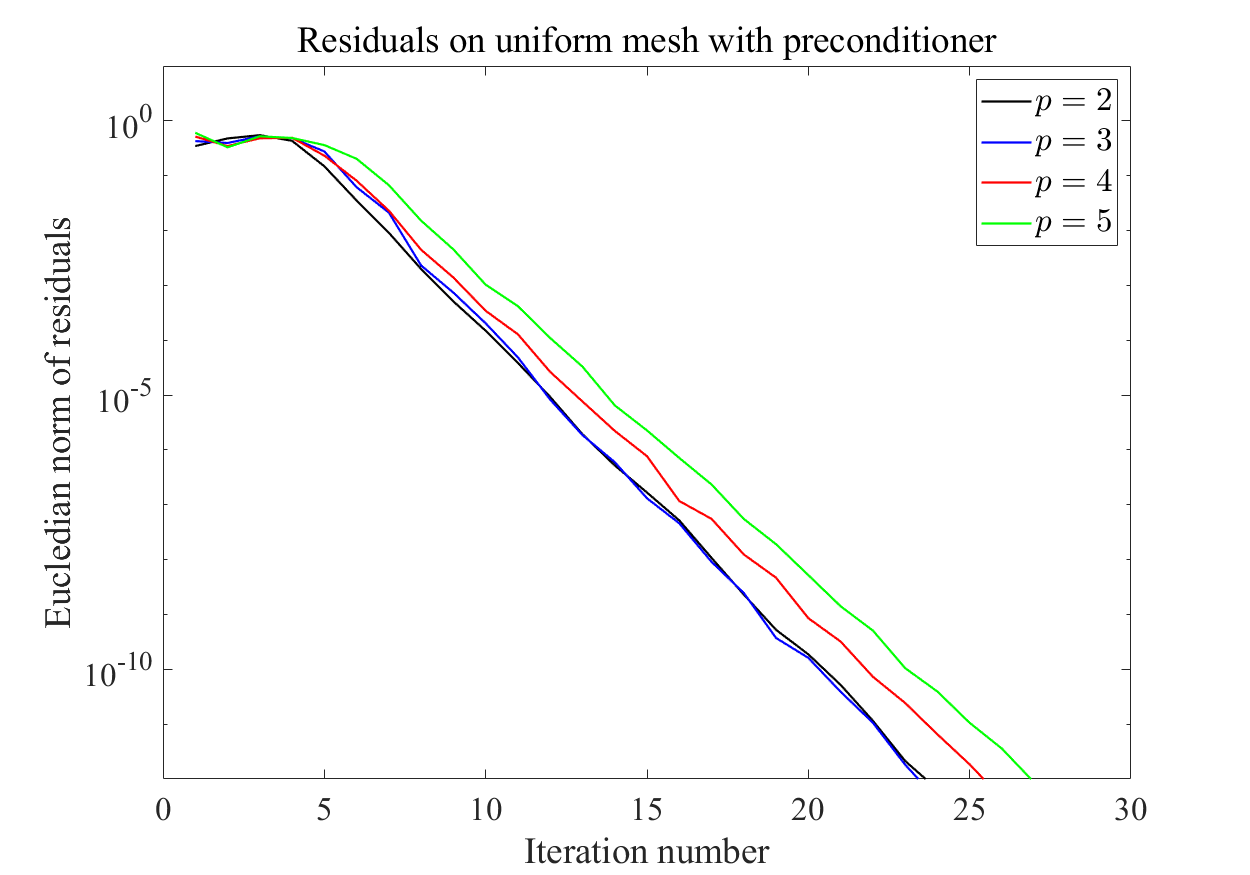}
    \end{minipage}
    \caption{The convergence of LOBPCG in solving a helium atom on the uniform mesh with different orders. Left: without preconditioner. Right: with preconditioner.}
    \label{fig:UIGKS_Helium_lobpcg}
\end{figure}

\begin{figure}[!h]
    \centering
    \begin{minipage}[h]{0.49\textwidth}
        \includegraphics[width=1\linewidth]{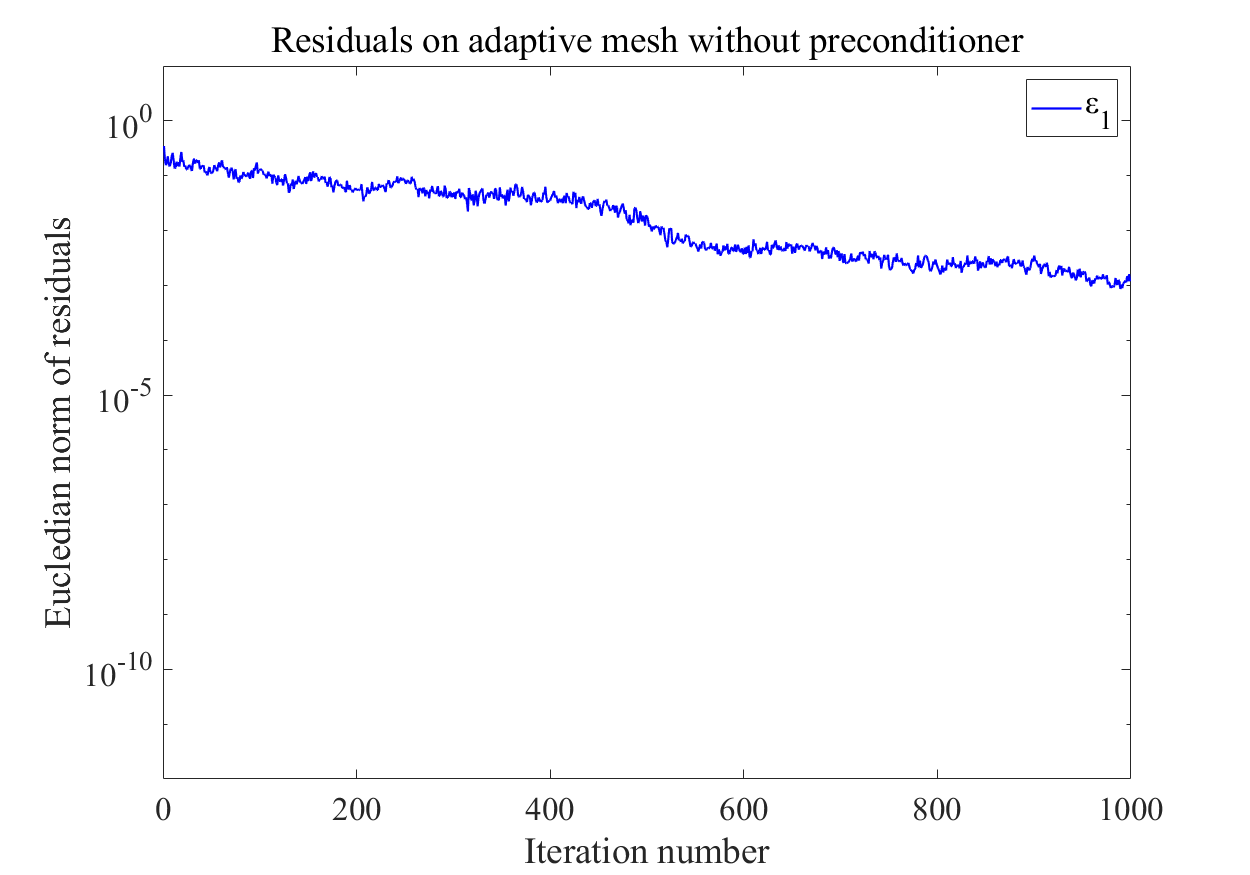}
    \end{minipage}
    \begin{minipage}[h]{0.49\textwidth}
        \includegraphics[width=1\linewidth]{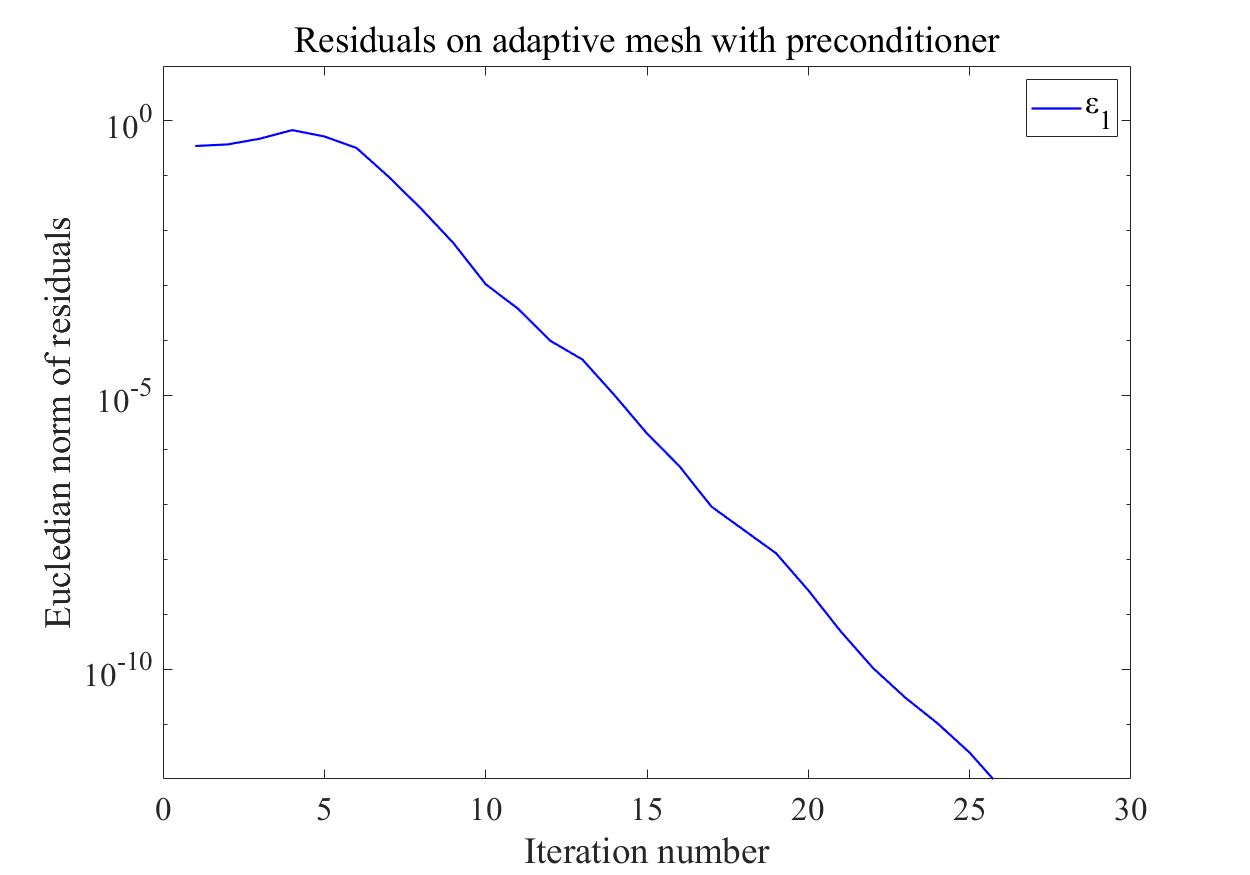}
    \end{minipage}
    \caption{The residual of LOBPCG in solving a helium atom on the adaptive mesh ($p=3$). Left: without preconditioner. Right: with preconditioner. }
    \label{fig:AIGKS_Helium_lobpcg}
\end{figure}

The results of the helium atom are shown in \Cref{fig:UIGKS_Helium_lobpcg} for the uniform mesh and \Cref{fig:AIGKS_Helium_lobpcg} for the adaptive mesh. We observe that on the uniform mesh, the number of iterations required for the convergence of LOBPCG increases as the order of the spline basis increases when the preconditioner is not applied. Specifically, for the spline order of $2$, achieving the set tolerance requires $70$ iterations, whereas for the order of $5$, LOBPCG fails to converge within $1000$ iterations.  However, with the preconditioner \eqref{eq:precon}, the iteration numbers for different orders are all below $30$. Remarkably, they vary slightly,  which means that the iteration number is independent of the basis order. The results for the case of the helium atom worsen on the adaptive mesh without a preconditioner, as seen in the left of \Cref{fig:AIGKS_Helium_lobpcg}, where residual of the occupied orbital decreases slowly and just reaches $10^{-3}$ within $1000$ steps. This occurs because the mesh sizes on the adaptive mesh vary widely, leading the Hamiltonian matrix to become ill-conditioned. In contrast, after applying the preconditioner, convergence is also achieved within $30$ steps, which demonstrates the robustness of the eigensolver. 

\begin{figure}[!h]
    \centering
    \begin{minipage}[h]{0.49\textwidth}
        \includegraphics[width=1\linewidth]{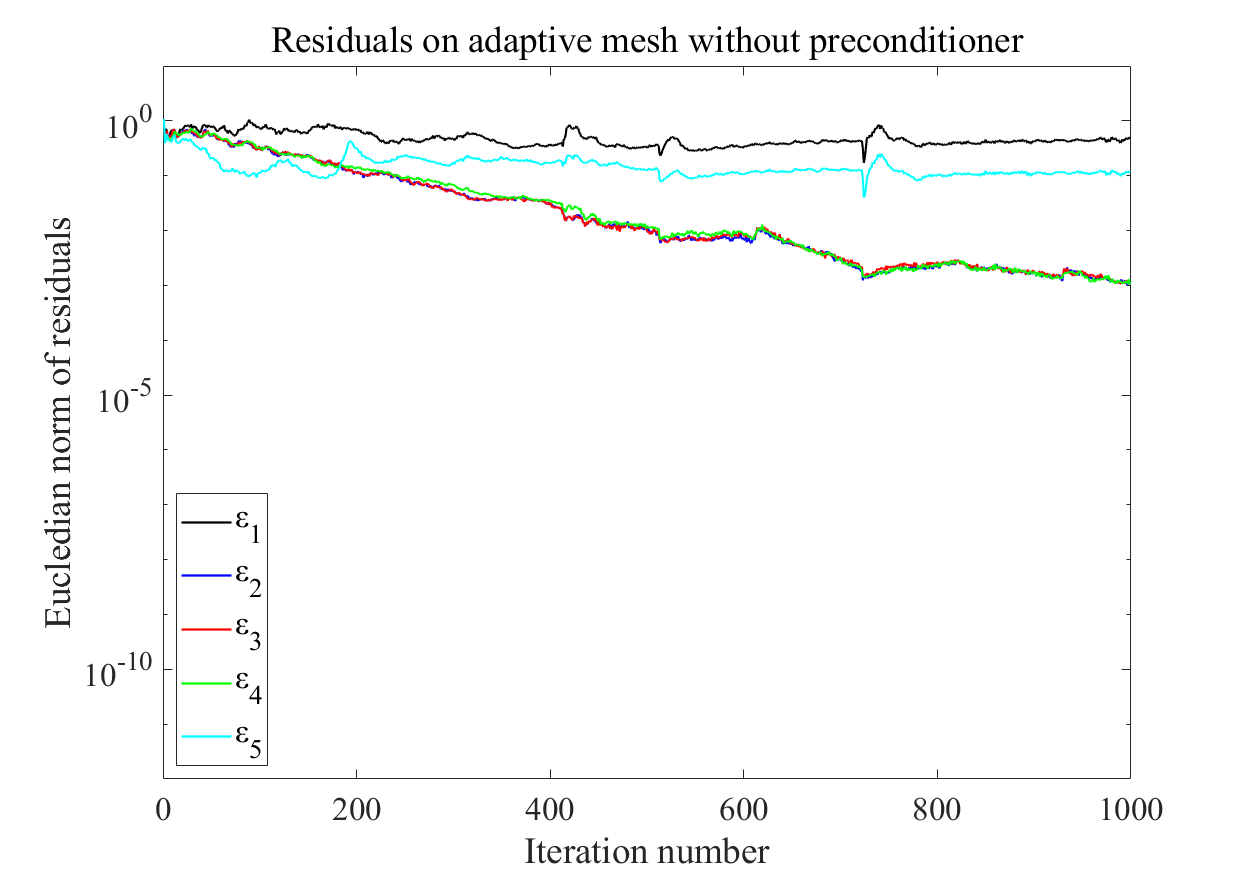}
    \end{minipage}
    \begin{minipage}[h]{0.49\textwidth}
        \includegraphics[width=1\linewidth]{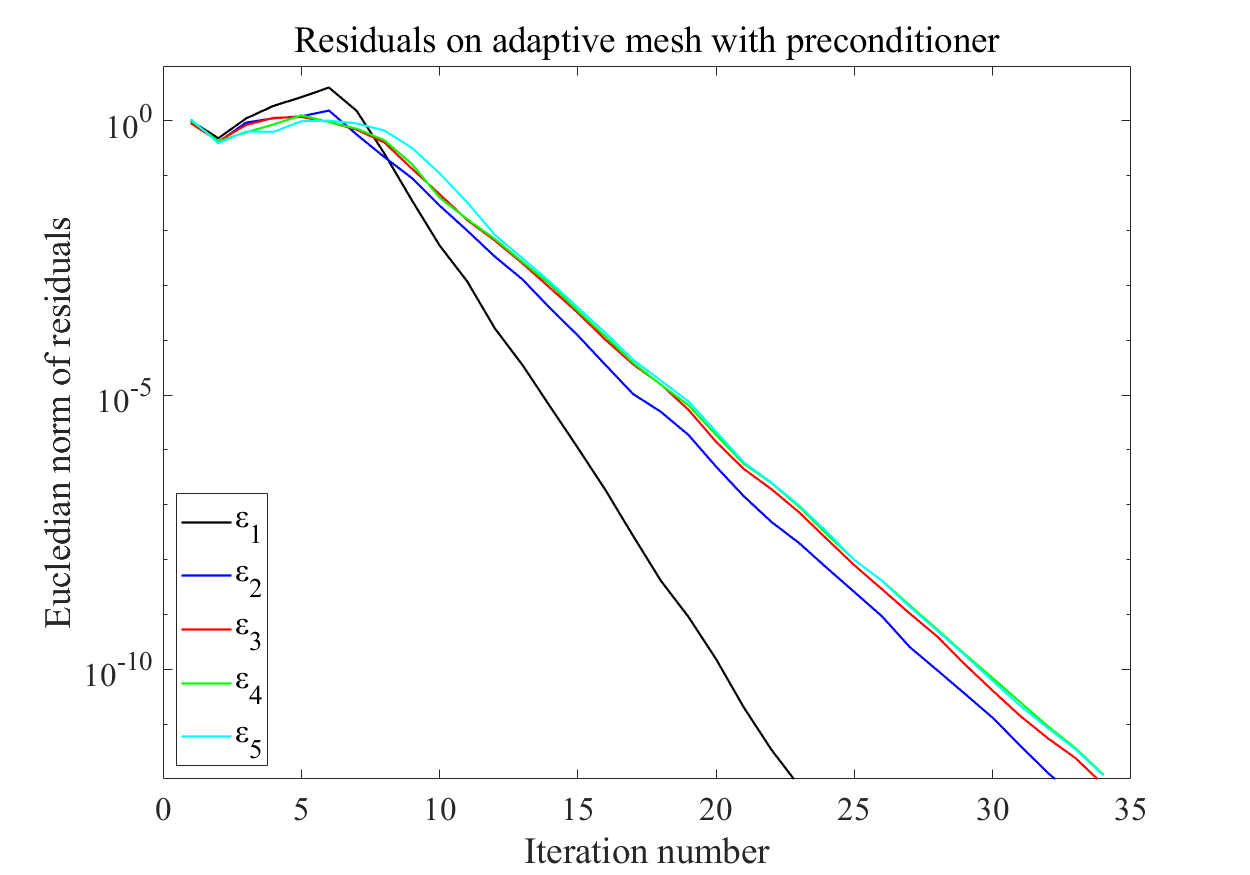}
    \end{minipage}
    \caption{The residual of LOBPCG in solving a methane molecule on the adaptive mesh ($p=3$). Left: without preconditioner. Right: with preconditioner. }
    \label{fig:AIGKS_CH4_lobpcg}
\end{figure}

Similar results in simulating the methane molecule are presented in \Cref{tab:UIGKS_CH4_lobpcg} on the uniform mesh and in \Cref{fig:AIGKS_CH4_lobpcg} on the adaptive mesh. In the uniform mesh case, it is obvious that more iteration numbers are required for the convergence of LOBPCG without the preconditioner as the spline order increases. This issue becomes worse for the order $p=4,5$ where LOBPCG fails to converge within $1000$ 
iterations. However, when the preconditioner is employed, an accuracy $10^{-12}$ is achieved within $40$ iterations for $p=2,3,4$ and within $170$ iterations for $p=5$ in the methane molecule simulations. On the adaptive mesh, the performance without the preconditioner is even worse, as displayed in \Cref{fig:AIGKS_CH4_lobpcg}, where the residual only reaches $10^{-3}$ within $1000$ iterations due to the ill-conditioned number of the Hamiltonian matrix. With the preconditioner, the residual of LOBPCG quickly decreases to $10^{-12}$ within $30$ iterations, demonstrating the effectiveness of the preconditioner \eqref{eq:precon}. Combined with the results for the helium atom, we highlight the robustness of the preconditioned LOBPCG solver, which remains effective regardless of the spline basis order.

\begin{table}[h!]
\centering
\caption{Required iteration numbers of the residual of LOBPCG in solving a methane molecule without and with the elliptic preconditioner utilizing different orders, where the number $1000$ means that the accuracy is not touched within maximum iteration numbers $1000$.}
\label{tab:UIGKS_CH4_lobpcg}
\begin{tabular}{cc|cccc}
\hline
 &          & $p=2$ & $p=3$ & $p=4$ & $p=5$ \\ \hline
\multicolumn{1}{c|}{\multirow{2}{*}{$\varepsilon_1$}} & without pre & $210$ & $1000$   & $1000$   & $1000$   \\
\multicolumn{1}{c|}{}                                 & with pre   & $23$  & $20$  & $23$  & $21$  \\ \hline
\multicolumn{1}{c|}{\multirow{2}{*}{$\varepsilon_2$}} & without pre & $142$ & $662$ & $1000$   & $1000$   \\
\multicolumn{1}{c|}{}                                 & with pre   & $34$  & $35$  & $35$  & $36$  \\ \hline
\multicolumn{1}{c|}{\multirow{2}{*}{$\varepsilon_3$}} & without pre & $148$ & $666$ & $1000$   & $1000$   \\
\multicolumn{1}{c|}{}                                 & with pre   & $34$  & $35$  & $35$  & $36$  \\ \hline
\multicolumn{1}{c|}{\multirow{2}{*}{$\varepsilon_4$}} & without pre & $147$ & $653$ & $1000$   & $1000$   \\
\multicolumn{1}{c|}{}                                 & with pre   & $35$  & $36$  & $36$  & $36$  \\ \hline
\multicolumn{1}{c|}{\multirow{2}{*}{$\varepsilon_5$}} & without pre & $503$ & $775$ & $1000$   & $1000$   \\
\multicolumn{1}{c|}{}                                 & with pre   & $165$ & $39$  & $102$ & $43$  \\ \hline
\end{tabular}
\end{table}

\section{Numerical experiments}\label{sec:numerical_experiments}

In this section, the performance of the $h$-adaptive isogeometric solver for Kohn--Sham density functional theory is validated through a series of numerical experiments. The experiments commence with the radial all-electron calculations for hydrogen, lithium, and aluminum atoms as benchmarks, and the reproduction of the atoms in the periodic table in one dimension. Subsequently, we focus on the all-electron calculation for the total energy of the ground state in the three dimensions, examining systems such as the helium atom, the lithium hydride (LiH), methane (CH$_4$), and benzene (C$_6$H$_6$) molecules. All subsequent experiments are conducted using the function-based indicator described in Algorithm \ref{alg:adaptivemesh}, and are performed by using the isogeometric Matlab/Octave GeoPDEs library \cite{de2011geopdes,garau2018algorithmsGeoPDEs} in the workstation with AMD Ryzen Threadripper PRO 5975WX 32-Cores (3.6 GHz and 256 GB memory).

The linear mixing scheme is adopted in the SCF iteration, in which the mixing parameter is set as $\alpha = 0.618$. The stop criterion for SCF iteration is $10^{-8}$. The stop criterion for mesh adaption is $10^{-4}$ for the single atoms and $10^{-3}$ for the molecules.
For the radial calculations, the Vosko-Wilk-Nusair (VWN4) correlation potential \cite{vosko1980accuratevmn4} is used. For the three-dimensional simulations, the Perdew-Zunger (PZ) correlation potential \cite{perdew1981pztypexc} is adopted.

\subsection{Radial calculations}
In this subsection, we examine the convergence and effectiveness of the presented $h$-adaptive isogeometric solver described in 
\Cref{alg:adaptivemesh} through the radial all-electron calculations for the single atoms in the periodic table. In virtue of the spherical symmetric property for the atomic systems, the Kohn--Sham equation \eqref{KSeq_infty} can be simplified to the following form
\begin{equation}
    \label{Radial_KS_calculations}
    \left(-\frac{1}{2}\nabla^2_r + V_{\mathrm{ext}} + V_{\mathrm{Har}} + V_{\mathrm{xc}} + \frac{l(l+1)}{2r^2}\right)\psi_i = \varepsilon_i \psi_i,
\end{equation}
where $l$ is the azimuthal quantum number for different occupied orbitals. The derivation and implementation details of the above radial Kohn--Sham equation can be found in many literature, such as  \cite{romanowski2007numericalradialcalculation, vcertik2013dftatom,luokuang2024highorderaccuratemovingmesh}. In the following, we first evaluate the convergence of the presented solver with respect to the eigenvalues for a hydrogen atom, whose analytical solution is known. Next, we illustrate the effectiveness of the adaptive solver by conducting a detailed comparison with the uniform mesh method for the lithium and aluminum examples. Finally, we conclude this subsection by presenting a reproduction of the NIST database results for elements in the periodic table, covering atomic numbers from 1 to 92.

\subsubsection{Hydrogen atom}
We first test a hydrogen atom whose solutions are analytically known to verify the convergence of the presented $h$-adaptive solver. The first three states, i.e., the ground state, the first excited state and the second excited state are studied, and the exact energies are given as
\begin{equation*}
    \varepsilon_n = -\frac{1}{2n^2},\quad n=1,2,3.
\end{equation*}
It is noted that the computational domain in this example is set as $[0,100]$ since the slow decay of the wavefunction for the third state. 

\begin{figure}[h!]
    \centering
    \begin{minipage}[h]{0.325\textwidth}
        \includegraphics[width=1\linewidth]{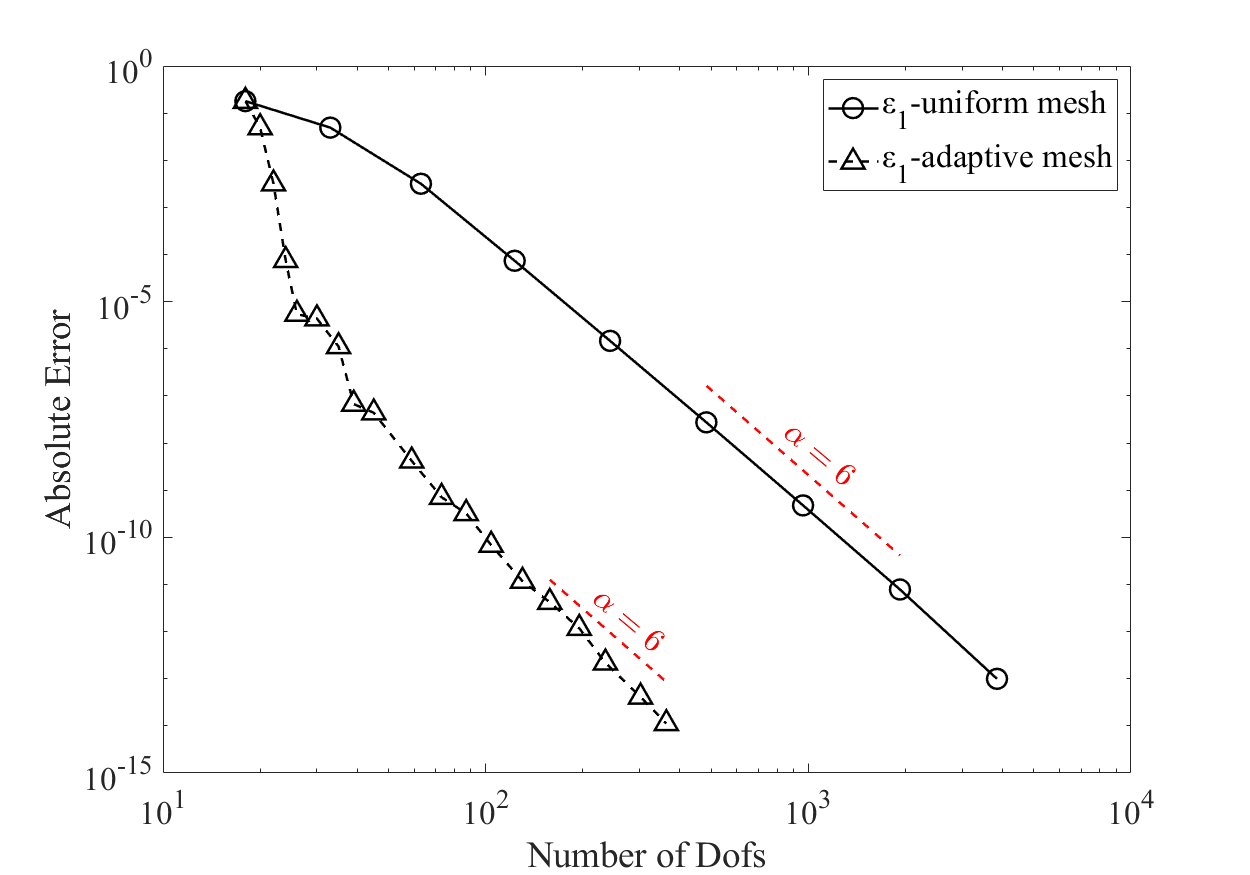}
    \end{minipage}
    \begin{minipage}[h]{0.325\textwidth}
        \includegraphics[width=1\linewidth]{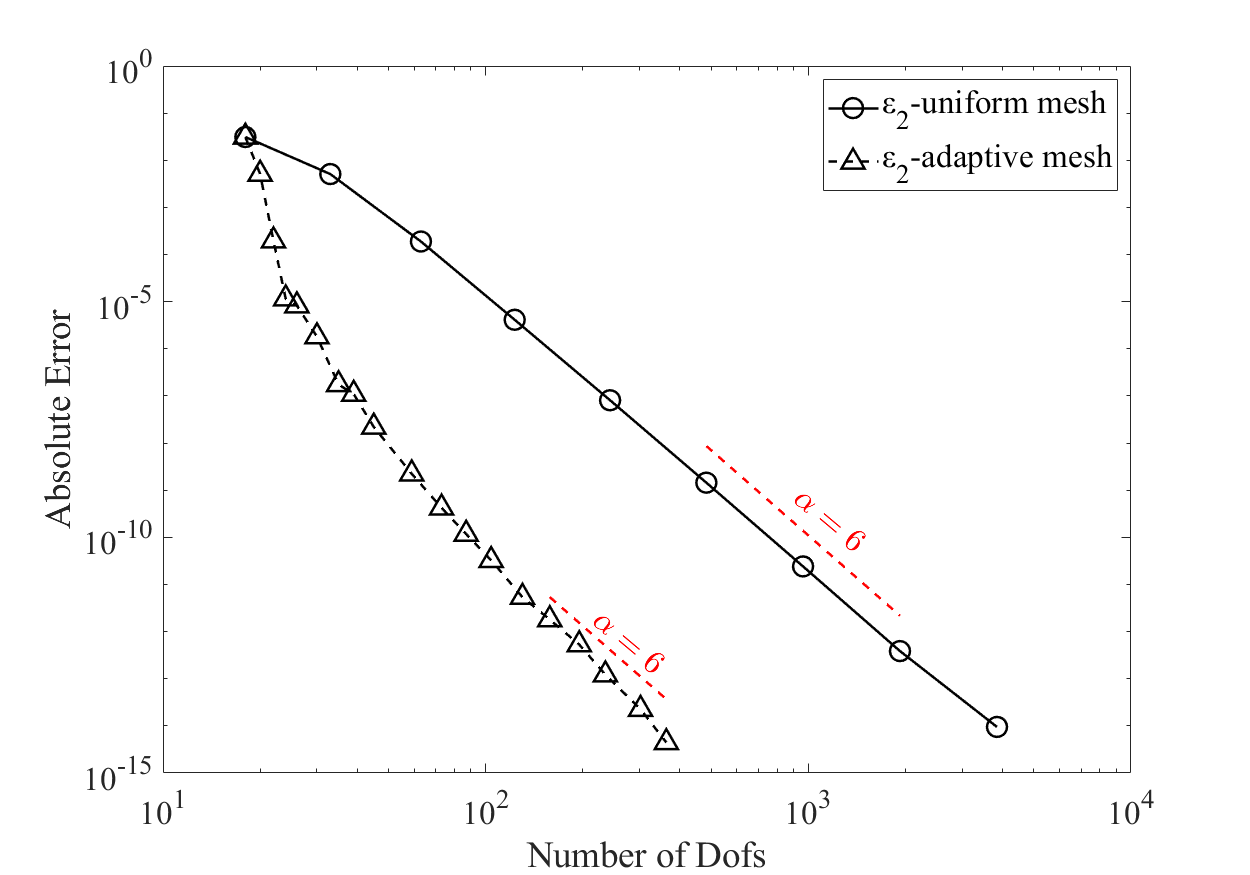}
    \end{minipage}
    \begin{minipage}[h]{0.325\textwidth}
        \includegraphics[width=1\linewidth]{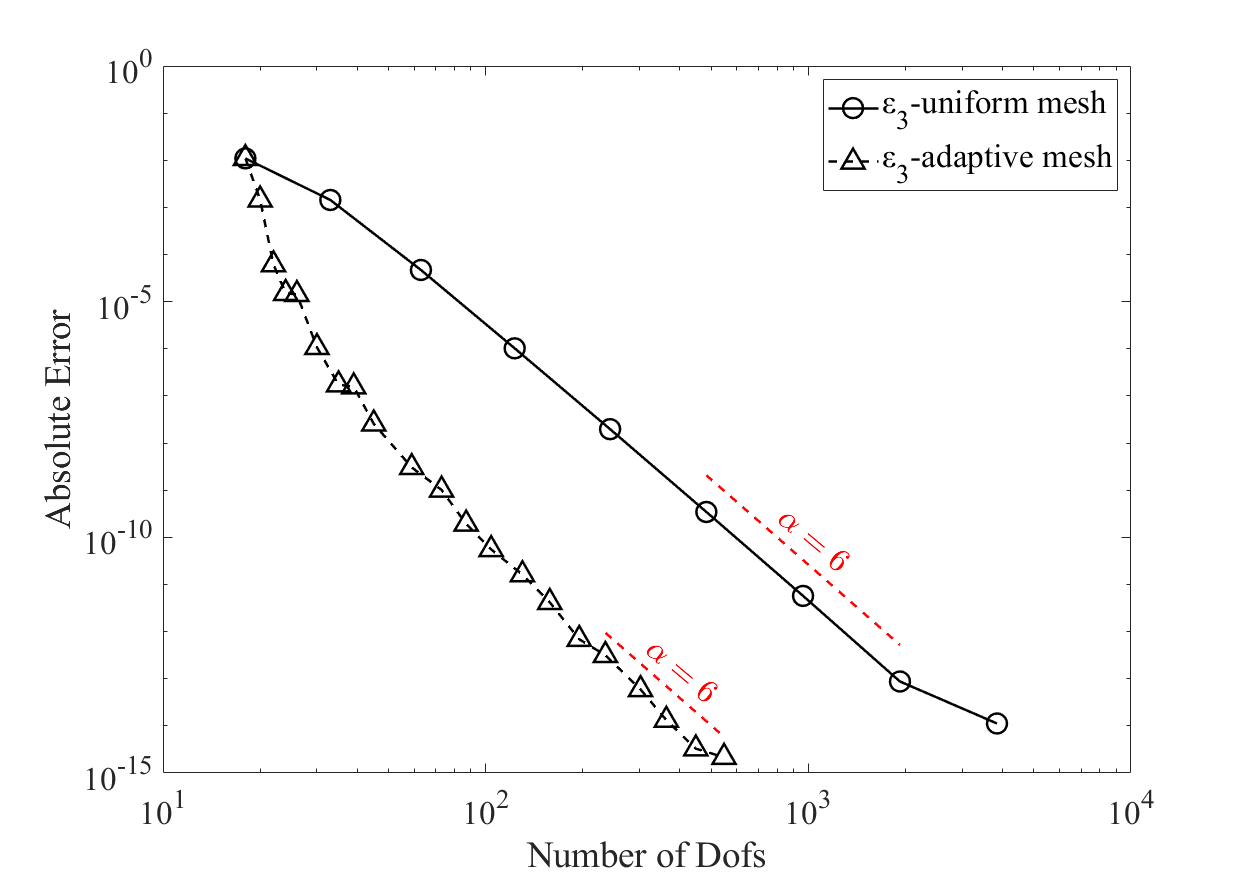}
    \end{minipage}
    \caption{Convergence history of the three states for the hydrogen atoms for $p=3$ under uniform mesh and $h$-adaptive mesh, respectively.}
    \label{fig:Radial_Cal_Hydrogen}
\end{figure}

The convergence of the energies of three states with respect to the number of degrees of freedom (Dofs) is illustrated in \Cref{fig:Radial_Cal_Hydrogen}. A cubic HB-spline basis is utilized in the simulations, and the mesh refinement process is terminated once machine accuracy in the range of $10^{-14}$ to $10^{-15}$ is achieved. As shown in \Cref{fig:Radial_Cal_Hydrogen}, a uniform mesh requires $4000$ Dofs to reach machine accuracy for all three states. In contrast, an $h$-adaptive mesh requires only $400$ Dofs, representing a tenfold reduction compared to the uniform mesh. Additionally, it is observed that the uniform mesh achieves an accuracy of approximately $10^{-5}$ within 100 Dofs, while the $h$-adaptive mesh surpasses an accuracy of $10^{-10}$, highlighting the superior computational performance of \Cref{alg:adaptivemesh}. In addition, the convergence rate for the eigenvalues on either uniform mesh or adaptive mesh is $\alpha=6$, which agrees with the theoretical rate $2p=6$.
\subsubsection{Lithium and Aluminum atoms}
\begin{figure}[h!]
    \centering
    \begin{minipage}[h]{0.49\textwidth}
        \includegraphics[width=1\linewidth]{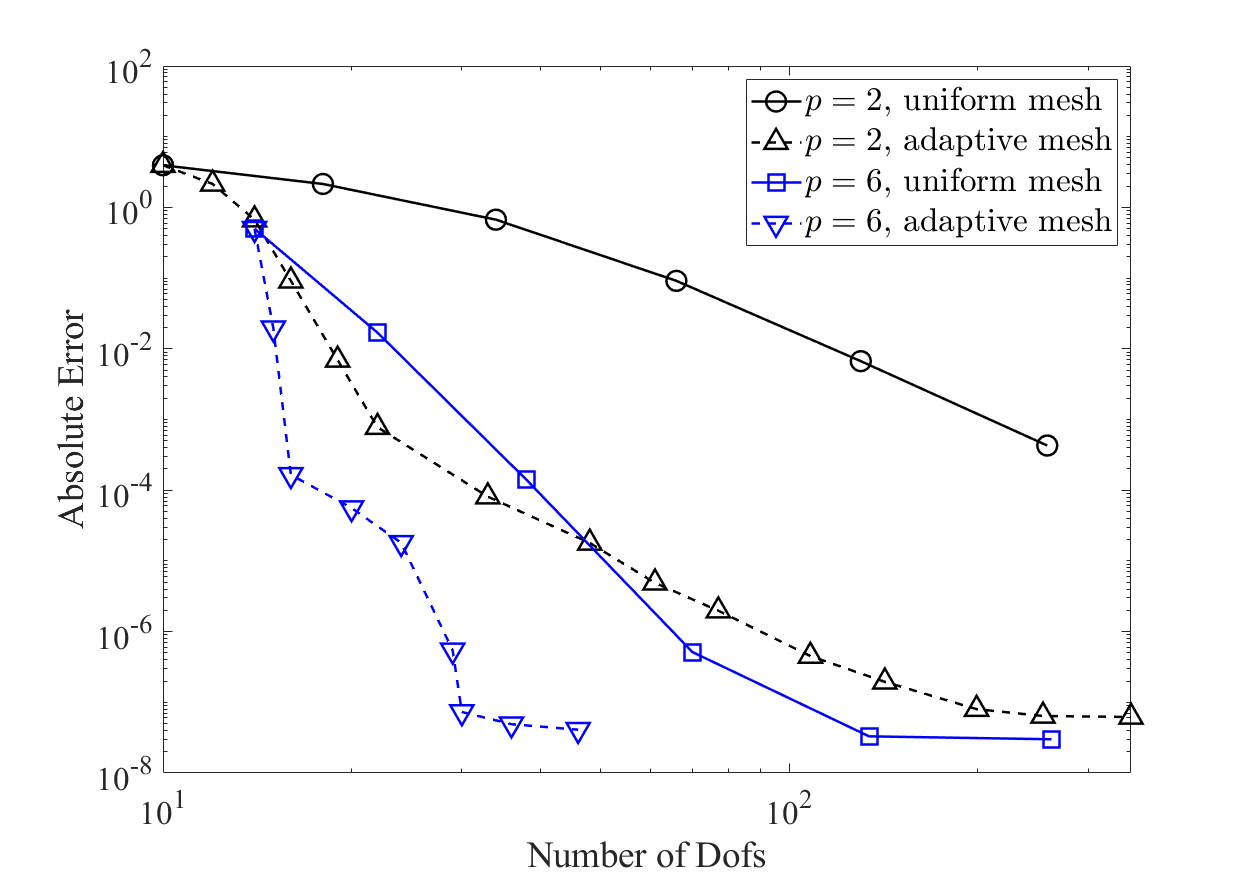}
    \end{minipage}
    \begin{minipage}[h]{0.49\textwidth}
        \includegraphics[width=1\linewidth]{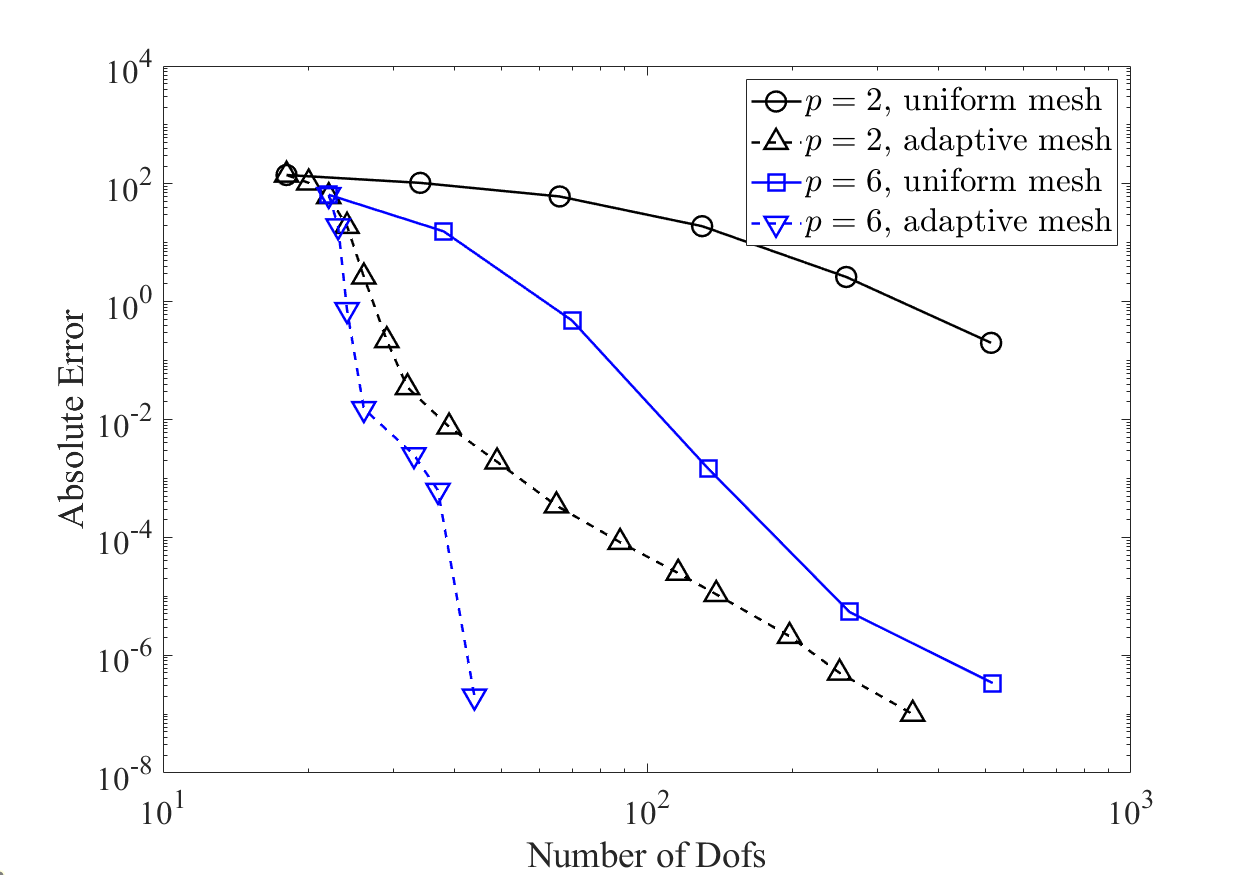}
    \end{minipage}
    \caption{Convergence history of the total energy for lithium (left) and aluminum (right) atoms for $p=2$ and $p=6$ under uniform mesh and $h$-adaptive mesh, respectively.}
    \label{fig:Radial_Cal_Li_and_Al}
\end{figure}

In the simulations, the computational domain is set as $[0,20]$. The referenced ground state energies are obtained from the NIST database \cite{kotochigova2009atomicnist}. Specifically, for the for the lithium atom is $E_\mathrm{ref}=-7.335195~\mathrm{a.u.}$, and for the aluminum atom  $E_\mathrm{ref}=-241.315573~\mathrm{a.u.}$. We test the presented method with order $p=2$ and $p=6$. To illustrate the effectiveness of the mesh adaptivity, we also perform simulations on the uniform mesh. Results are demonstrated in \Cref{fig:Radial_Cal_Li_and_Al}.

The convergence of the energy with respect to the number of Dofs for the lithium atom is shown on the left of \Cref{fig:Radial_Cal_Li_and_Al}. For $p=2$, we observe that a uniform mesh achieves only about $10^{-3}$ a.u. accuracy with $256$ Dofs. In contrast, the $h$-adaptive mesh reaches an accuracy of $10^{-7}$ with a similar number of Dofs, effectively reproducing the NIST database result. This indicates that the adaptive method significantly outperforms the uniform mesh in terms of accuracy. A similar trend is observed for $p=6$, where the adaptive method reproduces the NIST result using only around $30$ Dofs, demonstrating the high accuracy achievable with the high-order adaptive approach and underscoring its superiority over the lower-order method. This conclusion also holds true for the aluminum atom, as shown on the right side of \Cref{fig:Radial_Cal_Li_and_Al}, where fewer than $50$ Dofs are sufficient to reach the desired accuracy with $p=6$, compared to more than $300$ Dofs required for $p=2$ on the adaptive mesh.

\subsubsection{Ground state energy in the periodic table}
In the one-dimensional radial calculation, the Dofs used for $p=3$ to reproduce the ground state energy of elements in the periodic table are presented in \Cref{tab:Radial_periodic_table}. Reference data is obtained from the NIST database \cite{kotochigova2009atomicnist} for single atoms with atomic numbers ranging from $1$ to $92$. The computational domain is set to $[0, 20]$ for the first three rows and $[0, 50]$ for the subsequent four rows. It is observed that an accuracy of $10^{-6}$ is achieved using at most $120$ Dofs for the first three rows and at most $300$ Dofs for the remaining rows, demonstrating the effectiveness of our adaptive isogeometric solver.

\subsection{All-electron calculations}
In this subsection, the $h$-adaptive isogeometric solver is validated through a series of experiments involving all-electron calculations in three dimensions. We begin by thoroughly testing the solver using a helium atom as a benchmark example. Subsequently, the method is applied to various molecules, including lithium hydride, methane, and benzene.

\subsubsection{Helium atom}
\begin{figure}[h!]
\centering
\includegraphics[width=.5\linewidth]{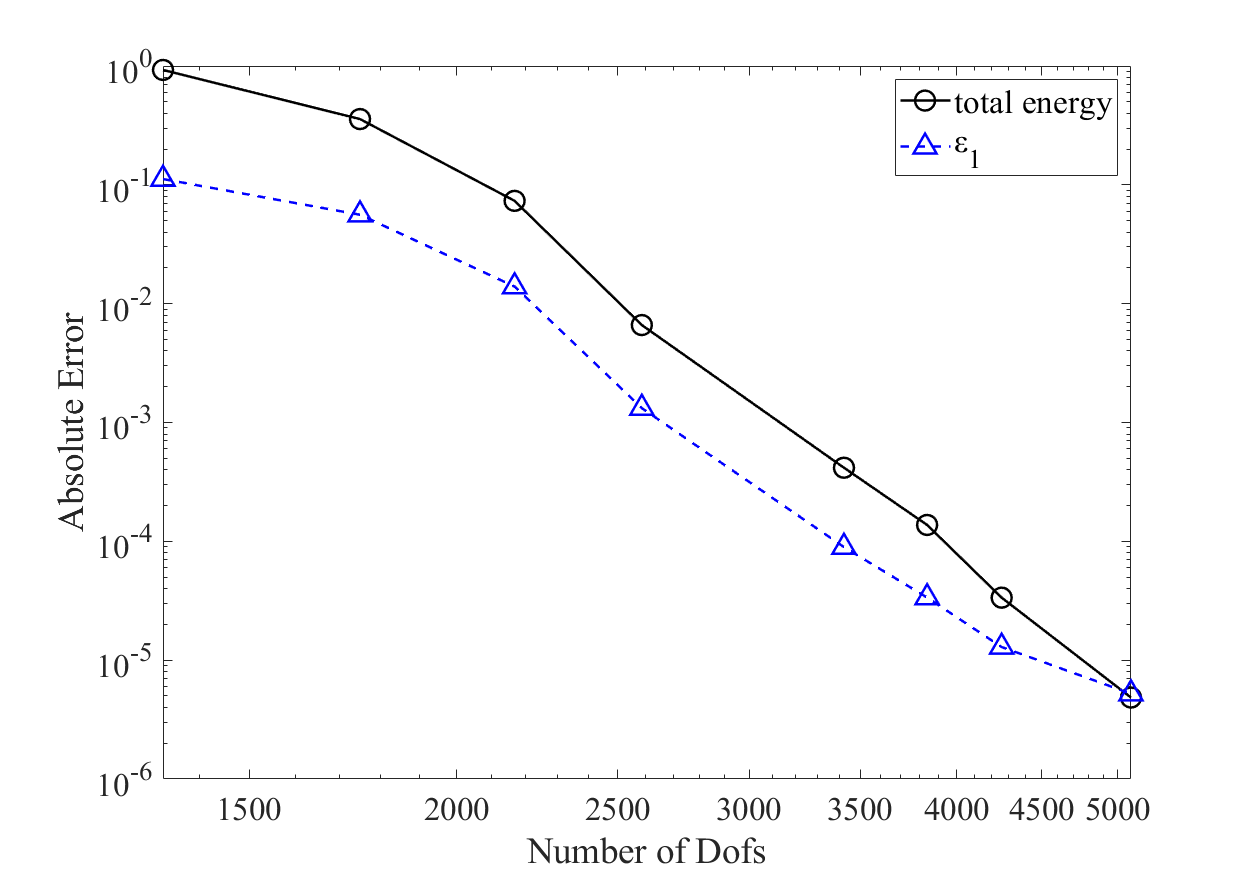}
\includegraphics[width=.4\linewidth]{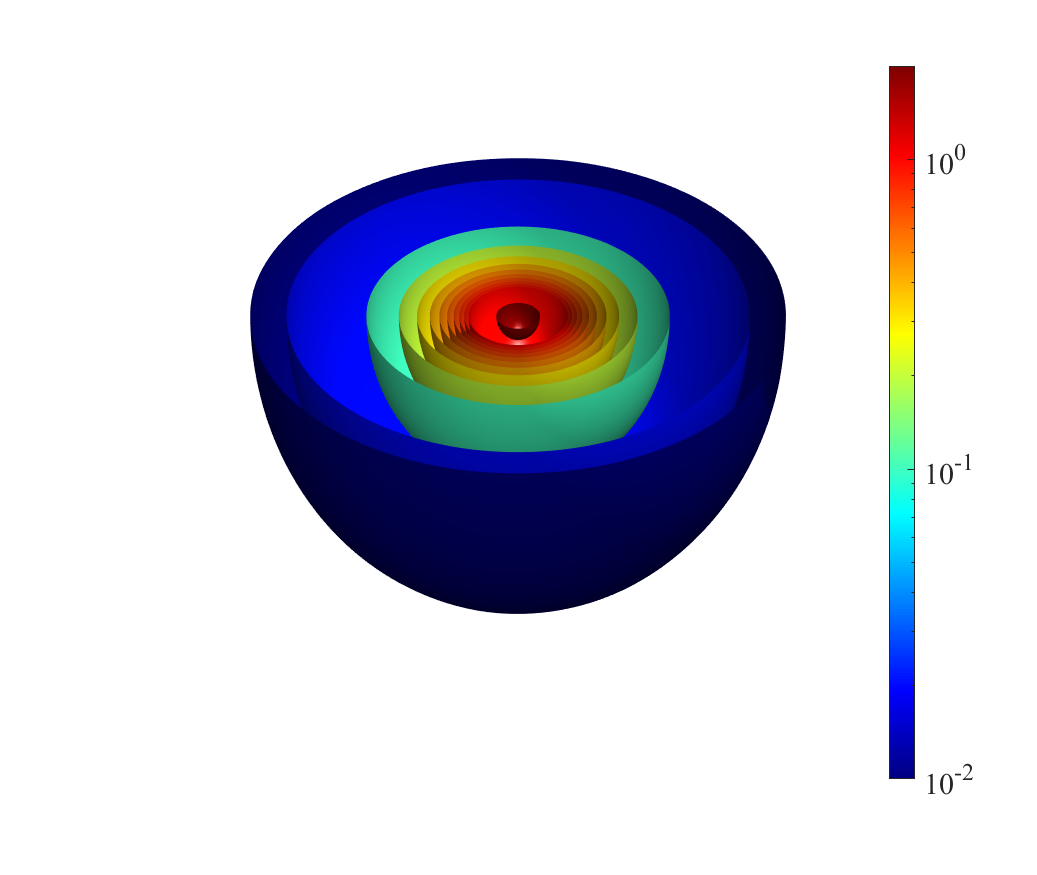}
\caption{Left: Convergence history of the total energy and first eigenvalue for a helium atom ($p=3$). Right: The sliced density in the $X$-$Y$ plane for a helium atom $(p=3)$.}
\label{fig:Helium_p3_Aerror}
\end{figure}

The $h$-adaptive isogeometric solver is used to simulate a helium atom, employing the function-based indicator in equation \eqref{indicator_function_based}. The computational domain is defined as $\Omega = [-10,10]^3$. Through adaptive local refinement, we compute the ground state total energy and the eigenvalue $\varepsilon_1$ for the helium atom. The reference values, generated from the state-of-art package \texttt{NWChem} using the \emph{aug-cc-pv6z} basis set \cite{valiev2010nwchem}, are a total energy of $E_{\mathrm{ref}} = -2.834289 \, \mathrm{a.u.}$ and an eigenvalue of $\varepsilon_{\mathrm{ref}} = -0.570209$. The convergence history of the ground state total energy and the eigenvalue is shown on the left side of \Cref{fig:Helium_p3_Aerror}, illustrating that the numerical accuracy $10^{-3}$ is achieved within $3000$ Dofs. Meanwhile, an accuracy of $10^{-5}$ Hartree is reached in the final stage of mesh adaptation, involving approximately $5000$ Dofs. The right side of \Cref{fig:Helium_p3_Aerror} shows a sliced isosurface of the electron density on the $X$-$Y$ plane for the helium atom during the final iteration step, highlighting its spherical symmetry.

\begin{figure}[!h]
    \centering
    \begin{minipage}[h]{0.49\textwidth}
        \includegraphics[width=.9\linewidth]{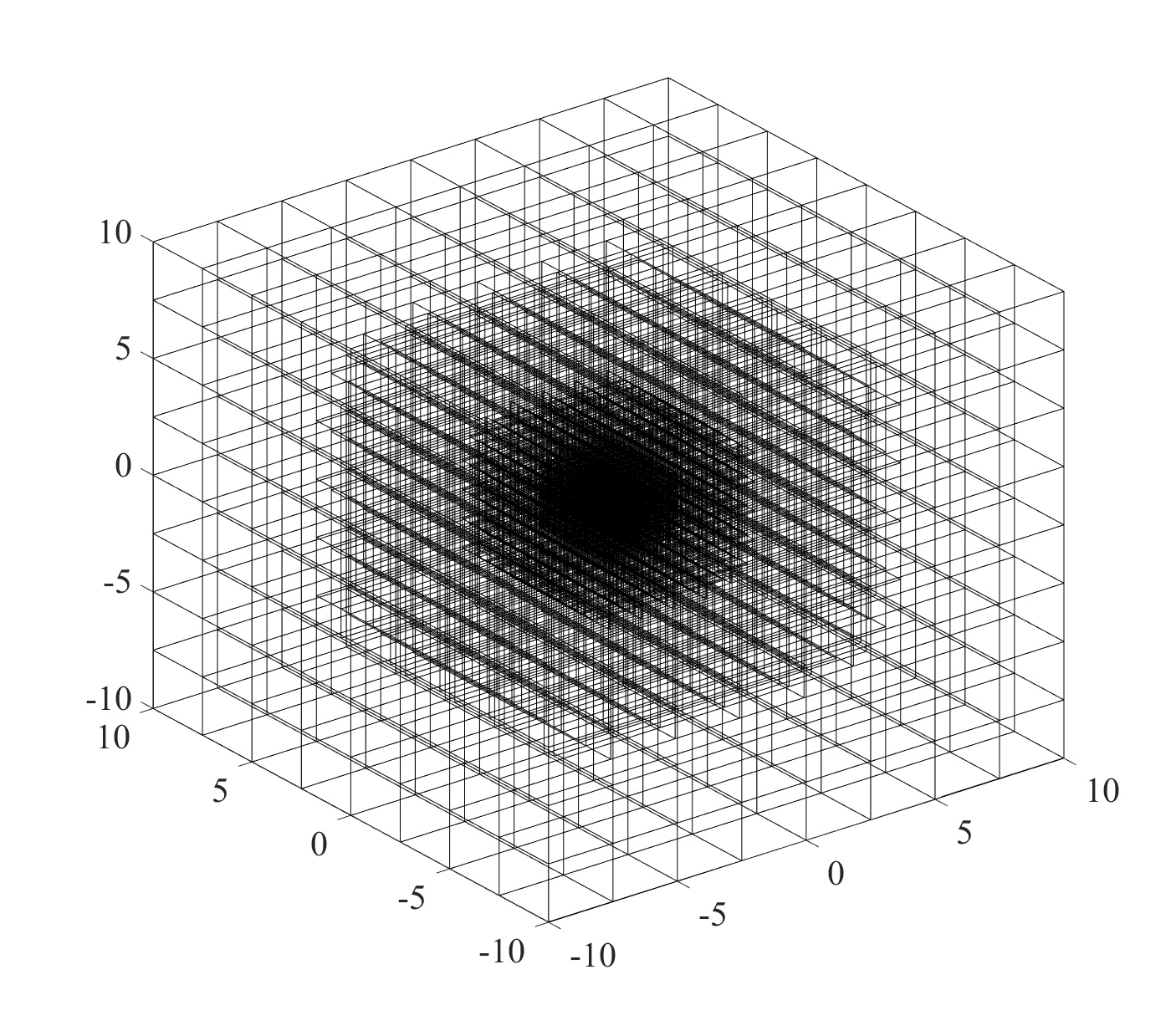}
    \end{minipage}
    \begin{minipage}[h]{0.49\textwidth}
        \includegraphics[width=.9\linewidth]{figure/Helium_p3/Helium_p3_2Dmesh.png}
    \end{minipage}
    \caption{The three-dimensional mesh (left) and the sliced mesh in the $X$-$Y$ plane (right) for a helium atom ($p=3$).}
    \label{fig:Helium_p3_mesh}
\end{figure}

The final mesh for the helium atom is depicted in \Cref{fig:Helium_p3_mesh}, with a three-dimensional visualization on the left and a sliced mesh on the $X$-$Y$ plane shown on the right. It is evident that the mesh grids are significantly denser around the singularity compared to regions farther away. This distribution aligns with the behavior of the wavefunction, which exhibits large values and steep variations near the singularity, while approaching zero in regions distant from it. Specifically, the smallest mesh size is $0.0012$ a.u., whereas the largest mesh size is $2.5$ a.u., demonstrating the capability of the $h$-adaptive solver to generate meshes with considerable variation in grid size. This adaptability enables the solver to effectively resolve the problem with a small number of Dofs, substantially reducing computational cost.

\begin{figure}[h!]
    \centering
    \begin{minipage}[h]{0.49\textwidth}
        \includegraphics[width=1.1\linewidth]{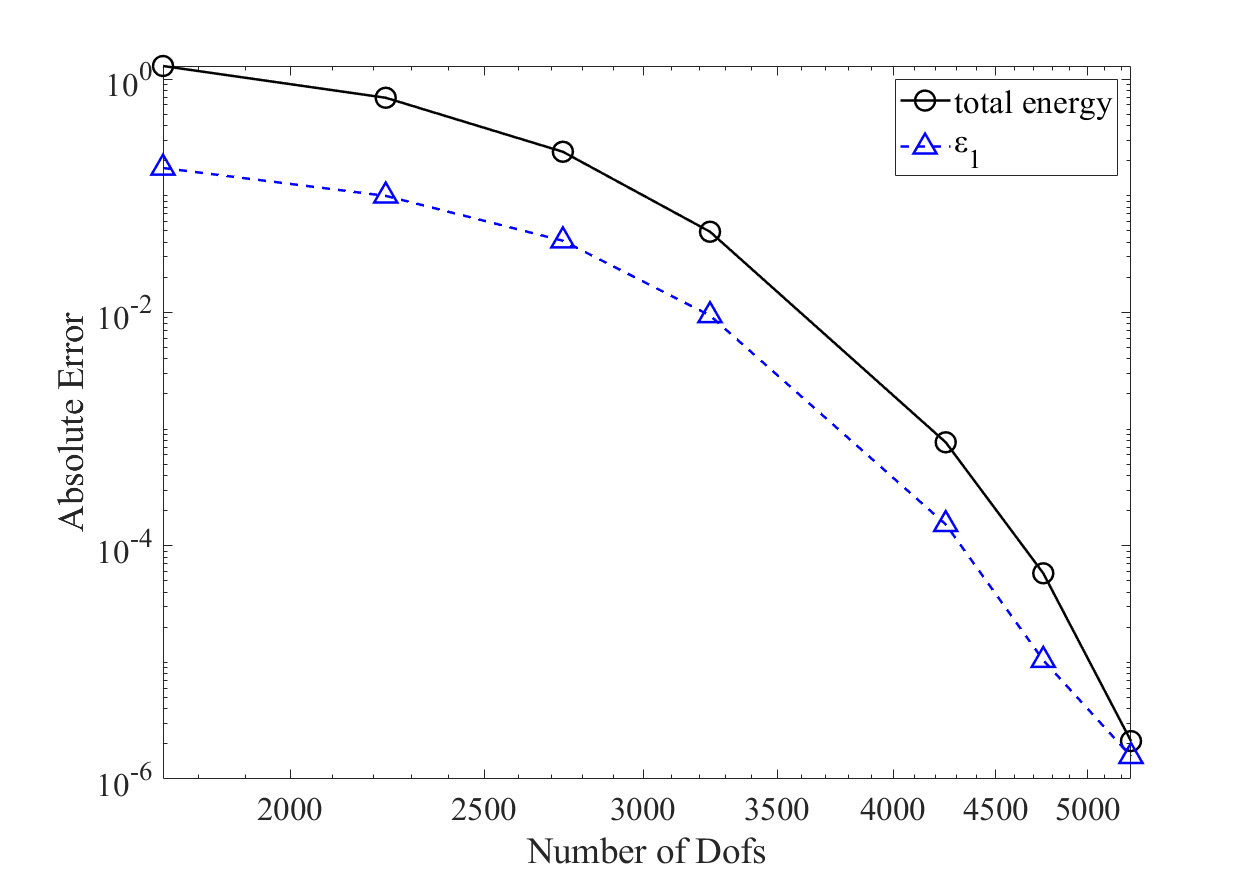}
    \end{minipage}
    \begin{minipage}[h]{0.49\textwidth}
        \includegraphics[width=0.9\linewidth]{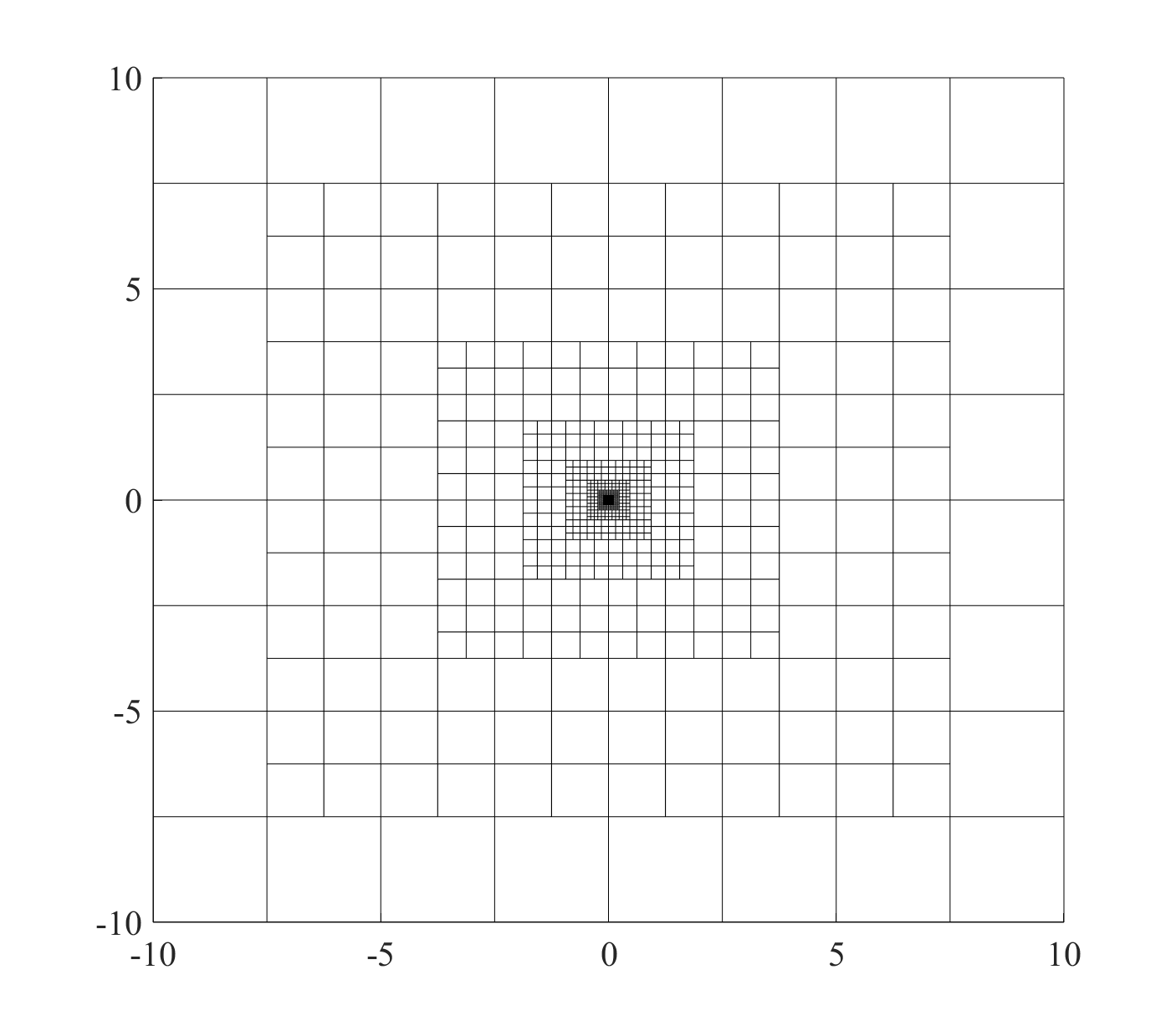}
    \end{minipage}
    \caption{Left: Convergence history of the total energy and first eigenvalue for a helium atom ($p=4$). Right: The sliced mesh in the $X$-$Y$ plane for a helium atom ($p=4$).}
    \label{fig:Helium_p4_result}
\end{figure}

We further analyze the helium atom using a THB-spline function space with $p=4$ to evaluate the advantages of high-order splines. The sliced mesh on the $X$-$Y$ plane and the absolute error for $p=4$ are shown in Figure \ref{fig:Helium_p4_result}. Both the total energy of the ground state and the first eigenvalue for $p=4$ exhibit a decreasing trend, demonstrating the effectiveness of the $h$-adaptive isogeometric solver, as depicted in Figure \ref{fig:Helium_p4_result}. However, compared to the sliced mesh for $p=3$, the refined region is broader, which is caused by the increased local support of splines with higher order. This results in a slightly larger number of Dofs for higher degrees, as shown in Figure \ref{fig:Helium_p4_result}. Therefore, for subsequent simulations, we will use the THB-spline basis with $p=3$.

\subsubsection{Lithium hydride}
\begin{figure}[h!]
    \centering
    \begin{minipage}[h]{0.49\textwidth}
        \includegraphics[width=1\linewidth]{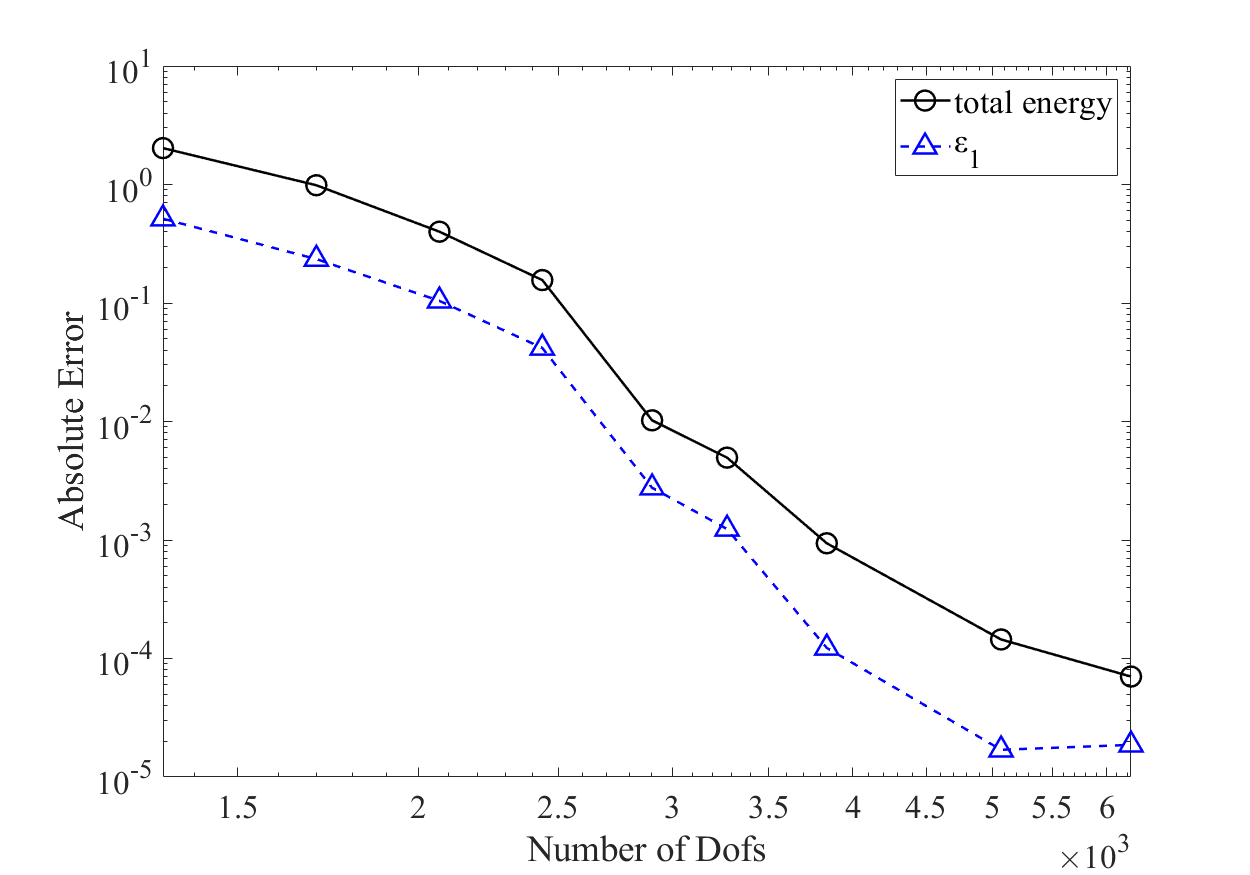}
    \end{minipage}
    \begin{minipage}[h]{0.49\textwidth}
        \includegraphics[width=1\linewidth]{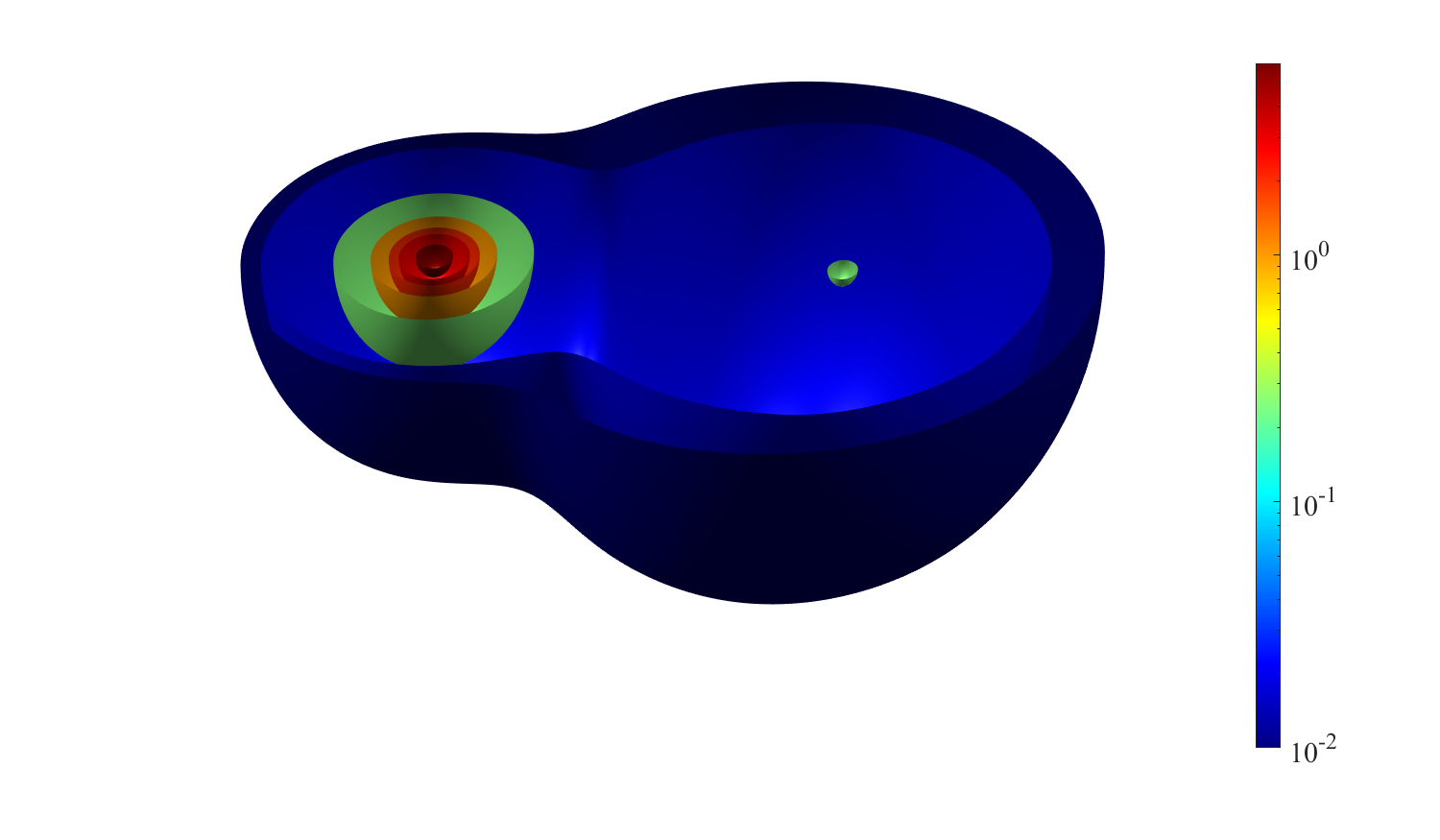}
    \end{minipage}
    \caption{Left: Convergence history of the total energy and first eigenvalue for a LiH molecule ($p=3$). Right: The sliced density in the $X$-$Y$ plane for a LiH molecule $(p=3)$.}
    \label{fig:LiH_p3_Aerror_and_isosurface}
\end{figure}

We then apply the presented $h$-adaptive isogeometric solver to find the ground state of the lithium hydride (LiH) molecule, with the computational domain set as $\Omega=[-10,10]^3$. The total energy of the ground state and the first eigenvalue $\varepsilon_1$ are calculated by our \Cref{alg:adaptivemesh}. The referenced value is obtained \texttt{NWChem} with the \emph{aug-cc-pv6z} basis set, where $E_{\mathrm{ref}} = -7.918733~ \mathrm{a.u.}$ and the first eigenvalue $\varepsilon_{\mathrm{ref}} = -1.841358$. The convergence history of the total energy of the ground state and the eigenvalue is demonstrated on the left side of \Cref{fig:LiH_p3_Aerror_and_isosurface}, obtaining the accuracy $10^{-3} \mathrm{~Hartree/particle}$ within $4000$ Dofs. Moreover, the accuracy lower than $10^{-4} \mathrm{~Hartree/particle}$ is achieved in the final adaptive local refinement, utilizing around $6000$ Dofs in the simulations. The sliced electron density for the LiH molecule is drawn on the right side of \Cref{fig:LiH_p3_Aerror_and_isosurface}, from which the positions of lithium atom (left) and hydrogen atom (right) can be seen.

\begin{figure}[h]
    \centering
    \begin{minipage}[h]{0.49\textwidth}
        \includegraphics[width=.85\linewidth]{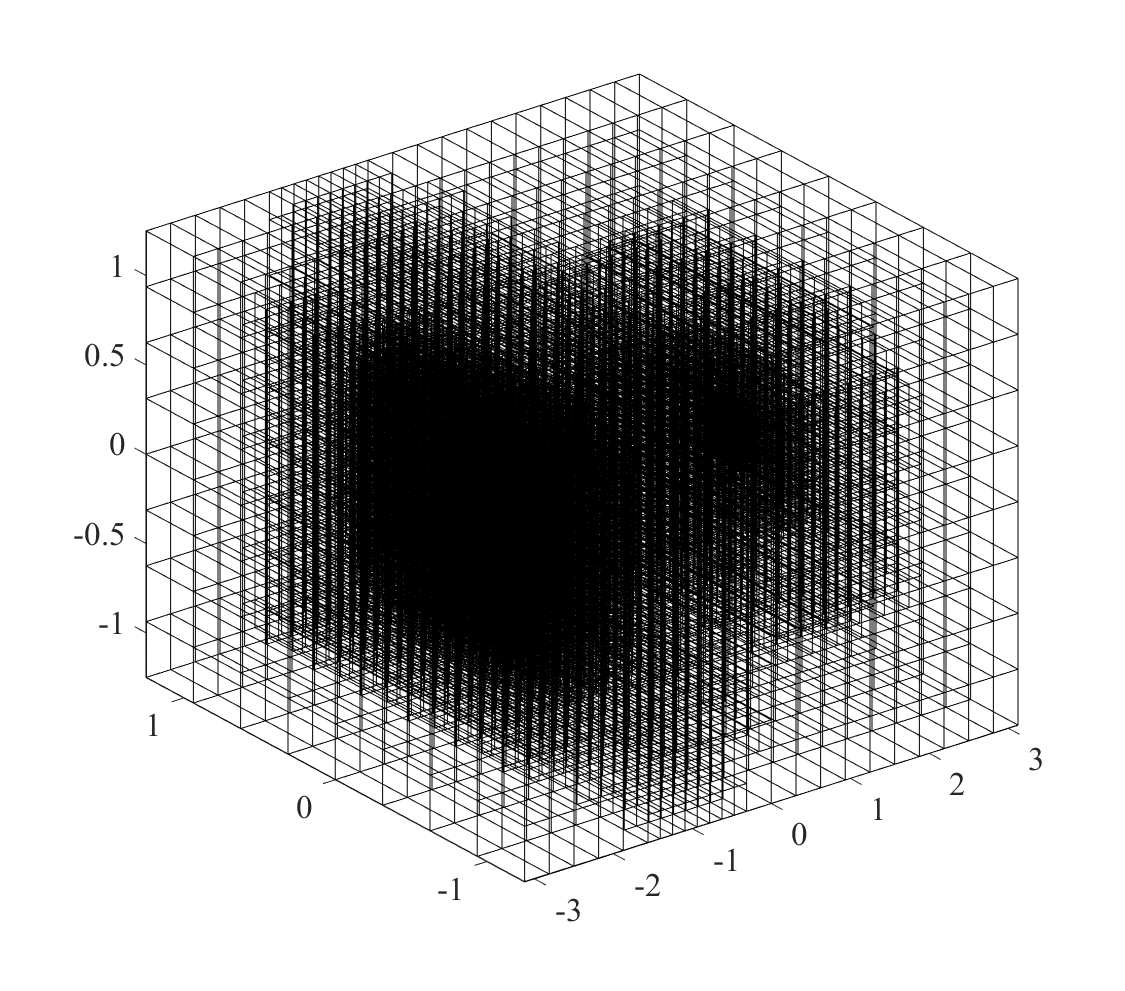}
    \end{minipage}
    \begin{minipage}[h]{0.49\textwidth}
        \includegraphics[width=.85\linewidth]{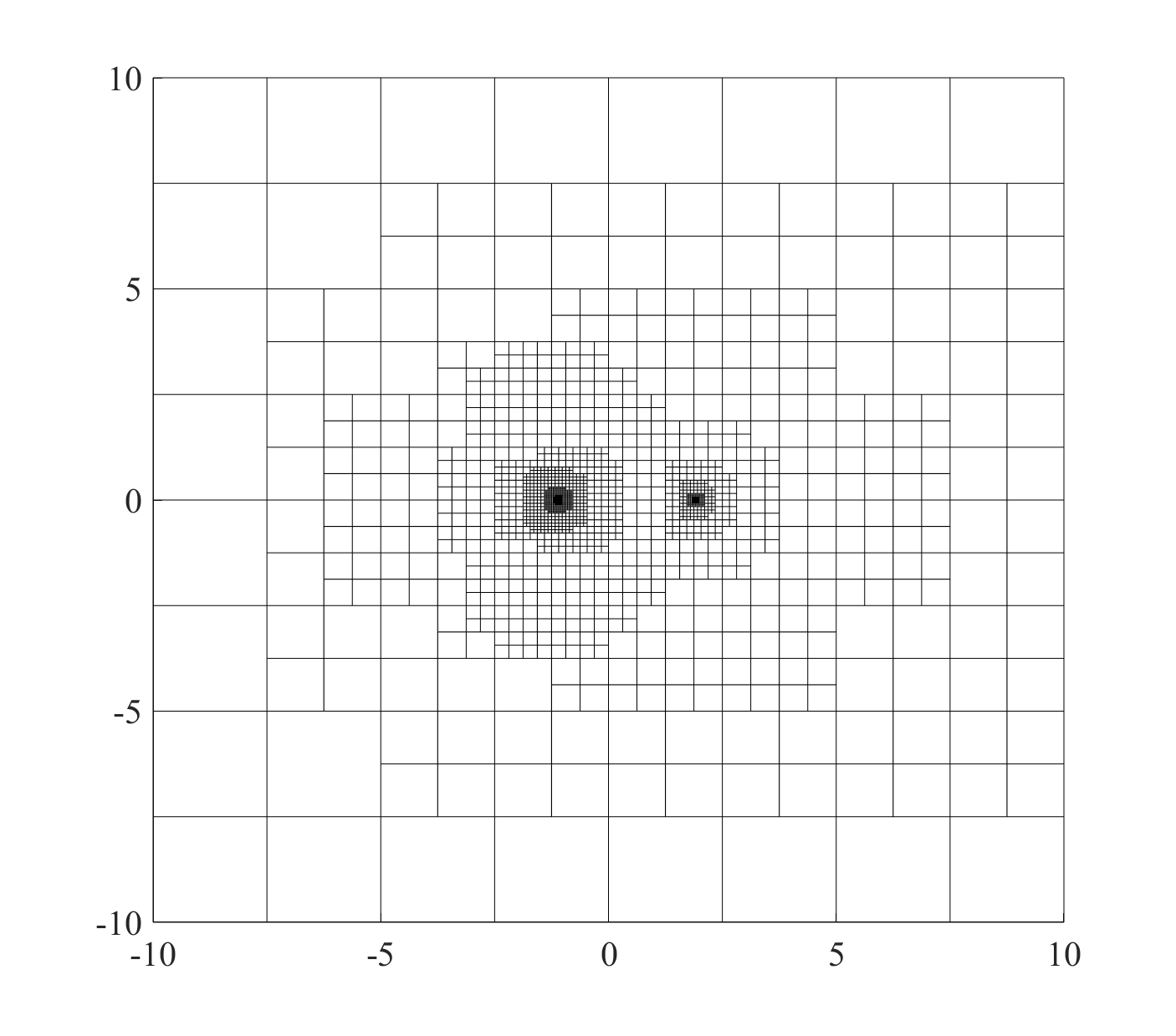}
    \end{minipage}
    \caption{The internal three-dimensional mesh (left) and the sliced mesh in the $X$-$Y$ plane (right) for a LiH molecule ($p=3$).}
    \label{fig:LiH_mesh}
\end{figure}

The three-dimensional visualization for the mesh is displayed on the left  of \Cref{fig:LiH_mesh} and the sliced mesh on the $X$-$Y$ plane is displayed on the right. We observe the finer grids are concentrated around the lithium atom ($-1.109786,0,0$) and the hydrogen atom ($1.919776,0,0$), which aligns with the characteristics of the wavefunction in all-electron calculations. To be specific, the mesh size near the lithium atom is $0.0012~\mathrm{a.u.}$ noticeably denser compared to that $0.0195~\mathrm{a.u.}$ near the hydrogen atom, which is consistent with our expectations. 

\subsubsection{Methane}
\begin{figure}[h!]
\centering
\includegraphics[width=.55\linewidth]{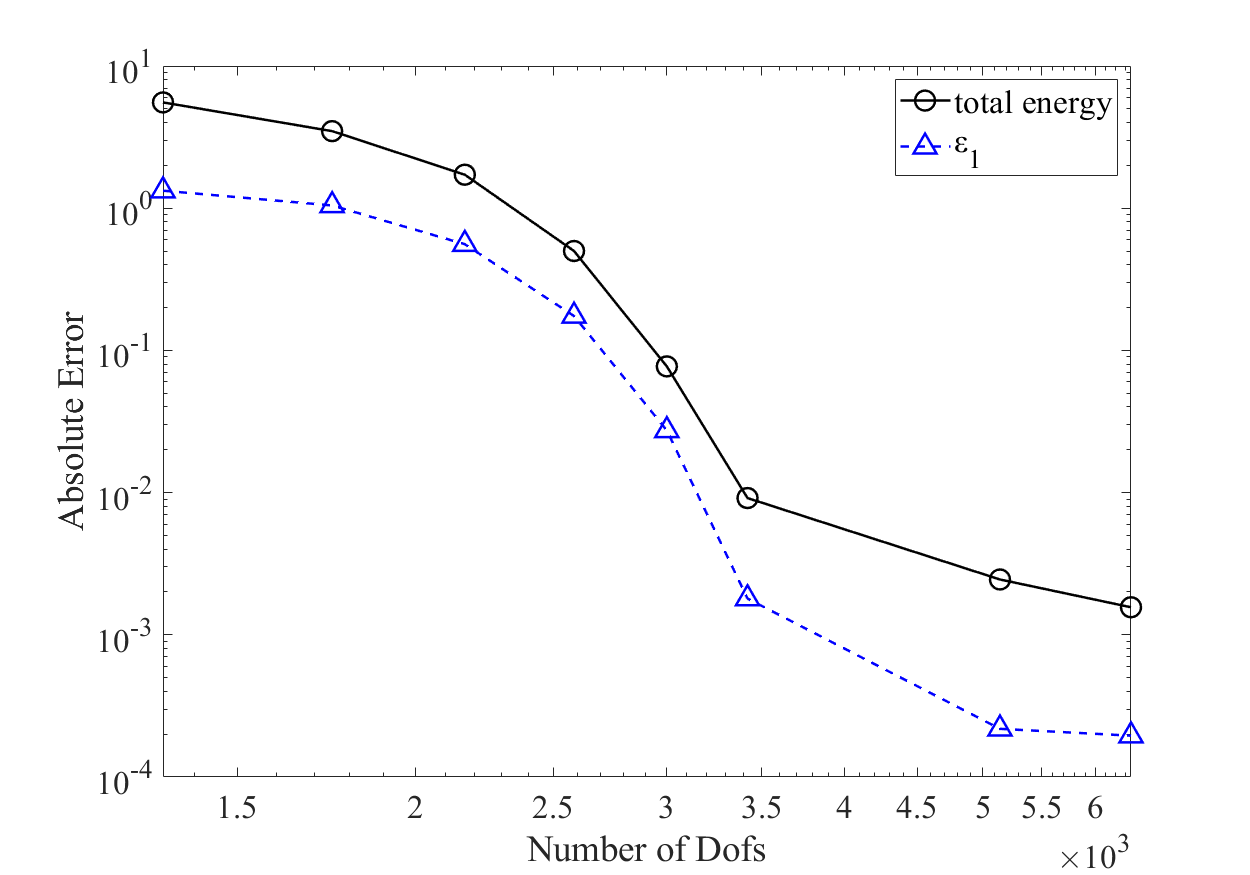}
\caption{Convergence history of the total energy and first eigenvalue for a methane molecule ($p=3$).}
\label{fig:CH4_p3_Aerror}
\end{figure}

The $h$-adaptive isogeometric solver is also applied to the simulation of a methane molecule, serving as the second experiment for the molecules. In contrast to the lithium hydrogen molecule, the atoms of a methane molecule do not lie in a plane. We define the computational domain as $\Omega = [-10,10]^3$. The total energy of the ground state and the first eigenvalue are computed with \Cref{alg:adaptivemesh}. Based on the reference of the ground state total energy $E_{\mathrm{ref}} = -40.119811 ~\mathrm{a.u.}$ and the first eigenvalue $\varepsilon_{\mathrm{ref}} = -9.761248$ from the package \texttt{NWChem} with \emph{aug-cc-pv6z} basis set \cite{valiev2010nwchem}, the numerical accuracy $10^{-3} \mathrm{~Hartree/particle}$ for the total energy of the ground state is touched within $6355$ Dofs, which is a significantly reduction compared to the more than $100000$ Dofs required with $p=3$ in \cite{temizer2020nurbsfordft}. Moreover, the accuracy $10^{-3}$ is achieved for the first eigenvalue $\varepsilon_1$ within $4000$ Dofs in \Cref{fig:CH4_p3_Aerror}, demonstrating the effective all-electron calculations for the methane molecule.

\begin{figure}[h!]
    \centering
    \begin{minipage}[h]{0.49\textwidth}
        \includegraphics[width=.8\linewidth]{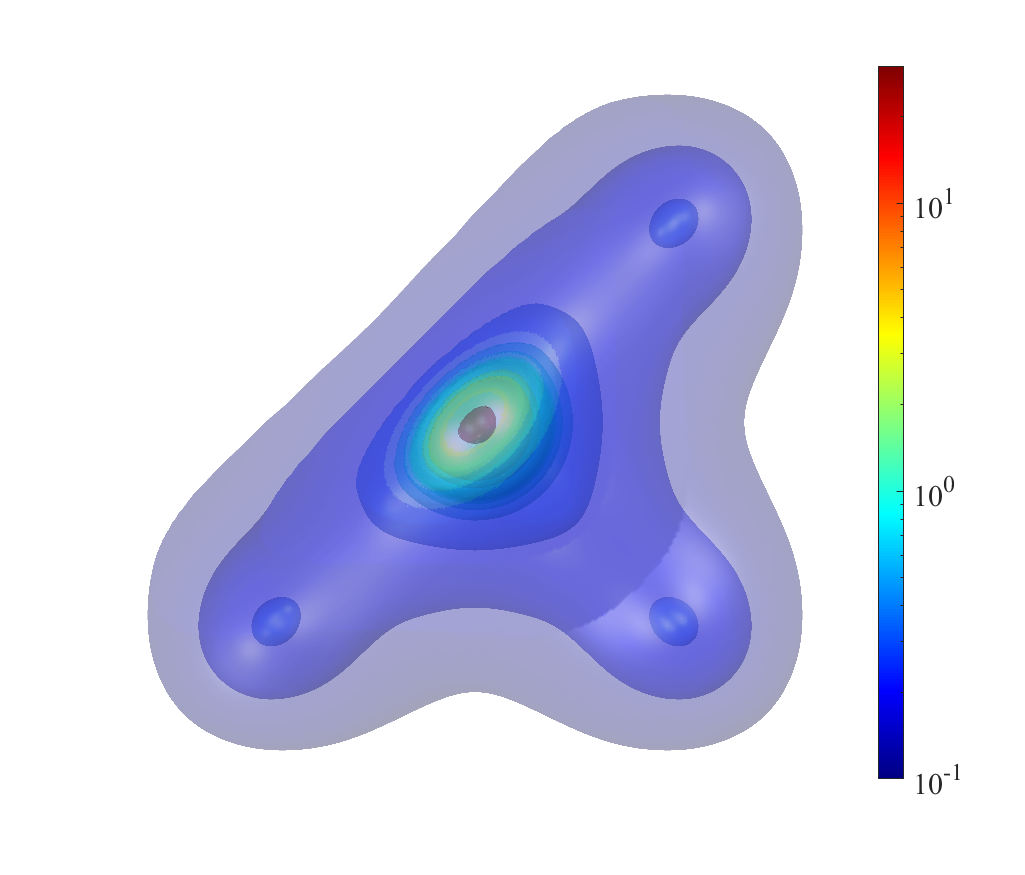}
    \end{minipage}
    \begin{minipage}[h]{0.49\textwidth}
        \includegraphics[width=.8\linewidth]{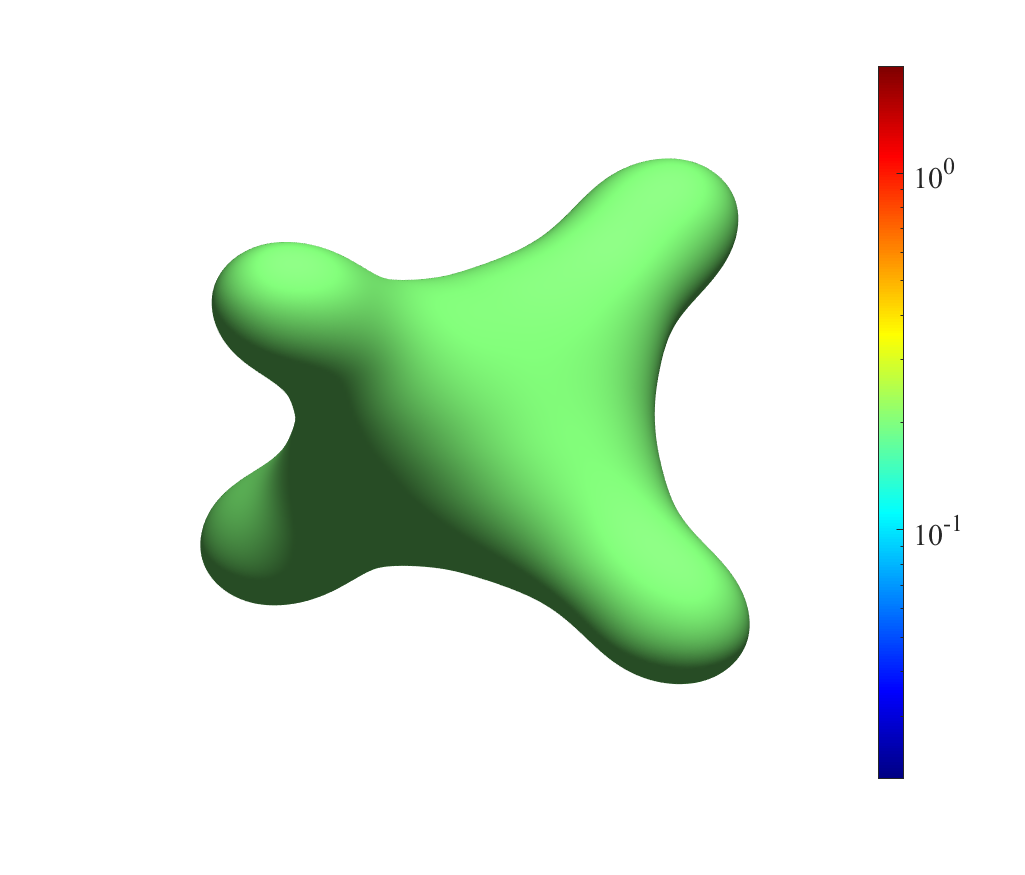}
    \end{minipage}
    \caption{The sliced density cut by a plane (left) and the three-dimensional density (right) for a methane molecule ($p=3$).}
    \label{fig:CH4_isosurface}
\end{figure}

\begin{figure}[h!]
    \centering
    \includegraphics[width=.5\linewidth]{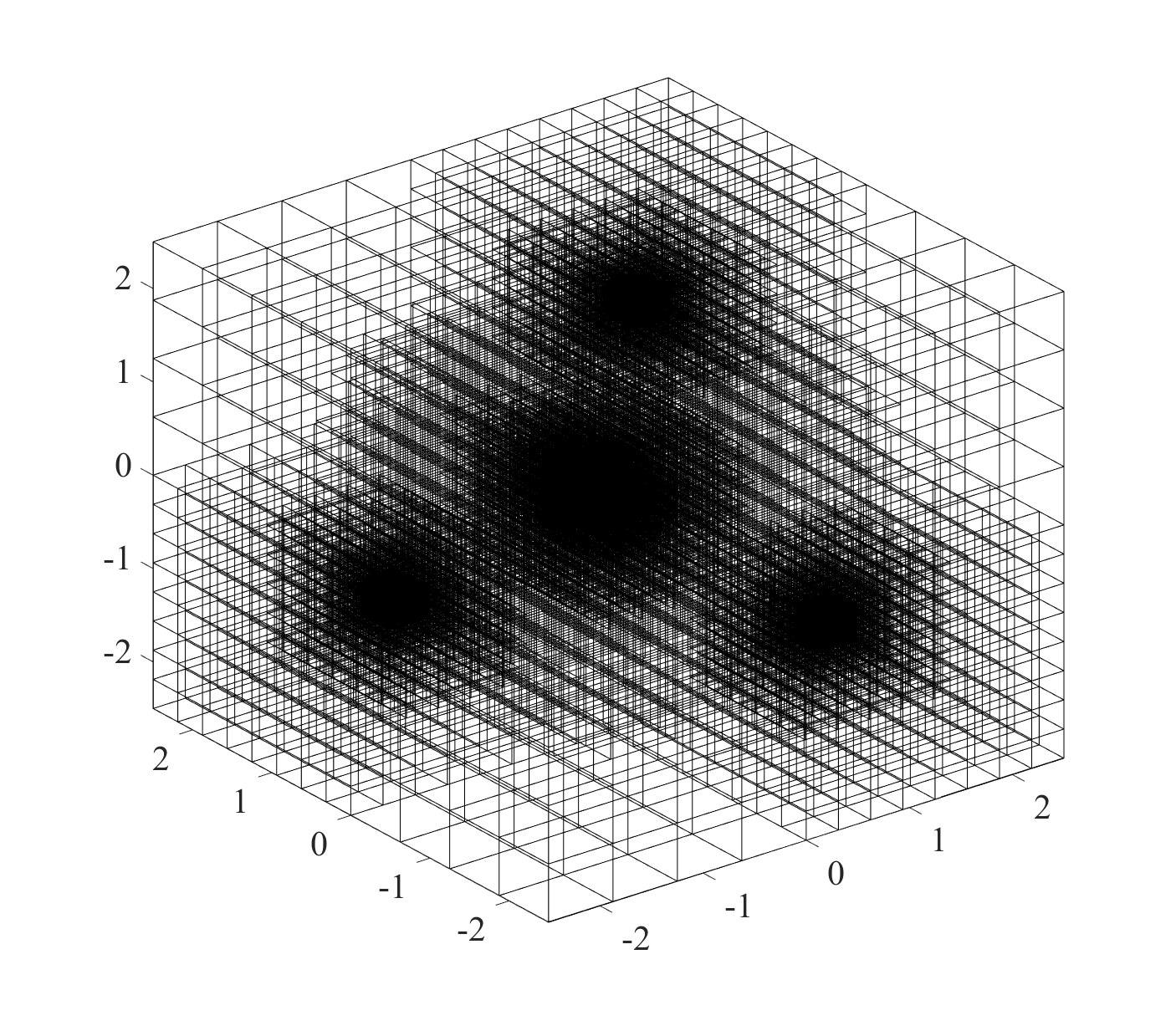}
    \caption{The internal three-dimensional mesh for a methane molecule ($p=3$).}
    \label{fig:CH4_mesh}
\end{figure}

The sliced density cut by a plane with its normal vector ($1,1,-1$) is drawn on the left side in \Cref{fig:CH4_isosurface}, where the electron density of three hydrogen atoms and the carbon atom is presented. Meanwhile, the isosurface of the density is displayed on the right side in \Cref{fig:CH4_isosurface}, showing the density of four hydrogen atoms. The final adaptive local refinement result for the methane molecule is presented in \Cref{fig:CH4_mesh}, constrained within finite three-dimensional coordinates axes. We observe that the refinement is assembled around the carbon and hydrogen atoms, while the coarse mesh sizes are reached in the far-field region, which is consistent with the behavior of the wavefunctions in all-electron calculations. More specifically, compared to the hydrogen atoms, the carbon atom, due to its higher atomic number and the greater number of electrons, exhibits a denser mesh size at $0.0006~\mathrm{a.u.}$ in \Cref{fig:CH4_mesh}, demonstrating the capability of the $h$-adaptive isogeometric solver in simulating the all-electron KS equation for the methane molecule. Consequently, we anticipate that the solver can be applied to the molecule with more atoms (e.g., benzene). with the corresponding results to be presented in the next experiment.

\subsubsection{Benzene}
\begin{figure}[h!]
    \centering
    \begin{minipage}[h]{0.49\textwidth}
        \includegraphics[width=1.1\linewidth]{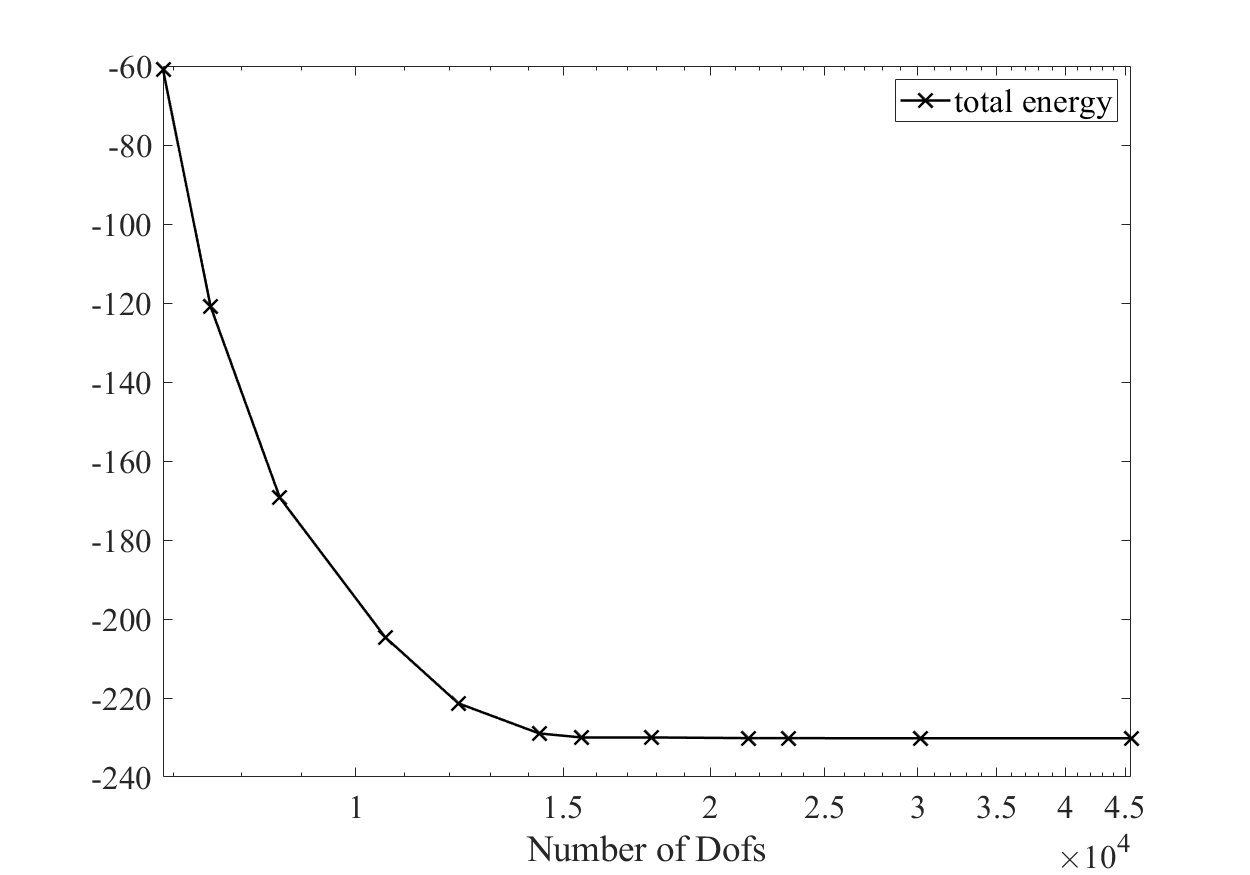}
    \end{minipage}
    \begin{minipage}[h]{0.49\textwidth}
        \includegraphics[width=.9\linewidth]{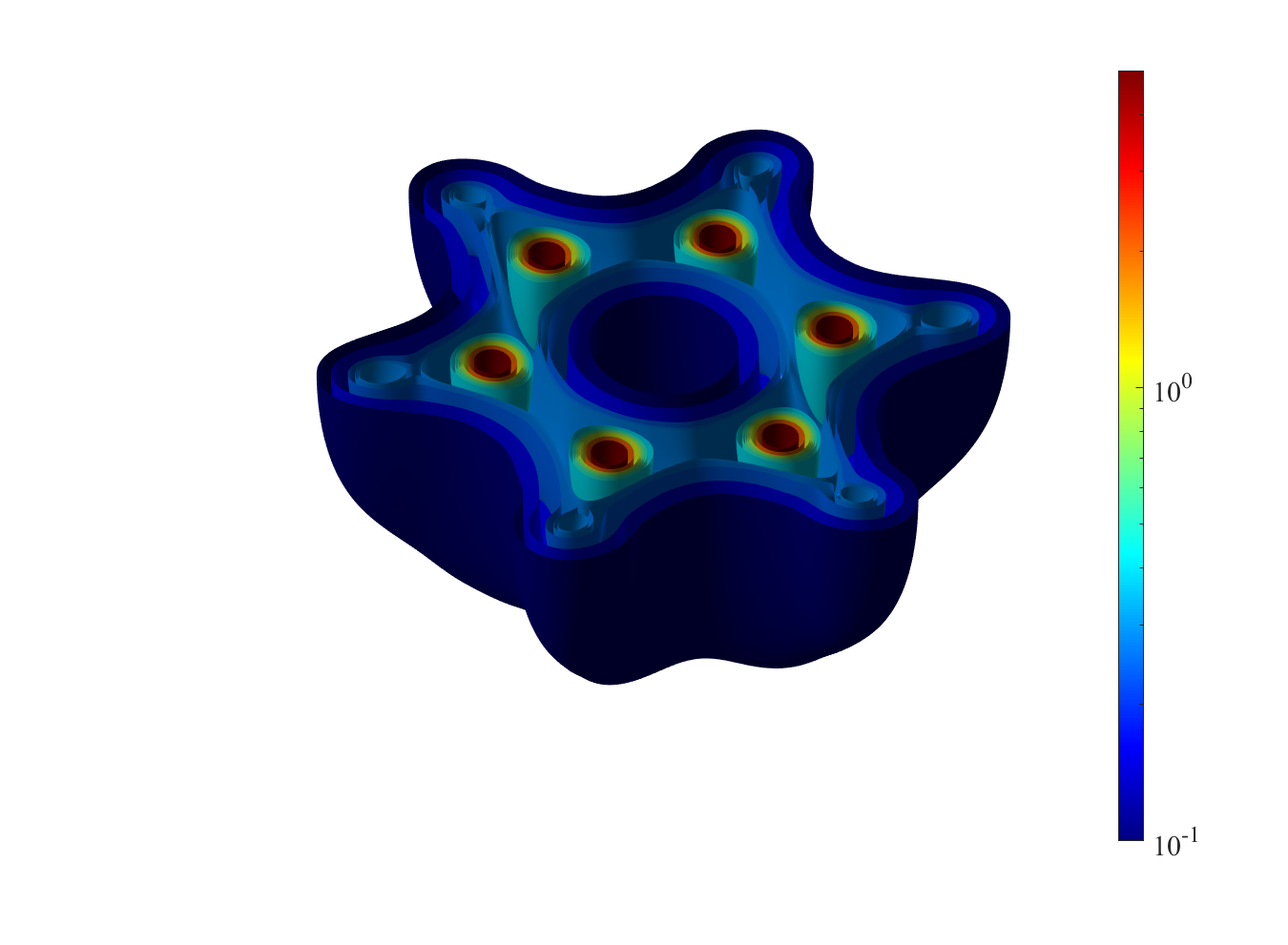}
    \end{minipage}
    \caption{Left: Convergence history of the total energy for a benzene molecule ($p=3$). Right: The sliced density in the $X$-$Y$ plane for a benzene molecule $(p=3)$.}
    \label{fig:Benzene_convergence_and_isosurface}
\end{figure}

In the all-electron calculations for the molecules, the $h$-adaptive isogeometric solver is utilized for the simulation of the benzene (C$_6$H$_6$) molecule. We specially set the computational domain as $\Omega=[-20,20]^3$, based on the multi-atomic structure of the benzene molecule. The calculated ground state total energy for the benzene molecule is $-230.193754$ a.u. from our solver and the convergence history of the total energy is presented on the left side in \Cref{fig:Benzene_convergence_and_isosurface}. To be specific, the convergence history indicates that the ground state energy begins to show a converging trend around $15000$ Dofs and continues to converge at $45000$ Dofs, demonstrating the potential of our solver for simulating the all-electron KS equation of multi-atomic molecules. Additionally, the right side of \Cref{fig:Benzene_convergence_and_isosurface} displays the sliced density of the benzene molecule in the $X$-$Y$ plane based on our solver. It is obvious that different carbon and hydrogen atoms contribute to the overall electron density of the benzene molecule.

\begin{figure}[h!]
\centering
\includegraphics[width=.5\linewidth]{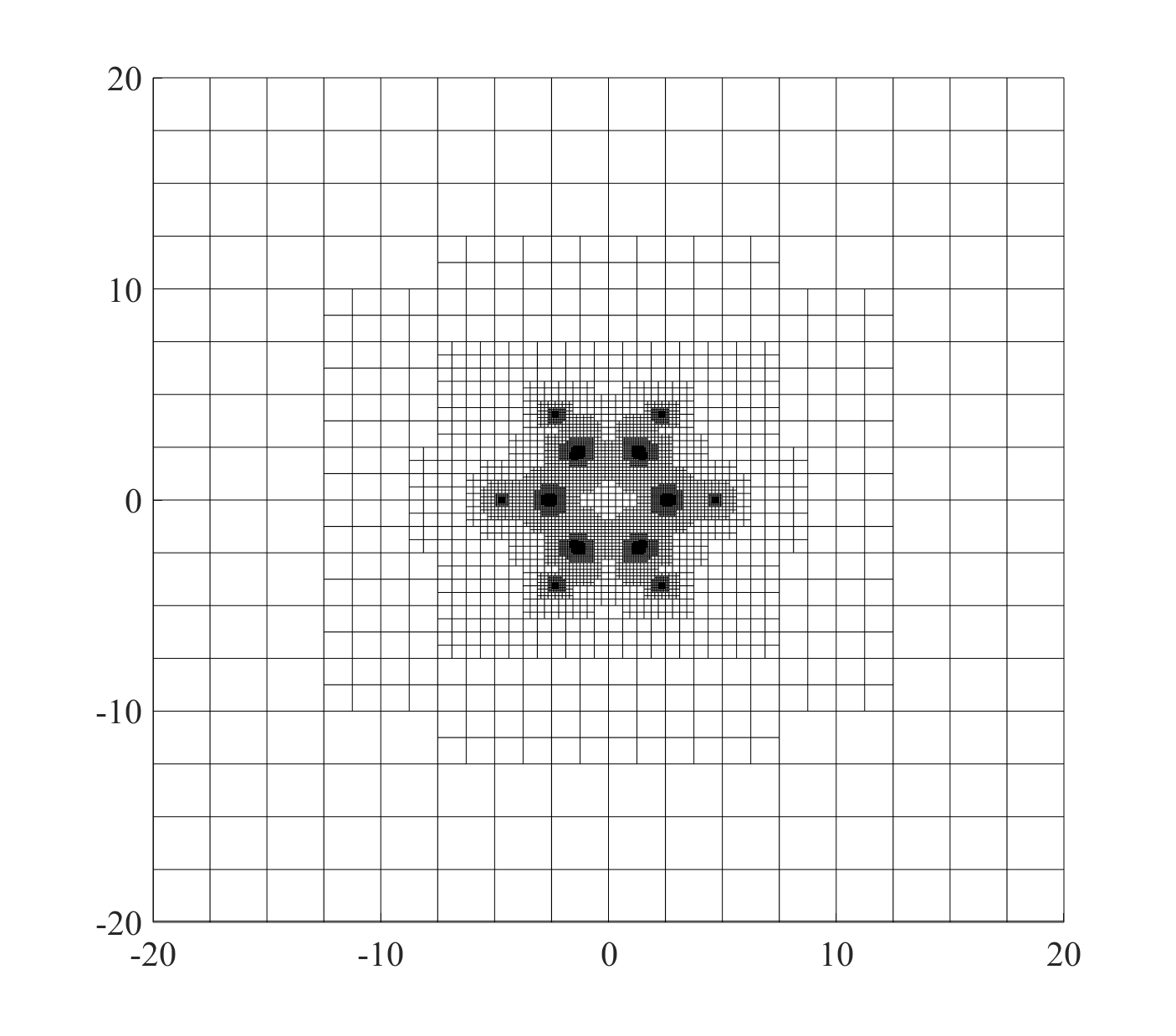}
\caption{The sliced mesh in the $X$-$Y$ plane for a benzene molecule ($p=3$).}
\label{fig:Benzene_2D_mesh}
\end{figure}

In \Cref{fig:Benzene_2D_mesh}, we present the final sliced mesh of the benzene molecule during the iteration of $h$-adaptive solver. The near-nucleic regions of the six carbon atoms and six hydrogen atoms are refined to varying degrees, while the far-field regions exhibit relatively large mesh sizes. Specifically, the densest mesh size near the carbon atom is approximately $0.0006~\mathrm{a.u.}$, while that near the hydrogen atoms is around $0.0391~\mathrm{a.u.}$, illustrating the capability of the $h$-adaptive isogeometric solver in generating high-quality mesh for all-electron KS equation.

\subsection{Binding energy and atomic force}
\begin{figure}[h!]
\centering
\includegraphics[width=.6\linewidth]{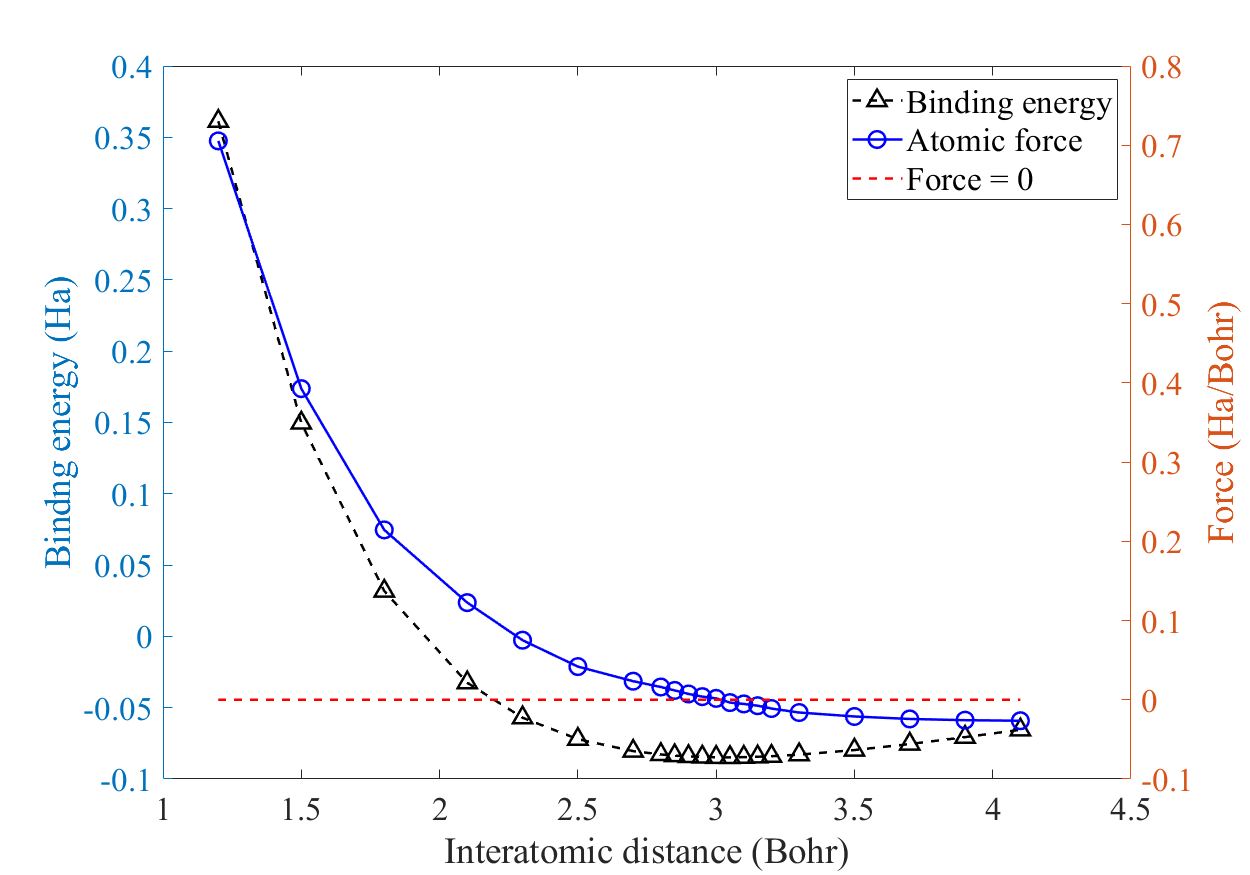}
\caption{The results of the binding energy and the atomic force for a LiH molecule ($p=3$).}
\label{fig:LiH_p3_force}
\end{figure}

In the final numerical experiment, we performed a numerical simulation of the intermolecular chemical bond in the lithium hydride (LiH) molecule. According to valence bond theory, there is only one chemical bond in the LiH molecule. Using our adaptive isogeometric solver, the initial bond length is set to $1.2$ atomic unit, and the evolutionary results of the numerical simulation are presented in \Cref{fig:LiH_p3_force} as the bond length increases. We observe that the binding energy of the LiH molecule reaches a minimum of approximately $-0.0848$ at around $3.05$ atomic units in \Cref{fig:LiH_p3_force}, which is consistent with reference ground state total energy of lithium atom $E_{\mathrm{ref}} = -7.334089$ a.u. and the hydrogen atom $E_{\mathrm{ref}} = -0.5$ a.u. from the software \texttt{NWChem} \cite{valiev2010nwchem}. Furthermore, based on the numerical simulation of the binding energy, we also computed the atomic forces in the LiH molecule. Notably, the atomic force is zero at the minimum position of the binding energy, aligning with the theoretical expectation that the energy is at its lowest in the ground state of a LiH molecule.

\section{Concluding Remarks}\label{sec:concluding_remarks}
In this paper, we propose a novel $h$-adaptive isogeometric solver based on high-order hierarchical splines for the all-electron Kohn--Sham equation. The framework consists of four main modules. The first module, \textit{Solve}, is developed for simulating the Kohn--Sham equation on the current hierarchical mesh by the self-consistent iteration outlined in Algorithm \ref{alg:SCFiteration}. Within this module, we discretize the Kohn--Sham equation using the hierarchical splines function space. The generalized eigenvalue system is effectively solved by LOBPCG with an elliptic preconditioner, where the convergence of the eigensolver is independent of the spline basis order.

Upon completion of the \textit{Solve} module, a residual-type error indicator is specially designed to estimate the residual on the hierarchical mesh in the \textit{Estimate} module. Unlike the residual-type indicators commonly employed in finite element methods, we emphasize that the jump term vanishes and the Laplacian term can be precisely calculated due to the global regularity of high-order splines. In the \textit{Mark} module, cells are marked by the maximum strategy, and in the \textit{Refine} module, the hierarchical mesh is refined with the marked cells, providing a high-quality mesh for the simulation of the Kohn--Sham equation. Simultaneously, an approximated electron density is generated by projecting from the previous hierarchical mesh to the refined hierarchical mesh,  thereby accelerating the convergence of the SCF iteration in the \textit{Solve} module. The effectiveness and accuracy of the $h$-adaptive isogeometric solver are demonstrated through a series of numerical experiments. With our adaptive algorithm, the numerical accuracy $10^{-3} \mathrm{~Hartree/particle}$ is achieved in the simulations of both single atoms and molecules. We note that the accuracy $10^{-3} \mathrm{~Hartree/particle}$ is obtained within $6355$ Dofs for a methane molecule in the all-electron KS equation. This indicates that few thousands Dofs per atom are sufficient to reach a high numerical accuracy using the presented $h$-adaptive isogeometric solver, thereby demonstrating the potential of our solver for addressing large-scale all-electron KS equation.

\begin{figure}[h!]
\centering
    \begin{minipage}[h]{0.49\textwidth}
        \includegraphics[width=1\linewidth]{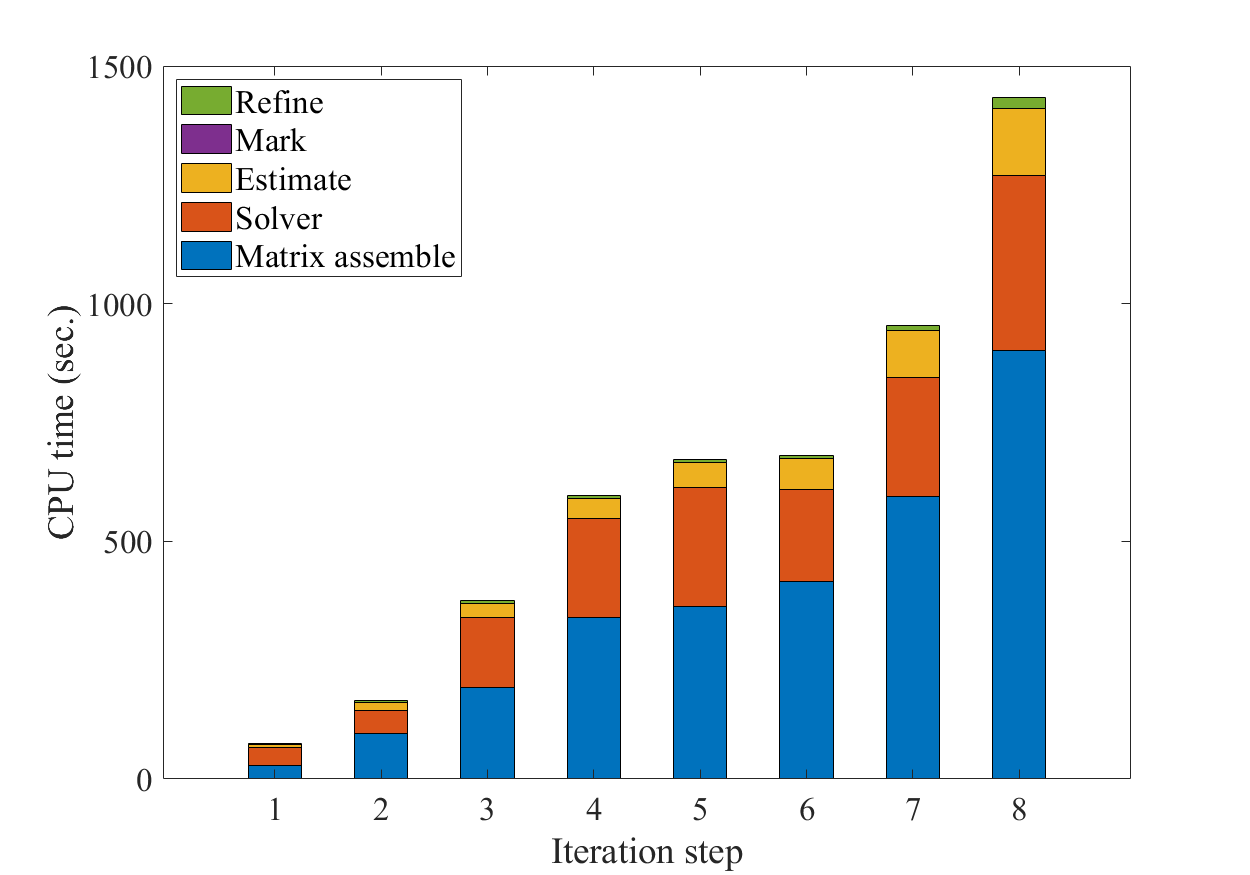}
    \end{minipage}
    \begin{minipage}[h]{0.49\textwidth}
        \includegraphics[width=1\linewidth]{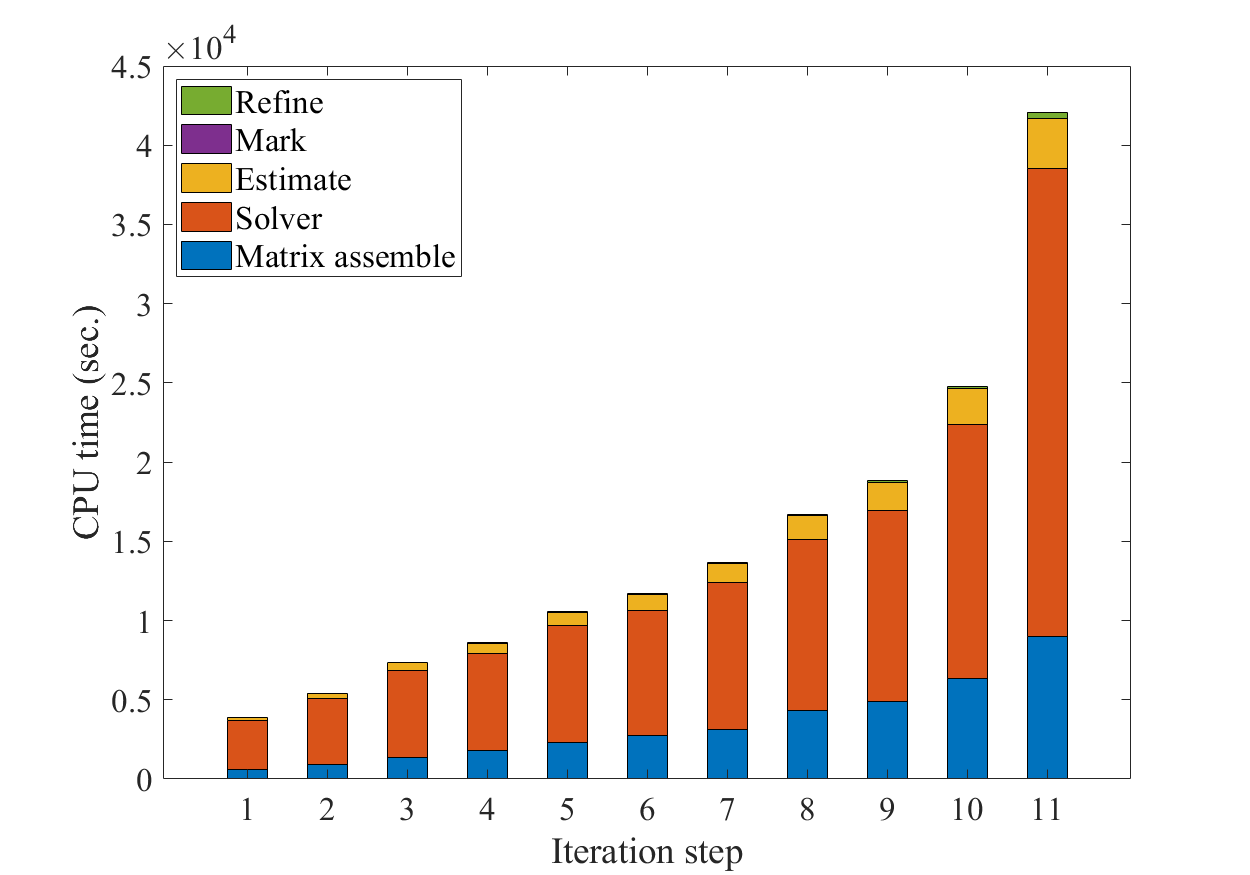}
    \end{minipage}
\caption{The CPU time for simulating a methane molecule (left) and a benzene molecule (right) on five parts ($p=3$).}
\label{fig:CH4_benzene_cpu_time}
\end{figure}

However, computation time remains a significant challenge in our experiment due to the dense stiffness matrix associated with the high-order basis. A detailed comparison of CPU time for iterative steps in the adaptive algorithm, based on the numerical simulations of the methane molecule and the benzene molecule, is presented in \Cref{fig:CH4_benzene_cpu_time}. The CPU time is categorized into five components based on the four modules of \Cref{alg:adaptivemesh}: Matrix assemble, Solver, Estimate, Mark, and Refine, where the Solver records the CPU time on solving the eigenvalue system. In the left of \Cref{fig:CH4_benzene_cpu_time}, as the adaptive mesh for the methane molecule is refined, although Matrix assemble accounts for a significant portion of the computational cost, the Solver's contribution is non-negligible since the Solver needs to solve five occupied orbitals. Furthermore, it is evident that Solver dominates the CPU time, primarily due to the requirement to approximate the wavefunctions of $21$ occupied orbitals for the benzene molecule, as demonstrated in the right of \Cref{fig:CH4_benzene_cpu_time}. Therefore, for larger systems, we anticipate that solving the eigenvalue system will become the most time-consuming component. This highlights the necessity of developing an efficient and robust eigenvalue solver specifically tailored for high-order splines such as the multigrid solver \cite{hofreither2016multigridforTHB,de2019robustmgiga} to address such challenges effectively. Moreover, we aim to apply our adaptive framework to tackle more challenging practical issues, including the large-scale all-electron Kohn--Sham equation and time-dependent Kohn--Sham equation \cite{marques2006timedependentDFT}, which will be a key focus of our future work.

\section*{Acknowledgement}
The authors thank Prof. Tao Tang (Division of Science and Technology, BNU-HKBU United International
College) for his valuable guidance and insightful suggestions on this work. The work of Y. Kuang is supported by the National Natural Science Foundation of China (Nos. 12326362 and 12201130) and the Guangzhou Municipal Science and Technology Bureau (No. 2023A04J1321). The work of G. Hu was supported by The Science and Technology Development Fund, Macao SAR (No. 0068/2024/RIA1), National Natural Science Foundation of China (No. 11922120), MYRG of University of Macau (No. MYRG-GRG2023-00157-FST-UMDF). The work of R. Zhang was supported in part by the Natural National Science Foundation of China (Grant No. 22341302), the National Key Research and Development Program of China (Grant No. 2020YFA0713602, 2023YFA1008803), and the Key Laboratory of Symbolic Computation and Knowledge Engineering of Ministry of Education of China housed at Jilin University.


\bibliographystyle{plain}
\bibliography{manuscript}


\appendix
\newpage
\begin{table}[h!]
    \centering
    \caption{The number of used Dofs for $p=3$ to reproduce the energies in the NIST database \cite{kotochigova2009atomicnist}. The atomic number ranges from $1$ to $92$.}
    \begin{adjustbox}{angle = 90}
    \newcommand{\ElemLabel}[4]{
  \begin{minipage}{2.6cm}
    \centering
      {\hfill \textbf{#2}}%
      \linebreak \linebreak
      {\textbf{#3}}%
      \linebreak \linebreak
      {{#4}}
  \end{minipage}
}

\newcommand{\NaturalElem}[4]{\ElemLabel{#1}{#2}{\Huge {#3}}{\large{#4}}}

\newcommand{\SyntheticElem}[4]{\ElemLabel{#1}{#2}{\color{gray}{\huge #3}}{#4}}
\resizebox{.9\textheight}{!}{
\begin{tikzpicture}[font=\sffamily]

  \tikzstyle{ElementFill} = [fill=yellow!15]
  \tikzstyle{AlkaliMetalFill} = [fill=blue!55]
  \tikzstyle{AlkalineEarthMetalFill} = [fill=blue!40]
  \tikzstyle{MetalFill} = [fill=blue!25]
  \tikzstyle{MetalloidFill} = [fill=orange!40]
  \tikzstyle{NonmetalFill} = [fill=teal!40]
  \tikzstyle{HalogenFill} = [fill=yellow!40]
  \tikzstyle{NobleGasFill} = [fill=green!55]
  \tikzstyle{LanthanideActinideFill} = [fill=red!40]

  \tikzstyle{Element} = [ElementFill,
    minimum width=2.6cm, minimum height=2.6cm, node distance=3.05cm]
  \tikzstyle{AlkaliMetal} = [Element, AlkaliMetalFill]
  \tikzstyle{AlkalineEarthMetal} = [Element, AlkalineEarthMetalFill]
  \tikzstyle{Metal} = [Element, MetalFill]
  \tikzstyle{Metalloid} = [Element, MetalloidFill]
  \tikzstyle{Nonmetal} = [Element, NonmetalFill]
  \tikzstyle{Halogen} = [Element, HalogenFill]
  \tikzstyle{NobleGas} = [Element, NobleGasFill]
  \tikzstyle{LanthanideActinide} = [Element, LanthanideActinideFill]
  \tikzstyle{PeriodLabel} = [font={\sffamily\LARGE}, node distance=2.0cm]
  \tikzstyle{GroupLabel} = [font={\sffamily\LARGE}, minimum width=2.75cm, node distance=2.0cm]

  \node[Element] (H) {\NaturalElem{1} {31}{H}{-0.445671}};
  \node[below of=H, AlkaliMetal] (Li) {\NaturalElem{3}{42}{Li}{-7.335195}};
  \node[below of=Li, AlkaliMetal] (Na) {\NaturalElem{11}{57}{Na}{-161.440060}};
  \node[below of=Na, AlkaliMetal] (K) {\NaturalElem{19}{100}{K}{-598.200590}};
  \node[below of=K, AlkaliMetal] (Rb) {\NaturalElem{37}{139}{Rb}{-2936.337293}};
  \node[below of=Rb, AlkaliMetal] (Cs) {\NaturalElem{55}{179}{Cs}{-7550.557710}};
  \node[below of=Cs, AlkaliMetal] (Fr) {\NaturalElem{87}{204}{Fr}{-22470.319655}};

  \node[right of=Li, AlkalineEarthMetal] (Be) {\NaturalElem{4}{52}{Be}{-14.447209}};
  \node[below of=Be, AlkalineEarthMetal] (Mg) {\NaturalElem{12}{70}{Mg}{-199.139406}};
  \node[below of=Mg, AlkalineEarthMetal] (Ca) {\NaturalElem{20}{115}{Ca}{-675.742283}};
  \node[below of=Ca, AlkalineEarthMetal] (Sr) {\NaturalElem{38}{156}{Sr}{-3129.453161}};
  \node[below of=Sr, AlkalineEarthMetal] (Ba) {\NaturalElem{56}{211}{Ba}{-7880.111578}};
  \node[below of=Ba, AlkalineEarthMetal] (Ra) {\NaturalElem{88}{237}{Ra}{-23088.688083}};

  \node[right of=Ca, Metal] (Sc) {\NaturalElem{21}{121}{Sc}{-758.679275}};
  \node[below of=Sc, Metal] (Y) {\NaturalElem{39}{174}{Y}{-3329.520604}};
  \node[below of=Y, LanthanideActinide] (LaLu) {\NaturalElem{57-71}{}{La-Lu}{}};

  \node[right of=Sc, Metal] (Ti) {\NaturalElem{22}{112}{Ti}{-847.277216}};
  \node[below of=Ti, Metal] (Zr) {\NaturalElem{40}{196}{Zr}{-3536.737751}};
  \node[below of=Zr, Metal] (Hf) {\NaturalElem{72}{182}{Hf}{-14317.493720}};

  \node[right of=Ti, Metal] (V) {\NaturalElem{23}{140}{V}{-941.678904	}};
  \node[below of=V, Metal] (Nb) {\NaturalElem{41}{181}{Nb}{-3751.196175}};
  \node[below of=Nb, Metal] (Ta) {\NaturalElem{73}{176}{Ta}{-14795.884202}};

  \node[right of=V, Metal] (Cr) {\NaturalElem{24}{132}{Cr}{-1042.030238}};
  \node[below of=Cr, Metal] (Mo) {\NaturalElem{42}{164}{Mo}{-3973.013235}};
  \node[below of=Mo, Metal] (W) {\NaturalElem{74}{183}{W}{-15283.448822}};

  \node[right of=Cr, Metal] (Mn) {\NaturalElem{25}{131}{Mn}{-1148.449372}};
  \node[below of=Mn, Metal] (Tc) {\NaturalElem{43}{185}{Tc}{-4202.188857}};
  \node[below of=Tc, Metal] (Re) {\NaturalElem{75}{186}{Re}{-15780.236024}};

  \node[right of=Mn, Metal] (Fe) {\NaturalElem{26}{135}{Fe}{-1261.093056}};
  \node[below of=Fe, Metal] (Ru) {\NaturalElem{44}{179}{Ru}{-4438.981228}};
  \node[below of=Ru, Metal] (Os) {\NaturalElem{76}{185}{Os}{-16286.295408}};

  \node[right of=Fe, Metal] (Co) {\NaturalElem{27}{132}{Co}{-1380.091264}};
  \node[below of=Co, Metal] (Rh) {\NaturalElem{45}{191}{Rh}{-4683.301031}};
  \node[below of=Rh, Metal] (Ir) {\NaturalElem{77}{183}{Ir}{-16801.677471}};

  \node[right of=Co, Metal] (Ni) {\NaturalElem{28}{156}{Ni}{-1505.580197}};
  \node[below of=Ni, Metal] (Pd) {\NaturalElem{46}{213}{Pd}{-4935.368406}};
  \node[below of=Pd, Metal] (Pt) {\NaturalElem{78}{226}{Pt}{-17326.576377}};

  \node[right of=Ni, Metal] (Cu) {\NaturalElem{29}{210}{Cu}{-1637.785861}};
  \node[below of=Cu, Metal] (Ag) {\NaturalElem{47}{181}{Ag}{-5195.031215}};
  \node[below of=Ag, Metal] (Au) {\NaturalElem{79}{187}{Au}{-17860.790943}};

  \node[right of=Cu, Metal] (Zn) {\NaturalElem{30}{157}{Zn}{-1776.573850}};
  \node[below of=Zn, Metal] (Cd) {\NaturalElem{48}{197}{Cd}{-5462.390982}};
  \node[below of=Cd, Metal] (Hg) {\NaturalElem{80}{209}{Hg}{-18404.274220}};

  \node[right of=Zn, Metal] (Ga) {\NaturalElem{31}{145}{Ga}{-1921.846456}};
  \node[above of=Ga, Metal] (Al) {\NaturalElem{13}{98}{Al}{-241.315573}};
  \node[above of=Al, Metalloid] (B) {\NaturalElem{5}{68}{B}{-24.344198}};
  \node[below of=Ga, Metal] (In) {\NaturalElem{49}{231}{In}{-5737.309064}};
  \node[below of=In, Metal] (Tl) {\NaturalElem{81}{237}{Tl}{-18956.957627}};

  \node[right of=B, Nonmetal] (C) {\NaturalElem{6}{79}{C}{Carbon}};
  \node[below of=C, Metalloid] (Si) {\NaturalElem{14}{105}{Si}{-288.198397}};
  \node[below of=Si, Metalloid] (Ge) {\NaturalElem{32}{155}{Ge}{-2073.807332}};
  \node[below of=Ge, Metal] (Sn) {\NaturalElem{50}{241}{Sn}{-6019.953353}};
  \node[below of=Sn, Metal] (Pb) {\NaturalElem{82}{253}{Pb}{-19518.993145}};

  \node[right of=C, Nonmetal] (N) {\NaturalElem{7}{72}{N}{-54.025016}};
  \node[below of=N, Nonmetal] (P) {\NaturalElem{15}{86}{P}{-339.946219}};
  \node[below of=P, Metalloid] (As) {\NaturalElem{33}{142}{As}{-2232.534978}};
  \node[below of=As, Metalloid] (Sb) {\NaturalElem{51}{230}{Sb}{-6310.376268}};
  \node[below of=Sb, Metal] (Bi) {\NaturalElem{83}{217}{Bi}{-20090.414449}};

  \node[right of=N, Nonmetal] (O) {\NaturalElem{8}{62}{O}{-74.473077}};
  \node[below of=O, Nonmetal] (S) {\NaturalElem{16}{87}{S}{-396.716081}};
  \node[below of=S, Nonmetal] (Se) {\NaturalElem{34}{140}{Se}{-2398.111440}};
  \node[below of=Se, Metalloid] (Te) {\NaturalElem{52}{198}{Te}{-6608.631413}};
  \node[below of=Te, Metalloid] (Po) {\NaturalElem{84}{219}{Po}{-6608.631413}};

  \node[right of=O, Halogen] (F) {\NaturalElem{9}{77}{F}{-99.099648}};
  \node[below of=F, Halogen] (Cl) {\NaturalElem{17}{88}{Cl}{-458.664179}};
  \node[below of=Cl, Halogen] (Br) {\NaturalElem{35}{161}{Br}{-2570.620700}};
  \node[below of=Br, Halogen] (I) {\NaturalElem{53}{230}{I}{-6914.773092}};
  \node[below of=I, Halogen] (At) {\NaturalElem{85}{297}{At}{-21261.555215}};

  \node[right of=F, NobleGas] (Ne) {\NaturalElem{10}{69}{Ne}{-14.447209}};
  \node[above of=Ne, NobleGas] (He) {\NaturalElem{2}{43}{He}{-2.834836}};
  \node[below of=Ne, NobleGas] (Ar) {\NaturalElem{18}{107}{Ar}{-525.946195}};
  \node[below of=Ar, NobleGas] (Kr) {\NaturalElem{36}{138}{Kr}{-2750.147940}};
  \node[below of=Kr, NobleGas] (Xe) {\NaturalElem{54}{202}{Xe}{-7228.856107}};
  \node[below of=Xe, NobleGas] (Rn) {\NaturalElem{86}{292}{Rn}{-21861.346869}};
  
  \node[left of=H, PeriodLabel] (Period1) {1};
  \node[left of=Li, PeriodLabel] (Period2) {2};
  \node[left of=Na, PeriodLabel] (Period3) {3};
  \node[left of=K, PeriodLabel] (Period4) {4};
  \node[left of=Rb, PeriodLabel] (Period5) {5};
  \node[left of=Cs, PeriodLabel] (Period6) {6};
  \node[left of=Fr, PeriodLabel] (Period7) {7};

  \node[above of=H, GroupLabel] (Group1) {1 \hfill IA};
  \node[above of=Be, GroupLabel] (Group2) {2 \hfill IIA};
  \node[above of=Sc, GroupLabel] (Group3) {3 \hfill IIIB};
  \node[above of=Ti, GroupLabel] (Group4) {4 \hfill IVB};
  \node[above of=V, GroupLabel] (Group5) {5 \hfill VB};
  \node[above of=Cr, GroupLabel] (Group6) {6 \hfill VIB};
  \node[above of=Mn, GroupLabel] (Group7) {7 \hfill VIIB};
  \node[above of=Fe, GroupLabel] (Group8) {8 \hfill VIIIB};
  \node[above of=Co, GroupLabel] (Group9) {9 \hfill VIIIB};
  \node[above of=Ni, GroupLabel] (Group10) {10 \hfill VIIIB};
  \node[above of=Cu, GroupLabel] (Group11) {11 \hfill IB};
  \node[above of=Zn, GroupLabel] (Group12) {12 \hfill IIB};
  \node[above of=B, GroupLabel] (Group13) {13 \hfill IIIA};
  \node[above of=C, GroupLabel] (Group14) {14 \hfill IVA};
  \node[above of=N, GroupLabel] (Group15) {15 \hfill VA};
  \node[above of=O, GroupLabel] (Group16) {16 \hfill VIA};
  \node[above of=F, GroupLabel] (Group17) {17 \hfill VIIA};
  \node[above of=He, GroupLabel] (Group18) {18 \hfill VIIIA};

  \node[below of=LaLu, LanthanideActinide] (Ac) {\NaturalElem{89}{216}{Ac}{-23716.496952}};
  \node[right of=Ac, LanthanideActinide] (Th) {\NaturalElem{90}{227}{Th}{-24353.832231}};
  \node[right of=Th, LanthanideActinide] (Pa) {\NaturalElem{91}{261}{Pa}{-25001.291382}};
  \node[right of=Pa, LanthanideActinide] (U) {\NaturalElem{92}{250}{U}{-25658.417889}};

  \node[below of=Th, LanthanideActinide, yshift=-1cm] (La) {\NaturalElem{57}{225}{La}{-8217.575230}};
  \node[right of=La, LanthanideActinide] (Ce) {\NaturalElem{58}{229}{Ce}{-8563.360285}};
  \node[right of=Ce, LanthanideActinide] (Pr) {\NaturalElem{59}{222}{Pr}{-8917.664369}};
  \node[right of=Pr, LanthanideActinide] (Nd) {\NaturalElem{60}{224}{Nd}{-9280.311037}};
  \node[right of=Nd, LanthanideActinide] (Pm) {\NaturalElem{61}{199}{Pm}{-9651.484134}};
  \node[right of=Pm, LanthanideActinide] (Sm) {\NaturalElem{62}{198}{Sm}{-10031.259090}};
  \node[right of=Sm, LanthanideActinide] (Eu) {\NaturalElem{63}{227}{Eu}{-10419.710775}};
  \node[right of=Eu, LanthanideActinide] (Gd) {\NaturalElem{64}{222}{Gd}{-10816.653877}};
  \node[right of=Gd, LanthanideActinide] (Tb) {\NaturalElem{65}{217}{Tb}{-11222.941975}};
  \node[right of=Tb, LanthanideActinide] (Dy) {\NaturalElem{66}{198}{Dy}{-11637.869664}};
  \node[right of=Dy, LanthanideActinide] (Ho) {\NaturalElem{67}{200}{Ho}{-12061.770549}};
  \node[right of=Ho, LanthanideActinide] (Er) {\NaturalElem{68}{206}{Er}{-12494.718304}};
  \node[right of=Er, LanthanideActinide] (Tm) {\NaturalElem{69}{256}{Tm}{-12936.786494}};
  \node[right of=Tm, LanthanideActinide] (Yb) {\NaturalElem{70}{201}{Yb}{-13388.048594}};
  \node[right of=Yb, LanthanideActinide] (Lu) {\NaturalElem{71}{251}{Lu}{-13848.230375}};

  \draw[thick,dotted] (LaLu.north east) -- (La.north west)
        (LaLu.south west) -- (La.south west);
   \fill[AlkaliMetalFill] ($(La.north -| Ra.west) + (0,1em)$)
     rectangle +(1em, 1em) node[right, yshift=-1.2ex]  (AlkaliMetal) {Alkali Metal};
   \fill[AlkalineEarthMetalFill] ($(AlkaliMetal.west) - (1em,2em)$)
     rectangle +(1em, 1em) node[right, yshift=-1.2ex] (AlkalineEarthMetal) {Alkaline Earth Metal};
   \fill[MetalFill] ($(AlkalineEarthMetal.west) - (1em,2em)$)
     rectangle +(1em, 1em) node[right, yshift=-1.2ex] (Metal) {Metal};
   \fill[MetalloidFill] ($(Metal.west) - (1em,2em)$)
     rectangle +(1em, 1em) node[right, yshift=-1.2ex] (Metalloid) {Metalloid};
   \fill[NonmetalFill] ($(Metalloid.west) - (1em,2em)$)
     rectangle +(1em, 1em) node[right, yshift=-1.2ex] (Non-metal) {Non-metal};
   \fill[HalogenFill] ($(Non-metal.west) - (1em,2em)$)
     rectangle +(1em, 1em) node[right, yshift=-1.2ex] (Halogen) {Halogen};
   \fill[NobleGasFill] ($(Halogen.west) - (1em,2em)$)
     rectangle +(1em, 1em) node[right, yshift=-1.2ex] (NobleGas) {Noble Gas};
   \fill[LanthanideActinideFill] ($(NobleGas.west) - (1em,2em)$)
     rectangle +(1em, 1em) node[right, yshift=-1.2ex] (Lanthanide/Actinide) {Lanthanide/Actinide};

  \node at (La -| Fr) [draw, Element, fill=white] (legend) {\NaturalElem{Z}{$N_{\mathrm{Dof}}$}{\LARGE Symbol}{Energy}};
\end{tikzpicture}}
    \end{adjustbox}
    \label{tab:Radial_periodic_table}
\end{table}
\end{document}